\documentclass[usenatbib]{mn2e}
\usepackage{times}
\usepackage{graphicx}
\usepackage{dcolumn}
\usepackage{lscape}
\usepackage{amsmath}
\usepackage{amssymb}
\usepackage{mathptm}
\usepackage{epsfig}
\usepackage{psfrag}
\usepackage{natbib}
\usepackage{longtable}
\begin{document}
\newcolumntype{d}[1]{D{.}{.}{#1}}
\def\dhead#1{\multicolumn{1}{c}{#1}}
\title[Observed properties of FRII quasars and radio galaxies at $z < 1.0$]{Observed properties of FRII quasars and radio galaxies at $z < 1.0$}
\author[L. M. Mullin, J. M. Riley and M. J. Hardcastle]{L. M. Mullin$^{1}$\thanks{E-mail:
mullin@extragalactic.info (LMM)}, J. M. Riley$^{1}$\thanks{E-mail: julia@mrao.cam.ac.uk (JMR)}  and M. J. Hardcastle$^{2}$\thanks{m.j.hardcastle@herts.ac.uk (MJH)}\\
$^{1}$Astrophysics Group, Cavendish Laboratory, University of Cambridge, J J Thomson Avenue, Cambridge CB3 0HE\\
$^{2}$School of Physics, Astronomy and Mathematics, University of Hertfordshire, College Lane, Hatfield AL10 9AB}

\date{}

\pagerange{\pageref{firstpage}--\pageref{lastpage}} \pubyear{2008}

\maketitle

\label{firstpage}

\begin{abstract}
In a long-term observing project we have imaged a complete sample of
FRII quasars and radio galaxies with $z < 1.0$ at high resolution and
high sensitivity with the VLA and MERLIN. This sample of 98 sources
includes 15 quasars, 11 broad line radio galaxies and 57 narrow line
radio galaxies, allowing unification to be considered in terms of
source morphological properties. Radio maps of all the targets have
been presented in earlier papers. Here we carry out a systematic
analysis of the properties of the jets, cores, lobes and hotspots of
objects in the sample. The majority of the tests that we perform show
that the data are consistent with a model in which quasars and
broad-line radio galaxies are unified with narrow-line objects.
Relativistic beaming is the main effect that determines the properties
of kiloparsec-scale jets, and it may also have some effect on
hotspots. However, some properties of the sample are difficult to
account for in simple unified models.
\end{abstract}

%
%
\section{Introduction}\label{sec:intro}
Fanaroff \& Riley (1974) type II radio sources (hereafter FRIIs) are
powerful sources associated with bipolar outflows that extend great
distances from the central engine, remaining highly collimated as they
do so. They can be divided into different classes based on features of
their optical spectra: radio-loud quasars (Qs), broad line radio
galaxies (BLRGs), narrow line radio galaxies (NLRGs) and low
excitation radio galaxies (LERGs) can all be FRIIs. A principal
defining characteristic is the presence, or absence, of broad line
emission, with the Qs and BLRGs having both broad and narrow line
emission lines, the NLRGs having narrow line emission only and the
LERGs lacking strong high-excitation lines of either type (Hine \&
Longair 1979; Laing et al. 1994).

The current standard unification scheme proposes that the Qs, BLRGs
and (at least some of) the NLRGs are intrinsically part of the same
population (Scheuer 1987; Barthel 1987, 1989). In this model, the
broad emission line region lies close to the very compact central
engine and is surrounded by a dusty torus, whereas the narrow line
emission region lies further out. Sources that are viewed along or
close to the axis of the torus show both broad and narrow line
emission -- these are the Qs and BLRGs, which we refer to collectively
as broad-line objects -- but the broad line emission
region is obscured for those that are oriented closer to the plane
of the sky, the NLRGs. Thus differences in the orientation of the source
axis to the observer's line of sight are the origin of the three
spectral classes. The LERGs lie outside of this scheme; it has been
suggested (e.g. Barthel 1994) that LERGs form part of the parent
population of BL Lac objects rather than core-dominated Qs and should
not show broad line emission at any angle to the line-of-sight, a
model consistent with their nuclear properties at other wavebands
(Chiaberge, Capetti \& Celotti 2002; Hardcastle, Evans \& Croston 2006).

An important detail of the model arises from the fact that the
observed luminosity distributions of Qs and BLRGs are not the same. Qs
are more powerful and found at higher redshifts (or, equivalently in a
flux-limited sample, higher radio luminosities) than the BLRGs; for
example, in the 3CR sample (Bennett 1962) Qs are found only with $z \gtrsim 0.3$,
while BLRGs are found with $z \lesssim 0.3$. It has been suggested
that BLRGs may be the low-luminosity equivalents of Qs, or that they lie
near the critical angle dividing the quasars and radio galaxies
(Barthel 1989; Hardcastle et al. 1998, hereafter H98). While some
high-luminosity BLRGs may indeed be intermediate objects, it is clear
that at low luminosity, where there are no Qs, BLRGs are the only
candidate for the aligned counterpart of the population of
low-luminosity NLRGs.

Often FRIIs exhibit a bright linear feature called a jet that extends
at least some of the distance between the central feature, the core,
and the bright hotspot at the end of the lobe. The jets in most FRIIs
are one-sided: either no counterjet is seen or it is much fainter than
the feature that is identified as the jet. Relativistic beaming of the
jet emission is invoked to explain this asymmetry, as the large scale
lobe morphology appears otherwise roughly symmetric. The jet detection
rate is higher for Qs and BLRGs than for NLRGs; this can be explained
in unified models, since for the broad-line objects the beamed jet is
aligned closer to the line of sight and appears brighter. The jet
detection rate for LERGs is the highest of all classes (e.g. Mullin,
Hardcastle \& Riley 2006) which may be related to systematic
environmental differences between some of the LERGs and the other
emission-line types (Hardcastle 2004).

Further evidence in support of relativistic beaming in jets is
provided by the Laing-Garrington effect (Laing 1988; Garrington et al.
1988), which is the association of the jet side with the less
depolarized lobe. High-resolution multi-frequency observations
indicate that the depolarization occurs in an external Faraday
screen, so that the less depolarized lobe is expected to be the lobe
closer to us; any tendency for the (brighter) jet to be associated
with this lobe then implies that beaming is an important factor in jet
detection (Scheuer 1987).

While various aspects of the unification and beaming model have been
tested and discussed in the literature, there has been little work
using complete samples of radio sources free from orientation bias
that include sufficient numbers of objects of all spectral classes to
give statistically significant results. Good quality observations of
such a sample, with both high resolution and sensitivity, are
therefore vital, and this has been the rationale behind a long-term
observational project in which we have mapped a complete sample of the
brightest FRII radio sources with $z<1.0$. The sample, which is
defined in section \ref{sec:data}, includes 98 sources. Maps of these
have been presented in a series of papers: Black et al. (1992), Leahy
et al. (1997), Hardcastle et al. (1997), Gilbert et al. (2004) and
Mullin, Hardcastle \& Riley (2006). These maps are available
online\footnote{See http://zl1.extragalactic.info/} along with a
database of all measurements analysed and discussed in this paper. The
sample includes 15 Qs, 11 BLRGs and 57 NLRGs, thus enabling aspects of
unification to be tested along with trends in source properties over
the wide redshift and luminosity range spanned by the data. In section \ref{sec:data} we also define a number of morphological and
flux parameters corresponding to the observed source properties and
describe our measurement methods.

We examine the properties of the lobes, cores and jets and hotspots in
sections \ref{sec:lobes}, \ref{sec:cores_jets} and \ref{sec:hotspots}
respectively. For each feature, observational effects are considered
as well as trends across the power, redshift and source size range of
the sample and we quantify these where appropriate with statistical
tests. The significance of linear correlation is tested for using
Spearman's rank correlation coefficient. For the core and jet
prominence data, however, this is not possible as only upper limits on
these parameters are available for some sources: in statistical
terminology, the data are censored. Instead, a modified Kendall's
$\tau$ rank correlation coefficient as implemented in the
survival-analysis package {\sc asurv} (LaValley, Isobe \& Feigelson
1992) is used for these data. The Kolmogorov-Smirnov (hereafter K-S)
test determines if it is the case that the culmulative distribution
function of two samples differ and is used to address the question of
whether some property of two subsamples of the data (that is,
subsamples defined by power, redshift and size cutoffs or by spectral
class) differ significantly. It is sensitive to differences in both
location and shape of the functions. No modification of the K-S test
to take account of censoring is available to us, and so we do not use
it in situations where censoring is important. The
Wilcoxon-Mann-Whitney (hereafter W-M-W) test is also used to determine
if two defined subsamples differ, but in this case the null hypothesis
tested is that the probability of an observation of one population
exceeding an observation from the second is 0.5. Thus the W-M-W test
is used to determine whether there is a significant difference in the
magnitude of the quantity of interest between the two subsamples, that
is, if one dataset has significantly smaller or larger values than the
other. In order to treat censored data correctly when testing for such
differences, a generalized W-M-W test is used, the Peto-Prentice test,
which is implemented in {\sc asurv}. Finally, the binomial test is
used to determine the statistical significance of any correlation with
jet or longer lobe side for a number of properties. The significance
of all test results is discussed in the text and the results are
tabulated. We take a result to be significant enough to be discussed
if the null hypothesis is rejected at better than the 95 per cent confidence level.

The interpretation of the observed properties of our sample sources,
and the evidence for and against unified models, is discussed in
section \ref{sec:discussion}. The quantitative implications of our
results for beaming in the cores and jets of powerful radio galaxies
will be discussed in a separate paper.

The spectral index, $\alpha$, is defined throughout the paper in the sense that
$S = v^{-\alpha}$ (where $S$ is the flux and $v$ denotes the frequency) and  we
assume that $H_{0} = 70\ {\rm kms^{-1}Mpc^{-1}}$, $\Omega_{\rm
m}=0.3$ and $\Omega_{\Lambda}=0.7$.

%
%
\section{The Data}\label{sec:data}
%
%
\subsection{The Sample}\label{sec:sample}
%
%
The sample is selected from the complete flux-limited sample of
Laing, Riley \& Longair (1983, hereafter LRL), which is itself based on the
3CR survey. The LRL sample includes all the sources with total source flux densities measured at 178 MHz $S_{178} > 10.9$ Jy (on the scale
of Baars et al. 1977) with $\rm dec > 10^\circ$ and $ |b| >
10^\circ$. At this low frequency the source flux is dominated by the
emission from the large-scale lobe structure, so that little
contribution should be made by Doppler-boosted components, which
should ensure the sample is as free as possible from orientation bias. There are 173 LRL sources in total, including 29 FRI and
125 FRII objects. All 98 FRII radio galaxies and
quasars with $z < 1.0$ are listed in Table \ref{tab:the_sample}, which includes references to all the radio maps from which the data analysed in this paper have been obtained. 
%
%
\begin{table*}
\centering
\begin{minipage}{14cm}
\caption{The sample.}\label{tab:the_sample}
\begin{tabular}{lcccccccd{2}c}  \hline
Source & IAU Name & RA    & Dec & $z$ & Spectral & $S_{178}$ & $\alpha$ &\multicolumn{2}{c}{Maps}\\
       &          &[h m s]&$\left[\ ^{\circ} \ ' \ ''\ \right]$&     & class     &[Jy]&\tiny{($178-750\ \rm{MHz}$)}&\dhead{Freq. [GHz]} & Ref. \\ \hline
4C12.03 & 0007+124 & 00 07 18.25 & +12 27 23.1 & 0.156& L & 10.9 & 0.87 & 1.5  & 1      \\
3C6.1   & 0013+790 & 00 13 34.36 & +79 00 11.1 & 0.840& N & 13.7 & 0.68 & 8.4  & 2      \\
3C16    & 0035+130 & 00 35 09.16 & +13 03 39.6 & 0.405& L & 12.2 & 0.94 & 8.4  & 3      \\
3C19    & 0038+328 & 00 38 13.80 & +32 53 39.7 & 0.482& N & 13.2 & 0.63 & 4.5  & 3      \\
3C20 	& 0040+517 & 00 40 20.08 & +51 47 10.2 & 0.174& N & 46.8 & 0.66 & 8.4  & 4      \\
3C22 	& 0048+509 & 00 48 04.71 & +50 55 45.4 & 0.937& B & 12.1 & 0.78 & 8.5  & 2      \\
3C33 	& 0106+130 & 01 06 14.54 & +13 04 14.8 & 0.060& N & 59.3 & 0.76 & 1.5  & 1      \\
        &          &             &             &      &   &      &      & 4.8  & 5,6    \\
3C33.1 	& 0106+729 & 01 06 06.48 & +72 55 59.2 & 0.181& B & 14.2 & 0.62 & 4.9  & 7      \\
3C34 	& 0107+315 & 01 07 32.51 & +31 31 23.9 & 0.690& N & 11.9 & 1.06 & 4.8  & 2,8    \\
3C35 	& 0109+492 & 01 09 04.94 & +49 12 40.1 & 0.068& L & 11.4 & 0.77 & 0.61 & 9      \\
3C41 	& 0123+329 & 01 23 54.74 & +32 57 38.3 & 0.794& N & 10.6 & 0.51 & 8.5  & 2      \\
3C42 	& 0125+287 & 01 25 42.68 & +28 47 30.4 & 0.395& N & 13.1 & 0.73 & 8.5  & 3      \\
3C46 	& 0132+376 & 01 32 34.09 & +37 38 47.0 & 0.437& N & 11.1 & 1.13 & 8.5  & 3      \\
3C47 	& 0133+207 & 01 33 40.43 & +20 42 10.2 & 0.425& Q & 28.8 & 0.98 & 4.9  & 3,10   \\
3C55 	& 0154+286 & 01 54 19.50 & +28 37 04.8 & 0.735& N & 21.5 & 1.04 & 4.8  & 2,8    \\
3C61.1 	& 0210+860 & 02 10 37.10 & +86 05 18.5 & 0.186& N & 34.0 & 0.77 & 1.5  & 9      \\
        &          &             &             &      &   &      &      & 4.9  & 7      \\   
3C67 	& 0221+276 & 02 21 18.03 & +27 36 37.2 & 0.310& B & 10.9 & 0.58 & 4.8  & 3      \\
        &          &             &             &      &   &      &      & 8.4  & 11     \\ 
3C79 	& 0307+169 & 03 07 11.48 & +16 54 36.9 & 0.256& N & 33.2 & 0.92 & 8.4 & 4       \\
3C98 	& 0356+102 & 03 56 10.21 & +10 17 31.7 & 0.031& N & 51.4 & 0.78 & 8.4 & 12      \\
3C109 	& 0410+110 & 04 10 54.87 & +11 04 41.4 & 0.306& B & 23.5 & 0.85 & 8.3 & 3\\
4C14.11 & 0411+141 & 04 11 40.94 & +14 08 48.3 & 0.206& L & 12.1 & 0.84 & 8.4 & 4\\
3C123 	& 0433+295 & 04 33 55.21 & +29 34 12.6 & 0.218& L & 206.0& 0.70	& 8.4 & 4\\
3C132 	& 0453+227 & 04 53 42.18 & +22 44 43.9 & 0.214& L & 14.9 & 0.68	& 8.4 & 4\\
3C153 	& 0605+480 & 06 05 44.44 & +48 04 48.8 & 0.277& N & 16.7 & 0.66 & 8.4 & 4\\
3C171 	& 0651+542 & 06 51 10.83 & +54 12 47.6 & 0.238& N & 21.3 & 0.87	& 8.1 & 4\\
3C172 	& 0659+253 & 06 59 03.90 & +25 18 12.0 & 0.519& N & 16.5 & 0.86 & 8.5 & 3\\ 
3C173.1 & 0702+749 & 07 02 47.91 & +74 54 16.6 & 0.292& L & 16.8 & 0.88 & 8.4 & 4\\
3C175 	& 0710+118 & 07 10 15.38 & +11 51 24.0 & 0.768& Q & 17.6 & 0.98 & 8.4 & 2\\
3C175.1 & 0711+146 & 07 11 14.28 & +14 41 33.9 & 0.920& N & 11.4 & 0.91 & 4.9 & 2\\
3C184 	& 0733+705 & 07 33 59.01 & +70 30 01.1 & 0.990& N & 13.2 & 0.86 & 4.9 & 2\\
3C184.1 & 0734+805 & 07 34 25.05 & +80 33 24.1 & 0.119& N & 14.2 & 0.68 & 8.4 & 12\\
DA240 	& 0745+560 & 07 44 34.96 & +55 56 29.0 & 0.036& L & 23.2 & 0.77 & 0.61 & 9\\
3C192 	& 0802+243 & 08 02 35.50 & +24 18 26.4 & 0.060& N & 23.0 & 0.79 & 8.2 & 12\\
3C196 	& 0809+483 & 08 09 59.40 & +48 22 07.6 & 0.871& Q & 68.2 & 0.79 & 4.9 & 2\\
3C200 	& 0824+294 & 08 24 21.43 & +29 28 42.2 & 0.458& N & 12.3 & 0.84 & 8.5 & 3\\
4C14.27 & 0832+143 & 08 32 16.51 & +14 22 12.1 & 0.392& N & 11.2 & 1.15 & 8.5 & 3\\
3C207 	& 0838+133 & 08 38 01.72 & +13 23 05.6 & 0.684& Q & 13.6 & 0.90 & 4.9 & 2\\
3C215 	& 0903+169 & 09 03 44.14 & +16 58 16.1 & 0.411& Q & 12.4 & 1.06 & 4.9 & 10\\
3C217 	& 0905+380 & 09 05 41.42 & +38 00 29.9 & 0.898& N & 11.3 & 0.77 & 4.9 & 2\\
3C216 	& 0906+430 & 09 06 17.27 & +43 05 58.6 & 0.668& Q & 20.2 & 0.84 & 8.2 & 2,13\\
3C219 	& 0917+458 & 09 17 50.66 & +45 51 43.9 & 0.174& B & 44.9 & 0.81 & 4.8 & 14\\
3C220.1 & 0926+793 & 09 26 31.87 & +79 19 45.4 & 0.610& N & 15.8 & 0.93 & 8.4 & 2\\
3C220.3 & 0931+836 & 09 31 10.50 & +83 28 55.0 & 0.685& N & 15.7 & 0.75 & 4.9 & 2\\
3C223 	& 0936+361 & 09 36 50.87 & +36 07 35.0 & 0.137& N & 16.0 & 0.74 & 8.4 & 12\\
3C225B 	& 0939+139 & 09 39 32.21 & +13 59 33.3 & 0.582& N & 23.2 & 0.94 & 4.9 & 3\\
3C226 	& 0941+100 & 09 41 36.16 & +10 00 03.8 & 0.818& N & 15.0 & 0.88 & 8.5 & 2\\
4C73.08 & 0945+734 & 09 45 09.90 & +73 28 22.2 & 0.058& N & 15.6 & 0.85 & 0.61& 9\\
3C228 	& 0947+145 & 09 47 27.63 & +14 34 02.5 & 0.552& N & 23.8 & 1.00 & 8.5 & 3\\
3C234 	& 0958+290 & 09 58 57.42 & +29 01 37.4 & 0.185& N & 34.2 & 0.86 & 8.4 & 4\\
3C236 	& 1003+351 & 10 03 05.37 & +35 08 48.1 & 0.099& L & 15.7 & 0.51 & 0.61& 9\\
4C74.16 & 1009+748 & 10 09 49.81 & +74 52 29.5 & 0.810& N & 11.7 & 0.87 & 8.5 & 2\\
3C244.1 & 1030+585 & 10 30 19.75 & +58 30 05.2 & 0.428& N & 22.1 & 0.82 & 8.4 & 3\\
3C247 	& 1056+432 & 10 56 08.38 & +43 17 30.6 & 0.750& N & 10.6 & 0.61 & 4.9 & 2\\
3C249.1 & 1100+772 & 11 00 27.32 & +77 15 08.6 & 0.311& Q & 11.7 & 0.81 & 4.9 & 3\\
3C254 	& 1111+408 & 11 11 53.30 & +40 53 41.6 & 0.734& Q & 19.9 & 0.96 & 4.9 & 2\\
3C263 	&1137+660  & 11 37 09.30 & +66 04 27.0 & 0.656& Q & 15.2 & 0.82 & 4.9 & 2,10\\
3C263.1 & 1140+223 & 11 40 49.15 & +22 23 34.9 & 0.824& L & 18.2 & 0.87 & 8.1 & 2\\
3C265 	& 1142+318 & 11 42 52.39 & +31 50 29.1 & 0.811& N & 19.5 & 0.96 & 4.8 & 2,15\\
3C268.1 & 1157+732 & 11 57 48.12 & +73 17 30.6 & 0.950& N & 21.4 & 0.59 & 8.5 & 2\\
\hline
\end{tabular}
%
%
\end{minipage}
\end{table*}
\begin{table*}
\centering
\begin{minipage}{14cm}
%
%
\begin{tabular}{lcccccccd{2}c}  \hline
Source & IAU Name & RA    & Dec & $z$ & Spectral & $S_{178}$ & $\alpha$ &\multicolumn{2}{c}{Maps}\\
       &          &[h m s]&$\left[\ ^{\circ} \ ' \ ''\ \right]$&     & class     &[Jy]&\tiny{($178-750\ \rm{MHz}$)}&\dhead{Freq. [GHz]} & Ref. \\ \hline
3C268.3 & 1203+645 & 12 03 54.28 & +64 30 18.6 & 0.371& B & 11.7 & 0.50 & 5.0 & 3\\
3C274.1 & 1232+216 & 12 32 56.74 & +21 37 05.8 & 0.422& N & 18.0 & 0.87 & 8.5 & 3\\
3C275.1 & 1241+166 & 12 41 27.58 & +16 39 18.0 & 0.557& Q & 19.9 & 0.96 & 8.5 & 3\\
3C277.2 & 1251+159 & 12 51 04.20 & +15 58 51.2 & 0.767& N & 12.0 & 1.02 & 4.9 & 2\\
3C280 	& 1254+476 & 12 54 41.66 & +47 36 32.7 & 0.996& N & 23.7 & 0.81 & 4.9 & 2\\
3C284 	& 1308+277 & 13 08 41.33 & +27 44 02.6 & 0.239& N & 12.3 & 0.95 & 8.1 & 4\\
3C285 	& 1319+428 & 13 19 05.22 & +42 50 55.7 & 0.079& L & 12.3 & 0.95 & 1.6 & 9\\
        &          &             &             &      &   &      &      & 4.9 & 16\\
3C289 	& 1343+500 & 13 43 27.38 & +50 01 32.0 & 0.967& N & 12.0 & 0.81 & 4.9 & 2\\
3C292 	& 1349+647 & 13 49 13.07 & +64 44 24.4 & 0.713& N & 10.1 & 0.80 & 8.5 & 2\\
3C295 	& 1409+524 & 14 09 33.44 & +52 26 13.6 & 0.461& N & 91.0 & 0.63 & 8.6 & 3\\
3C299 	& 1419+419 & 14 19 06.29 & +41 58 30.2 & 0.367& N & 12.9 & 0.65 & 4.5 & 3\\
3C300 	& 1420+198 & 14 20 39.96 & +19 49 13.2 & 0.272& N & 19.5 & 0.78 & 8.1 & 4\\
3C303 	& 1441+522 & 14 41 24.82 & +52 14 18.4 & 0.141& B & 12.2 & 0.76 & 1.5 & 1\\
3C319 	& 1522+546 & 15 22 43.90 & +54 38 38.4 & 0.192& L & 16.7 & 0.90 & 8.4 & 4\\
3C321 	& 1529+242 & 15 29 33.42 & +24 14 26.2 & 0.096& N & 14.7 & 0.60 & 4.8 & 17\\
3C325 	& 1549+628 & 15 49 13.99 & +62 50 20.0 & 0.860& Q & 15.6 & 0.70 & 4.9 & 2\\
3C326 	& 1549+202 & 15 49 56.13 & +20 14 18.2 & 0.089& B & 22.2 & 0.88 & 1.4 & 9\\
3C330 	& 1609+660 & 16 09 13.90 & +66 04 22.3 & 0.549& N & 30.3 & 0.71 & 8.4 & 3\\
3C334 	& 1618+177 & 16 18 07.33 & +17 43 29.6 & 0.555& Q & 11.9 & 0.86 & 4.9 & 10\\
3C336 	& 1622+238 & 16 22 32.21 & +23 52 02.0 & 0.927& Q & 11.5 & 0.73 & 4.9 & 2,10\\
3C341 	& 1626+278 & 16 26 02.42 & +27 48 13.9 & 0.448& N & 10.8 & 0.85 & 8.5 & 3\\ 
3C337 	& 1627+444 & 16 27 19.07 & +44 25 38.2 & 0.630& N & 11.8 & 0.63 & 4.9 & 2\\
3C340 	& 1627+234 & 16 27 29.41 & +23 26 42.6 & 0.760& N & 10.1 & 0.73 & 4.9 & 2\\
3C349 	& 1658+471 & 16 58 04.44 & +47 07 20.3 & 0.205& N & 14.5 & 0.74 & 8.4 & 4\\
3C351 	& 1704+608 & 17 04 03.49 & +60 48 30.9 & 0.371& Q & 14.9 & 0.73 & 8.3 & 3\\
3C352 	& 1709+460 & 17 09 18.00 & +46 05 06.0 & 0.806& N & 11.3 & 0.88 & 4.7 & 2\\
3C381 	& 1832+474 & 18 32 24.47 & +47 24 39.0 & 0.161& B & 18.1 & 0.81 & 8.4 & 4\\
3C382 	& 1833+326 & 18 33 11.97 & +32 39 18.2 & 0.058& B & 21.7 & 0.59 & 8.5 & 18\\
3C388  	& 1842+455 & 18 42 35.44 & +45 30 21.7 & 0.091& L & 26.8 & 0.70 & 4.9 & 19\\
3C390.3 & 1845+797 & 18 45 37.57 & +79 43 06.5 & 0.056& B & 51.8 & 0.75 & 8.4 & 20\\
3C401 	& 1939+605 & 19 39 38.81 & +60 34 33.5 & 0.201& L & 22.8 & 0.71	& 8.4 & 4\\
3C427.1 & 2104+763 & 21 04 44.80 & +76 21 09.5 & 0.572& L & 29.0 & 0.97 & 8.5 & 3\\
3C436 	& 2141+279 & 21 41 57.91 & +27 56 30.3 & 0.215& N & 19.4 & 0.86 & 8.4 & 4\\
3C438 	& 2153+377 & 21 53 45.51 & +37 46 12.8 & 0.290& L & 48.7 & 0.88 & 8.4 & 4\\
3C441 	& 2203+292 & 22 03 49.27 & +29 14 43.8 & 0.780& N & 12.6 & 0.83	& 4.9 & 2,8\\
3C452 	& 2243+394 & 22 43 32.79 & +39 25 27.3 & 0.081& N & 59.3 & 0.78 & 8.5 & 18\\
3C455 	& 2252+129 & 22 52 34.53 & +12 57 33.5 & 0.543& Q & 14.0 & 0.71 & 4.9 & 3\\
3C457 	& 2309+184 & 23 09 38.53 & +18 29 22.0 & 0.428& N & 14.3 & 1.01 & 8.5 & 3\\ \hline
\end{tabular}
\vskip 10pt
Notes for Table \ref{tab:the_sample}. All data from Laing, Riley \& Longair (1983) and subsequent updates.\\
Column  [1]: 3CR catalogue source name.\\ 
Column  [2]: IAU source name (B1950.0).\\ 
Column  [3]: Right Ascension $[\rm h \ m \ s]$ of the optical ID (B1950.0).\\
Column  [4]: Declination $[^{\circ} \ ' \ '']$ of the optical ID (B1950.0).\\
Column  [5]: Redshift, rounded to 3 decimal places.\\ 
Column  [6]: Optical type. B: broad emission line radio galaxy, L: low excitation radio galaxy, N: narrow emission line radio galaxy, Q: quasar.\\ 
Column  [9]: Total flux density for the source as measured at 178~MHz [Jy].\\ 
Column [10]: Low frequency spectral index ($178-750 \rm {MHz}$).\\
Column [11]: Largest angular size [arcsec].\\
Column [12]: Reference for data. (1): Leahy \& Perley (1991), (2): Mullin, Hardcastle \& Riley (2006), (3): Gilbert et al. (2004), (4): Hardcastle et al. (1997), (5): Rudnick (1988), (6) Rudnick \& Anderson (1990), (7): unpublished VLA archive, (8): Fernini, Burns \& Perley (1997), (9): 3CRR atlas, (10): Bridle et al. (1994), (11): Katz \& Stone (1997), (12): Leahy et al. (1997), (13): Taylor et al. (1995), (14): Clarke et al. (1992), (15): Fernini et al. (1993), (16): van Breugel \& Day (1993), (17): Hough et al. (2004), (18): Black et al. (1992), (19): Roettiger at al. (1994), (20): Dennett-Thorpe et al. (1999)
\end{minipage}
\end{table*}
%

%
%
%
\subsection{Parameter definitions and measurement methods}\label{sec:definitions}
%
%
%
\subsubsection{Lobe size}\label{sec:def_lobe_size}
%
%
Since many sources show distortion and bending in the jet and lobe
features, there is no obvious single definition of source size. Shocks
associated with the deceleration of the outflow are often assumed to
produce the observed hotspot features, in which case the core-hotspot
separation should represent a measure of the beam length; however,
multiple hotspots are commonly found so the core-hotspot separation as
a parametrization of beam length is not without ambiguity. The same
ambiguity will affect the lobe size measured along the core-hotspot
axis, which could represent the extent of the post-shock flow of beam
material along the beam axis. Finally, the largest angular size of the
lobe does not always lie along the core-hotspot axis or the apparent
flow direction, as a few sources appear distorted with considerable
lateral extension in the lobes.

Accordingly, three source length measurements have been made. The
angular core-hotspot separation is the angular distance between the
core and primary hotspot (defined in section \ref{sec:def_hotspots})
within a lobe, $\Theta_{c-hs}$, and is measured using the hotspot
positions obtained from the highest resolution map available for
the source. $\Theta_{l}$, the angular lobe length, is defined as the
maximum angular lobe size measured from the core along the
core-primary hotspot axis. The largest angular size of the lobe,
$\Theta_{LAS_{\rm l}}$, is the maximum angular distance of the lobe
edge from the core. $\Theta_{l}$ and $\Theta_{LAS_{\rm l}}$ are
measured from the core to the $3\sigma$ contour (where $\sigma$ is
the off-source root mean square noise level) from the lowest
resolution map available for the source. This criterion was chosen to
give a consistent measure across the sample.

The procedure above introduces a potential source of bias, as the
position of the $3\sigma$ contour will be dependent on observing
resolution. Taking size measurements from the highest resolution map
available would minimize this effect but, as the low level emission
from the lobes is often resolved out at high resolution, the extent of
the large scale structure might be underestimated if this approach
were taken. The $\Theta_{l}$ and $\Theta_{LAS_{\rm l}}$ measurements
have therefore all been made from the lowest resolution map available
and a correction factor has been applied to compensate for resolution
dependent beam-width smearing. The correction is that used by Gilbert \& Riley (1999, hereafter G99). The
maximum intensity, $m$, within two half-power beam widths of the
apparent lobe edge (measured at the $3\sigma$ contour level) is
found on the relevant axis. The half-width at the $3 \sigma$ level of
a Gaussian, of height $m$ and with a half-power beam width equal to
that of the restoring beam, can then be determined and subtracted from
the apparent lobe length to correct for the effect of finite beam
width.

All angular size measurements, $\Theta_{c-hs}$, $\Theta_{l}$ and $\Theta_{LAS_{\rm l}}$, are converted
respectively to linear sizes, $c-hs$, $l$ and $LLS_{\rm l}$. The resolution correction factor is first subtracted
from $\Theta_{LAS_{\rm l}}$ and $\Theta_{l}$; that is, $l$ and
$LLS_{\rm l}$ are quoted with the factor applied. (For this conversion the proper distance, $R$, is calculated
using the {\sc angsiz} code\footnote{http://ascl.net/angsiz.html}.)

\begin{figure}
\centerline{\epsfig{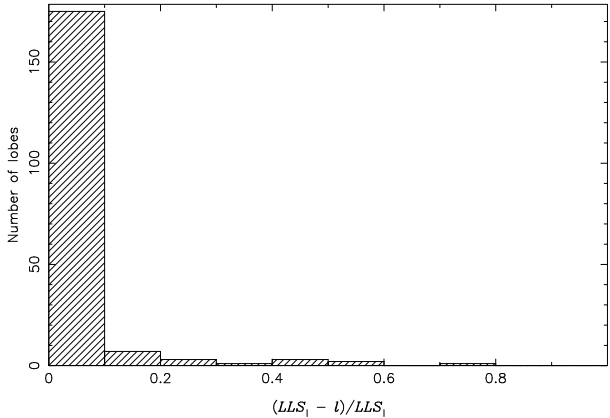}}
\caption{\label{fig:l_LLS}Histogram of ($LLS_{\rm l}-l)/LLS_{\rm l}$, for the entire sample.}
\end{figure}

Although $l$ and $LLS_{\rm l}$ are not always the same within a lobe,
as the core-hotspot axis may differ from that of the largest angular
extent of the lobe from the core, in practice the difference between these two parameters is usually small, as
illustrated in Fig. \ref{fig:l_LLS}. In the following analysis
$LLS_{\rm l}$ is therefore used as the lobe size measurement and in the evaluation of the lobe axial ratio and lobe size asymmetry, both of which
are discussed below in sections \ref{sec:def_axialratio} and
\ref{sec:def_lobeasym} respectively. The total linear source size,
$LLS_{\rm s}$, is defined as the sum of $LLS_{\rm l}$ of both lobes.

In order to consider jet detectability (see section
\ref{sec:obscoresjets}) we define the fractional observed lobe length, $f_{\rm
l}$, as the ratio of the observed extent of the
lobe emission measured along the $LLS_{\rm l}$ axis from the inner
lobe edge at the $3\sigma$ contour to the lobe extremity, to
$LLS_{\rm l}$.
%
%
%
%
%
\subsubsection{Lobe axial ratio}\label{sec:def_axialratio}
%
%
The definition of the lobe width, $\Theta_{\rm w}$, is problematic as
the morphologies of the lobes, both within individual sources and from
source to source, are often very different. As $\Theta_{\rm w}$ is to
be used to determine the lobe axial ratio, a measurement representing
the width at a set distance from the core, relative to the lobe
extent, is appropriate. Not all sources have lobes extending back to
the core but lobe emission is detected at least 2/3 along the
core-hotspot axis from the core in all but 6 lobes. $\Theta_{\rm w}$
has therefore been defined as the width of the lobe perpendicular to
the core-primary hotspot axis as measured from the core at the point
two-thirds along this axis. The $3\sigma$ contour is used to determine
the lobe edge and measurements are made from the lowest resolution map
available. This definition of lobe width was found to allow
greater consistency in the measurement of $\Theta_{\rm w}$ across the
sample than other definitions that have sometimes been used in the
literature, such as the Gaussian FWHM (e.g., Leahy \& Williams, 1994),
given that the data here are high-resolution 8-GHz maps often with
many beam widths across the lobes. While a Gaussian distribution
represents a reasonable model of a slice taken through many lobes in
the sample, a significant number would require multiple components to
be fitted, as structure is detected in the lobe, which reduces the
usefulness of this definition of width for our data.

The linear lobe width, $w$, is obtained from $\Theta_{\rm w}$ and the lobe axial ratio, $R_{\rm ax}$, is defined as $LLS_{\rm l}$ over $w$. 
%
%
\subsubsection{Lobe size asymmetry}\label{sec:def_lobeasym}
%
%
The lobe size asymmetry is defined by the fractional separation
difference, $x$, as defined by Banhatti (1980):
%
%
\begin{equation}
x = \frac{D_{1} - D_{2}}{D_{1} + D_{2}}
\end{equation}
%
%
where $D_{1}$ and $D_{2}$ are the two lobe sizes. $D_{1}$ may be taken
as the longer lobe, giving $x_{\rm lobe}$, or as the jet-side lobe
if a jet is detected, giving $x_{\rm jet}$. Previous
studies have argued that using $LLS_{\rm l}$ to define $x_{\rm lobe}$
and $x_{\rm jet}$ is preferable to $c-hs$ as from observations of
multiple hotspots, hotspots are inferred to be transient features in
the lobe (Scheuer 1995; Arshakian \& Longair 2000, hereafter AL00).
At the very least the physical region of the source to which hotspots
correspond is ambiguous. Therefore, $LLS_{\rm l}$ is used throughout
to define $x_{\rm lobe}$ and $x_{\rm jet}$.
%
%
\subsubsection{Cores}\label{sec:def_cores}
%
%
The core measurements were obtained using the {\sc aips} task JMFIT,
which fits an elliptical gaussian model of between one and four
components to a feature. One component was fitted and the peak
intensity was taken as the core flux. As most cores in the
sample were unresolved at all resolutions such a model fitted the data
well; three measurements were made in this way (with different
starting parameters) and averaged to give the final value. A
corresponding error was obtained from the square root of the average of the squared
formal errors returned from the fitting procedure. For around two
thirds of the sample this error is less than 2 per cent of the core
flux, so the calibration error (expected to be 2-3 per cent) will
dominate. Errors quoted therefore correspond to 3 per cent of the core
flux measurement, unless the formal error from JMFIT is greater, in
which case the latter is quoted.

Core measurements have been taken from the highest resolution
multi-array maps available for each source. For 7 sources, the core
feature was either not detected or not well defined in the map and a $3 \sigma$
upper limit for the core flux based on the local r.m.s. noise level was
obtained. A few sources had variable cores -- in these cases the core
flux quoted is the lowest value measured. For details, see the papers in which the observations are
presented as referenced in Table \ref{tab:the_sample}.
%
%
\subsubsection{Jets}\label{sec:def_jets}
%
%
A jet feature is defined by criteria based on those of Bridle \& Perley (1984). Thus, a jet is any feature that is
\begin{enumerate}
\item at least four times as long as it is wide;
\item separable at high resolution from other extended structures (if
any), either by brightness contrast or spatially (e.g. it should be a
	      narrow ridge running through more diffuse emission, or a
	      narrow feature in the inner part of the source entering
	      more extended emission in the outer part).
\end{enumerate}
In some sources jets appear to bend, in particular as they reach the
hotspot region. As discussed by Bridle et al.\ (1994), this could have
consequences for beaming model analysis. Thus, following H98, we also
define the straight jet, which fits the above criteria (i) and (ii)
but also must be aligned with the compact radio core where it is
closest to it (and is measured from the end closest the core along its
length only while the deviation from a straight line is less than the
jet radius), and the total jet, which fits the above criteria (i) and
(ii) and has no alignment restriction (and includes the entire feature
that is visible).

The method of measurement for both features is the same as that
adopted by H98. The straight jet was measured using the {\sc aips}
task TVSTAT to find the integrated flux within the region containing
the apparent jet emission, $F_{\rm obs}$. A background flux correction
was made by integrating two regions identical in size to the initial
jet measurement on either side of the feature. The average of these,
$B_{\rm obs}$, was then subtracted from the jet measurement to give
the observed jet flux, $J_{\rm obs} = F_{\rm obs} - B_{\rm obs}$. In
order to get the best estimate of $J_{\rm obs}$, three values of jet
flux were taken this way and averaged. The error in $J_{\rm obs}$ is
almost always dominated by the ambiguity in defining the jet emission
itself and so the errors quoted are half the measured maximum
range of the three jet measurements made.

For the total jet TVSTAT is used to measure the integrated flux of the
entire jet feature in the manner described for the straight jet, usually in straight sections that are then
combined to give the total jet measurement. There are only 4 sources for which the more prominent feature defined by the straight jet criteria is not in the same lobe as that defined by the total jet criteria. Otherwise, the total jet measurement is often the same as the straight
measurement (37 out of 65 sources with at least one possible or definite jet detection) or simply includes some further extension beyond a bend in the jet. In a few sources, the detected jet appears misaligned
with the source axis such that the feature is thought to be associated
with the flow downstream from some presumed bend in the beam. For
these sources, the total jet measurement then corresponds to this
feature. 

Apparent jet-like features that fail the jet criteria are classified
as possible jets and the fluxes of these are measured in the same way
as definite jets. Typically these are features that are not prominent
enough to be definite jets, though several fail on the length
criterion. For those sources with a visible jet on both sides of the
core, the brighter feature is defined as the jet, while the other jet
is referred to as the counterjet. Where no jet emission is detected,
an upper limit on the jet flux is estimated by measuring the
integrated flux of a region $\sim 2$ restoring beam widths across
the entire distance between the core and hotspot region. Background
flux is corrected for in the same manner as for the definite and
possible jets by taking two further integrated flux measurements
either side of the initial region. However, if the flux associated
with the central region is not the highest of the three, then the
upper limit estimate is the positive difference between the central
measure and the lower of the other two.

The straight jet measurement is used for considering beaming models
and is used to define the jet side for parameters such as $x_{\rm
jet}$ and hotspot ratios (defined in section \ref{sec:def_hotspots}).
The total jet is used when jet morphology is considered. This is
parametrized in the following way. The angular total jet length is
defined to be the angular length of the feature identified as the
total jet. The corresponding angular jet position and jet termination
are the angular separation of the base of the jet (that end of the
feature nearest the core) and the tip of the jet (the end of the jet
furthest from the core) from the core. The
fractional jet length, $f_{\rm j_{l}}$, fractional jet position,
$f_{\rm j_{p}}$ and fractional jet termination, $f_{\rm j_{t}}$ are
respectively the ratio of the linear jet length, position and
termination to the lobe length, $l$. Note that the jet axis is not
always the same as that along which the lobe length has been measured,
giving a source of scatter in all three parameters.
%
%
\subsubsection{Hotspots}\label{sec:def_hotspots}
%
%
Following H98, the hotspot is defined as any feature that is not part
of a jet and that has a largest dimension smaller than 10 per cent of
the main axis of the source as well as having a peak brightness
greater than ten times the off-source noise. It must be separated from
nearby peaks by a minimum falling to two-thirds or less of the
brightness of the fainter peak. Where more than one such feature is
observed, the most compact component is the primary hotspot while the
remaining components are secondary hotspots.

Measurements of the hotspots were taken from the highest resolution
multi-array map available for each source. The {\sc aips} task JMFIT
was used to give an integrated flux value as well as the major and
minor axes, $\Theta_{\rm maj}$ and $\Theta_{\rm min}$, the half-widths
of the fitted Gaussian. The angular hotspot size, $\Theta_{\rm h}$, was then
defined as the arithmetic mean of $\Theta_{\rm maj}$ and $\Theta_{\rm min}$.
The average angular hotspot size, $\Theta_{\rm h_{av}}$, is defined as the
arithmetic mean of the sizes of the primary hotspots in both lobes.

Fitting was carried out several times for each feature with varying
starting parameters and similar results were generally obtained.
However, for some of the most highly resolved features at lower
redshift this was not the case and an alternative to JMFIT was used. If the feature was too resolved, or convergence could not be
achieved in flux within a factor of 1.5 either way, a manual
measurement was made. Fluxes were estimated by integration from the
maps with TVSTAT; background emission was taken into account by
integrating over a surrounding region, normalising the flux to an area
equivalent to that of the hotspot and subtracting. In order to make
size measurements, the FWHM was estimated from slices taken through
the feature. Errors have not been quoted for hotspot flux density or size
since the parameters are subjective: the dominant error will
derive from the ambiguity in determining the hotspot region.

The linear hotspot size, $h$, is obtained from $\Theta_{\rm h}$ and the fractional hotspot size, $f_{\rm h}$, is the ratio of $h$ to $LLS_{\l}$.
%
%
\subsubsection{Hotspot recession}\label{sec:def_hotspotrec}
%
%
Three recession parameters are defined, $\eta,\ \zeta$ and $\Delta$. $\eta$ is the lobe hotspot recession: the ratio of $c-hs$ to
$LLS_{\rm l}$ for each lobe. $\zeta$ is the source hotspot recession: the ratio of the sum
of $c-hs$ for the two lobes to the total source size, $LLS_{\rm s}$. $\Delta$ quantifies
the recession asymmetry in a single source and is defined as the ratio
of the smaller to the larger value of $\eta$.
%
%
%
\subsection{Observing frequency and prominence}\label{sec:obsfreqprom}
%
%
The total source flux observed at 178~MHz is
K-corrected using the corresponding low frequency spectral index (both
parameters are taken from LRL), giving $S_{\rm total}$ at
178~MHz. $S_{\rm total}$ is used to obtain the source luminosity, $P_{178}$, using the relation
%
%
\begin{equation}
P_{178}=R^{2}(1+z)^{2}S_{\rm total}
\label{lumin}
\end{equation}
%
%
(where $R$ is the proper distance).
The source luminosity is determined from the low-frequency source flux as this ought to be dominated by the steep-spectrum, unbeamed emission
associated with the lobes, so that little contribution should be
made by any relativistic beaming.

All the flux densities of compact features of the sources discussed
above are extrapolated from the observed flux density to a common frequency of 8.4~GHz
%
%
\begin{equation}
S_{8.4} = S_{\nu_{\rm obs}} \left(\frac{\nu_{\rm obs}}{8.4} \right)^{\alpha}  
\end{equation}
and then K-corrected to an emitted frequency
\begin{equation}
S = S_{8.4} (1+z)^{\alpha - 1}
\end{equation}
%
%
where $\nu_{\rm obs}$ is the observing frequency in GHz and $\alpha$ the
spectral index, assumed to be 0 for core features and 0.5 for jet and
hotspot features. These flux densities are converted to luminosities using equation (\ref{lumin}).

The source luminosity is used as a normalization factor for these core, hotspot and jet luminosities to define prominence
parameters. The core, hotspot and straight jet prominence
($p_{\rm c},\ p_{\rm h},\ p_{\rm j}$) are respectively the ratio of
the core, hotspot and straight jet luminosity to $P_{178}$. (Note that
this normalization factor is different from that used by H98, so that our
prominences are different from theirs.)

A glossary of all parameters that
have been defined in this section (along with others that will be
defined subsequently) is given in Table \ref{tab:glossary}.
%
%
\subsection{Effective observing resolution}\label{sec:effobsres}
%
%
The sample extends to a redshift of 1. While the aim of the observing
program was to obtain data of a consistent quality across the sample,
there are inevitably instrumentational limits in achieving this. At
increasing redshift the angular resolution relative to the source size
must decrease for a fixed beam width, so that more distant sources are
observed at increasing linear scales for a given source size.
One key parameter here is the number of restoring beams across the
source, which we refer to throughout the paper as the effective
observing resolution. The high-resolution effective observing
resolution is defined as the ratio of the restoring beam size of the highest-resolution map for each source to $LLS_{\rm
s}$; the low-resolution effective observing resolution is defined similarly but using the restoring beam size from the lowest-resolution map.

\begin{figure}
\centerline{\epsfig{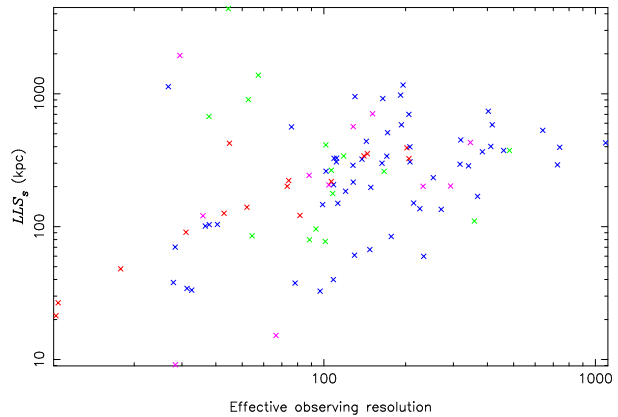}}
\caption{\label{fig:size_res} The largest linear source size, $LLS_{\rm s}$, plotted against the high-resolution effective observing resolution. Green: low excitation radio galaxies, blue: narrow line radio galaxies, magenta: broad line radio galaxies and red: quasars (on-line colour version).}
\end{figure}

In Fig. \ref{fig:size_res} the
linear source size, $LLS_{\rm s}$, is plotted against the {\it high-resolution} effective observing
resolution for the sample, to highlight the
range in this quantity that corresponds to the
sample's high-resolution maps, since this is the more
important quantity as regards source properties. It can be seen that
there is a range of effective observing resolutions associated with
the sources; this may have consequences, in particular, for the
consideration of jet and hotspot properties. In the following
sections, in which we discuss the lobe, core, jet and hotspot
properties, we consider the limitations imposed by our observing
strategy as well as trends with power, redshift and size.
%
%
%
%
%
\begin{table}
\centering
\caption{Glossary of symbols used.\label{tab:glossary}}
\begin{tabular}{lll}  \hline
Symbol  & Parameter & Reference \\\hline
$S_{178}$                & total source flux as measured at 178 MHz      & section \ref{sec:sample}\\
$S_{\rm total}$		& K-corrected total source flux density,	& section \ref{sec:obsfreqprom}\\
			& as measured at 178 MHz			& \\
$P_{178}$               & source luminosity, as measured at 178 MHz 			  & section \ref{sec:obsfreqprom}\\
$\alpha$		& spectral index						  & section \ref{sec:obsfreqprom}\\
 & & \\
$\Theta_{c-hs}$		& angular core-primary hotspot separation 	& section \ref{sec:def_lobe_size}\\
$\Theta_{l}$ 		& angular lobe length				& section \ref{sec:def_lobe_size}\\
$\Theta_{LAS_{\rm l}}$	& largest angular lobe size			& section \ref{sec:def_lobe_size}\\
$\Theta_{\rm w}$	& angular lobe width				& section \ref{sec:def_axialratio}\\
$c-hs$			& linear core-primary hotspot separation	& section \ref{sec:def_lobe_size}\\
$l$			& linear lobe length				& section \ref{sec:def_lobe_size}\\
$LLS_{\rm l}$		& largest linear lobe size			& section \ref{sec:def_lobe_size}\\
$LLS_{\rm s}$ 		& largest linear source size			& section \ref{sec:def_lobe_size}\\
$w$                     & linear lobe width                             & section \ref{sec:def_axialratio}\\
$f_{\rm l}$             & fractional observed lobe length               & section \ref{sec:def_lobe_size}\\  
$R_{\rm ax}$		& lobe axial ratio 				& section \ref{sec:def_axialratio}\\
$x_{\rm lobe}$		& fractional separation difference, 		& section \ref{sec:def_lobeasym}\\
			& as defined by the longer lobe			& \\
$x_{\rm jet}$		& fractional separation difference,	  	& section \ref{sec:def_lobeasym}\\
			& as defined by jet side			& \\
 & & \\
$F_{\rm obs}$		& measured jet flux						  & section \ref{sec:def_jets}\\
$B_{\rm obs}$		& jet background flux correction				  & section \ref{sec:def_jets}\\
$J_{\rm obs}$		& background-corrected jet flux					  & section \ref{sec:def_jets}\\
$f_{\rm j_{l}}$		& fractional jet length						  & section \ref{sec:def_jets}\\
$f_{\rm j_{p}}$		& fractional jet position					  & section \ref{sec:def_jets}\\
$f_{\rm j_{t}}$		& fractional jet termination					  & section \ref{sec:def_jets}\\
 & & \\
$\Theta_{\rm maj}$	& hotspot major axis						  & section \ref{sec:def_hotspots}\\
$\Theta_{\rm min}$	& hotspot minor axis						  & section \ref{sec:def_hotspots}\\
$\Theta_{\rm h}$	& hotspot size							  & section \ref{sec:def_hotspots}\\
$\Theta_{\rm h_{av}}$	& average primary hotspot size					  & section \ref{sec:def_hotspots}\\
$h$                     & linear hotspot size                                             & section \ref{sec:def_hotspots}\\
$f_{\rm h}$             & fractional hotspot size                                         & section \ref{sec:def_hotspots}\\
 & & \\
$\eta$			& lobe hotspot recession coefficient		& section \ref{sec:def_hotspotrec}\\
$\zeta$			& source hotspot recession coefficient		& section \ref{sec:def_hotspotrec}\\
$\delta$		& hotspot recession asymmetry			& section \ref{sec:def_hotspotrec}\\
 & & \\
$p_{\rm c}$		& core prominence						  & section \ref{sec:obsfreqprom}\\
$p_{\rm h}$		& hotspot prominence						  & section \ref{sec:obsfreqprom}\\
$p_{\rm j}$		& straight jet prominence				          & section \ref{sec:obsfreqprom}\\ 
 & & \\
$P_{\rm c}$             & luminosity cutoff of $5\cdot~10^{26}\ \rm{W\ Hz^{-1}\ sr^{-1}}$ & section \ref{sec:size_trends}\\
$\theta_{\rm c}$        & hypothetical angular spectral class cutoff                      & section \ref{sec:size_unif}\\\hline
\end{tabular}\\
\end{table}

%
%
%
\section{Lobes}\label{sec:lobes}
%
%
\subsection{Lobe size}
%
%
%
\subsubsection{Observational effects}\label{sec:lobesizobs}
%
%
Observational effects in lobe properties may be introduced by using
the $3\sigma$ contour as the criterion for defining $\Theta_{LAS_{\rm
l}}$ and $\Theta_{l}$ and this is addressed by the application of a
correction factor, as described in section \ref{sec:def_lobe_size}.
%
%
\subsubsection{Trends with $P_{178}$ and $z$}\label{sec:size_trends}
%
%
There is no straightforward physical correlation to be expected
between the beam kinetic power, the lobe size and $P_{178}$, although
sources are believed to decrease in luminosity as they expand and age
(Fanti et al., 1995; Kaiser \& Alexander, 1997; Kaiser, Dennett-Thorpe \& Alexander, 1997;
Blundell, Rawlings \& Willott, 1999). The higher-luminosity sources may be
observed at an earlier stage in their lifecycle as sources fall below
the sample flux limit as they move through the luminosity-source
linear size ($P-D$) diagram; statistically, therefore, they may be
expected to be smaller. In a flux-limited sample there is a $P-z$
degeneracy, so any tendency for $LLS_{\rm s}$ to decrease with
increasing $P_{178}$ may also be seen as a trend in redshift.

\begin{figure}
\centerline{\epsfig{file=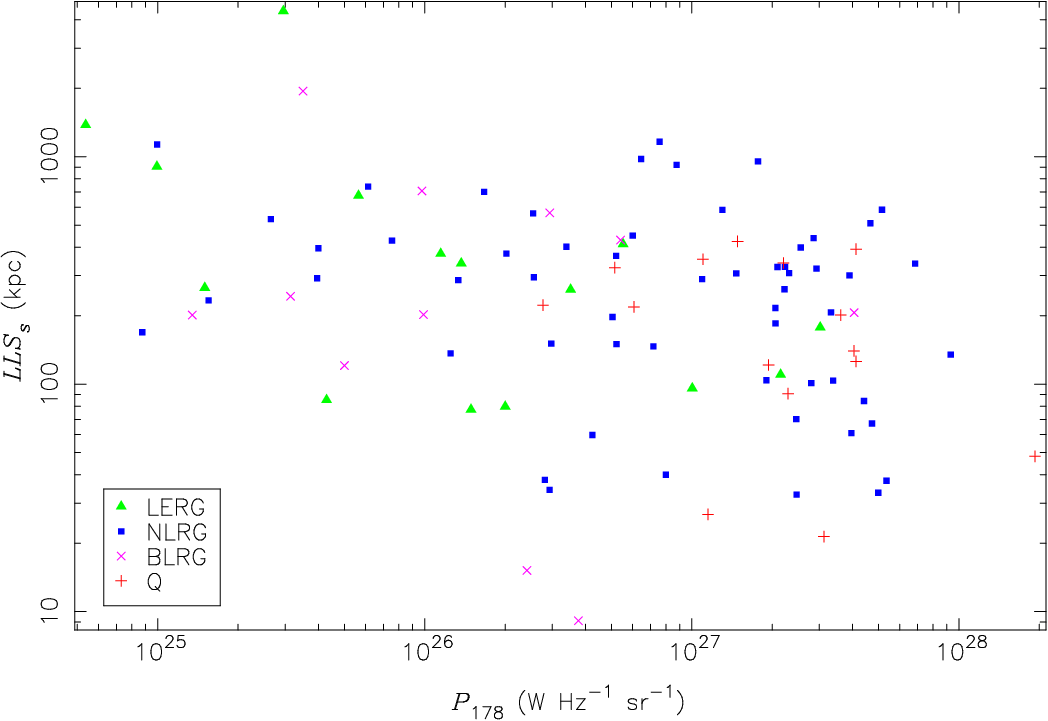,width=8cm}}
\caption{\label{fig:LLS_P}The largest linear source size, $LLS_{\rm s}$, plotted against the source luminosity as measured at 178~MHz, $P_{178}$.}
\centerline{\epsfig{file=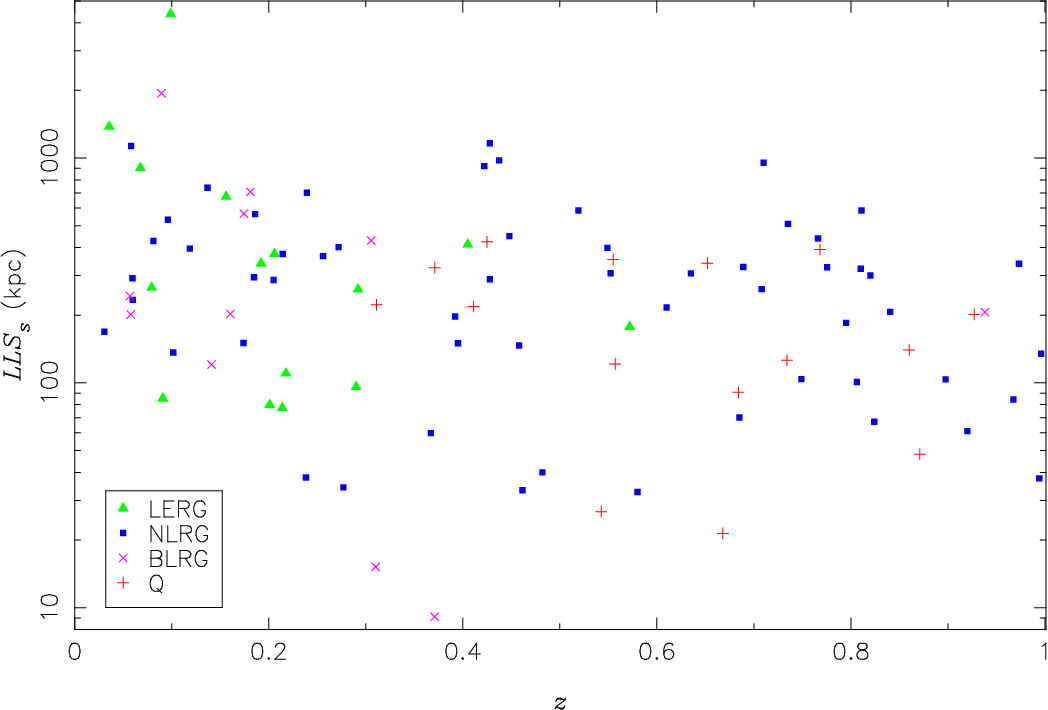,width=8cm}}
\caption{\label{fig:LLS_z}The largest linear source size, $LLS_{\rm s}$, plotted against redshift, $z$.}
\end{figure}

In Fig. \ref{fig:LLS_P} it can be seen that $LLS_{\rm s}$ tends to be
smaller for the higher luminosity sources; a similar but weaker effect
is shown in the plot of $LLS_{\rm s}$ against redshift in Fig.
\ref{fig:LLS_z}. Spearman rank correlation tests give $r_{\rm s} =
-0.31$ and $-0.29$ respectively for these two trends, implying a
correlation significant at better than the 99 per cent confidence level.
However, comparing subsamples of sources defined with a 178-MHz
luminosity cutoff, $P_{\rm c}=5 \times 10^{26}\ \rm{W\ Hz^{-1}\
sr^{-1}}$, inclusive of all spectral classes, a W-M-W test does not
show a significant difference in size between the high and low
luminosity populations; we can conclude that any trends with source
size found in other parameters for the sample should not then be
systematically biased across the power or redshift range.

The value of 178-MHz luminosity, $P_{\rm c}$, chosen above gives the
minimum overlap between the quasar and BLRG populations: these can be
seen from Figs \ref{fig:LLS_P} and \ref{fig:LLS_z} to occupy different
ranges of luminosity and redshift, with the
higher-luminosity quasar population found at higher redshift.
Throughout the paper we therefore make comparisons between the NLRG
and the BLRG, Q and LERG classes by dividing the NLRG data into low-
and high-luminosity samples at $P_{\rm c}$. This gives a low-luminosity
and high-luminosity NLRG subsample of 19 and 38 sources respectively:
the low-luminosity sample is very similar in luminosity to (and
contains many of the same sources as) the $z<0.3$ 3CRR/3CR sample used
by H98.
%
%
\subsubsection{Unification}\label{sec:size_unif}
%
%
Considering only those classes included in the standard FRII
unification scheme (BLRGs, NLRGs and Qs) the BLRGs constitute $31\pm11$ per cent of the sample at low luminosity and the Qs $26\pm7$
per cent of the sample at high luminosity (assuming errors of
$\sqrt{N}$), so the proportions are not significantly different for
the two subsamples. This lends support to a model in which the BLRGs
and Qs are equivalent populations and also implies no significant
variation in the opening angle of the torus.

The opening angle of the torus can be estimated from the number counts
of the Qs, BLRGs and NLRGs, assuming that the source axis must be
viewed at an angle less than the torus angle for the broad line
emission to be detected. For the simple unification model, this gives
$\theta_{\rm c}$ as a parameter that divides the classes, with
$\theta\le\theta_{\rm c}$ for the Qs and BLRGs and $\theta >
\theta_{\rm c}$ for the NLRGs, where $\theta$ is the angle the source
axis makes with the observer's line-of-sight. The expected fraction of
broad emission line objects detected is $\rm {P}(\le\theta_{\rm
c})=1-\cos\theta_{\rm c}$. This implies that for the lower luminosity
bin $\theta_{\rm c}\sim51^{\circ}$, while for the higher luminosity bin
$\theta_{\rm c}~\sim~45^{\circ}$, consistent with the findings of
Barthel (1989). An average value $\theta_{\rm c}=48^{\circ}$ will be
used hereafter.

According to the unification scheme, the Q and BLRG sources should be
orientated closer to the observer's line-of-sight than the NLRG
sources. Evidence consistent with this hypothesis would be a
difference in the size distributions of the spectral classes of the
Qs, BLRGs and NLRGs consistent with projection effects. The LERGs are believed to be randomly orientated with respect
to the observer. 

The relation between the true physical size, $LLS'_{\rm s}$, and observed source lengths, $LLS_{\rm s}$, is given by
%
%
\begin{equation}
LLS_{\rm s} \approx LLS'_{\rm s} \sin\theta\label{eq:LLS_theta}
\end{equation}
%
%
(the relation is not exact as the sum of the two lobes does not necessarily give a common axis). The ratio of the expected median size for the broad and narrow line
sources can be predicted from the ratio of
the median $\theta$,
%
%
\begin{equation}
\frac{<LLS_{\rm s,Q,B}>}{<LLS_{\rm s,N}>}=\frac{\sin(<\theta_{\rm Q,B}>)}{\sin(<\theta_{\rm N}>)}
\end{equation}
%
%
where $<X>$ denotes the median value of parameter $X$ and $<\theta>$ is evaluated by integrating over the
appropriate $\theta$ range. Using $\theta_{\rm c}=48^{\circ}$, $<\theta_{\rm Q,B}>=33^{\circ}$ and
$<\theta_{\rm N}>=70^{\circ}$, a predicted value of $<LLS_{\rm s,Q,B}>/<LLS_{\rm s,N}>=0.57$ is obtained. The
$<LLS_{\rm s}>$ values for the Qs, BLRGs and NLRGs are given in Table \ref{tab:LLS_median}; the data for the LERGs are included for comparison. The ratios of Q and BLRG median values to those of
the NLRGs in the respective luminosity bins is 0.73 and 0.70; if
we take all the objects together without binning by luminosity the ratio is 0.70. A W-M-W test does not show that this is statistically significant. According to the model, LERG sources are randomly
orientated so there is no predicted difference between them and the
low luminosity NLRG and BLRG population, that is, $<LLS_{\rm s,B,N}>/<LLS_{\rm s,E}>=1$ assuming that they have the same physical
size distribution. A ratio of 0.93 is found in the data, but a W-M-W test does not suggest that the LERGs are significantly smaller. (This is in contrast to the finding in H98 that LERGs were
significantly smaller than the BLRGs and NLRGs in the low redshift
sample they studied. That sample included those sources of this paper's sample
with $z<0.3$ along with a number of others. The difference may well
arise from the definition of source size used; H98 used the largest
linear source size as obtained from the largest angular source size,
whereas here $LLS_{\rm s}$ is used, the sum of the largest linear lobe
size for both lobes. For a good proportion of the sources in this sample with $z<0.3$ the largest angular source size is greater than
$LLS_{\rm s}$. In addition, the sample of H98 excluded a number of
giant sources that we include here.)

Whilst these results are in the sense expected in the
unification scheme, the effect is weaker than expected and no statistically significant difference in source size
between the Qs, BLRGs and NLRGs is found. 
%
%
\begin{table}
\scriptsize
\centering
\begin{minipage}{8cm}
\caption{\label{tab:LLS_median}The median largest linear source size, $<LLS_{\rm s}>$, for each of the spectral class distributions.}
\begin{tabular}{lclcc}  \hline
Spectral class & $<LLS_{\rm s}>$ &Spectral class & $<LLS_{\rm s}>$ &$<LLS_{\rm s}>$ ratio\\ \hline
Q          &  201  &  NLRG, high $P_{178 \rm MHz}$   & 275 &0.73\\
B          &  206  &  NLRG, low $P_{178 \rm MHz}$    & 295 &0.70\\
Q and BLRG &  204  &  NLRG                           & 292 &0.70\\
LERG       &  270  &  BLRG and low $P_{178 \rm MHz}$ & 289 &0.93\\\hline

\end{tabular}
\end{minipage}
\end{table}
%
%
\subsection{Lobe axial ratio}
%
%
%
\subsubsection{Observational effects}\label{sec:obsaxialratio}
%
%
Both $\Theta_{\rm w}$ and $\Theta_{LAS_{\rm l}}$ are taken from the
lowest resolution map available with the lobe edge determined by the
$3\sigma$ contour. However, as discussed in section
\ref{sec:lobesizobs}, this will be affected by observing resolution
and sensitivity. In the case of $\Theta_{LAS_{\rm l}}$ a correction factor was
applied in an attempt to compensate for any systematic bias introduced
by observing resolution effects. Whilst this factor is necessarily only
an order-of-magnitude correction, we felt that, as in almost all
sources the lobe extremity is associated with a bright emission peak,
the effect of beam-width smearing on the source structure is large
enough that the application of the correction factor as defined is
useful.

In the case of $\Theta_{\rm w}$, however, the emission at the lobe edges is usually at a low level and the validity of such a correction is less
clear. For example, orientation effects could affect the observed lobe width, with those sources observed with their axes closer to our line-of-sight having more extensive lobes if source
viewing angle allows a greater depth of emission to be detected near
the lobe edges. This effect, if it is significant, cannot be
compensated for by the correction factor in the way that we have
defined it for the lobe lengths. Accordingly, we have chosen not to
apply any correction factor to $\Theta_{\rm w}$. 

\begin{figure}
\centerline{\epsfig{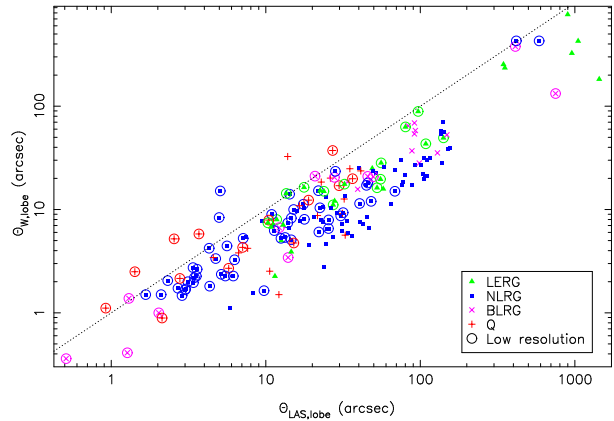}}
\caption{\label{fig:LAS_W_RES} The angular lobe width, $\Theta_{w}$,
  plotted against the resolution-corrected angular lobe size,
  $\Theta_{LAS_{\rm l}}$, binned by low-resolution effective observing
  resolution. Circled points: low-resolution effective observing resolution $\le 40$. The dotted line is the line of $\Theta_{w} = \Theta_{LAS_{\rm l}}$.}
\end{figure}

In Fig. \ref{fig:LAS_W_RES} we plot the angular lobe width against the
resolution-corrected angular largest lobe size, binned by {\it
low-resolution} effective observing
resolution. There is a tendency for those sources observed at relatively low resolution, that is with $\le 40$
restoring beams across the source, to have wider lobes. Dividing the
lobe width data points into two subsamples based on observing resolution, using a cutoff of 40
low-resolution restoring beams across the source, a W-M-W test
suggests this is significant above the 99.9 per cent confidence level. Furthermore, when we consider the
$R_{\rm ax}$ values themselves and divide them into two samples in the
same way, a W-M-W test shows that $R_{\rm ax}$ is significantly
higher in sources observed at high resolution compared to those observed
with an effective observing resolution of 40 or less, at above the 99.9 per
cent confidence level. The $R_{\rm ax}$ data, therefore, seem to be affected by observing resolution.
%
%
%
%
%
\subsubsection{Trends with $P_{178}$, $z$ and size}\label{sec:trendaxialratio}
\begin{figure}
\centerline{\epsfig{file=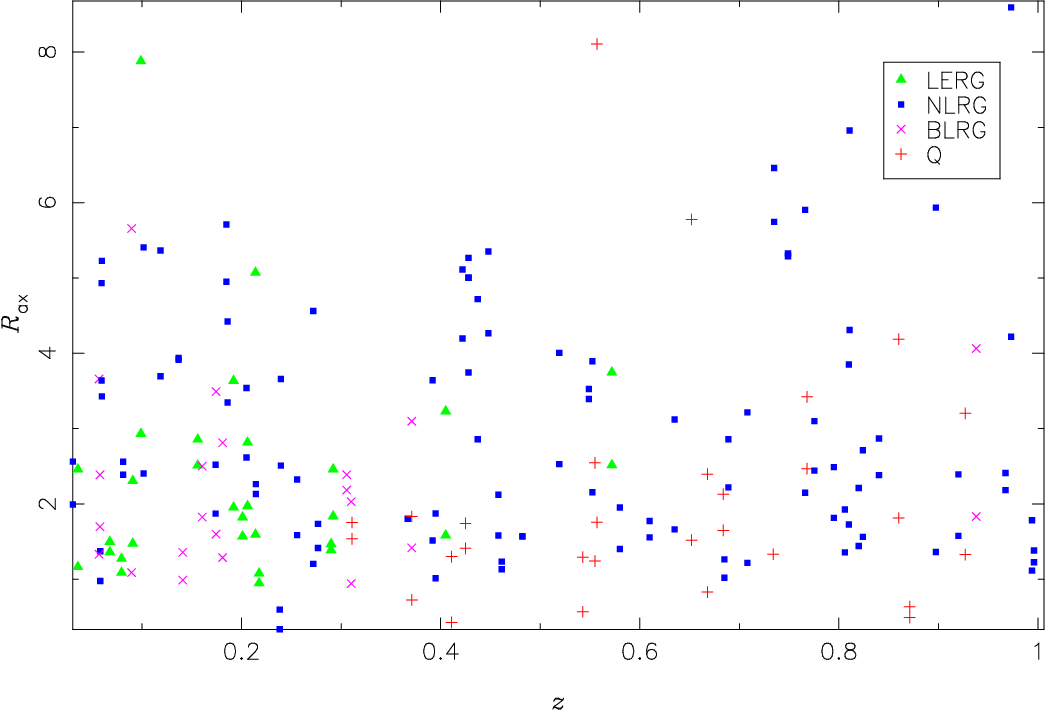,width=8cm}}
\caption{\label{fig:aspect_z} The lobe axial ratio, $R_{\rm ax}$, plotted against redshift, $z$ (two points per source).}
\centerline{\epsfig{file=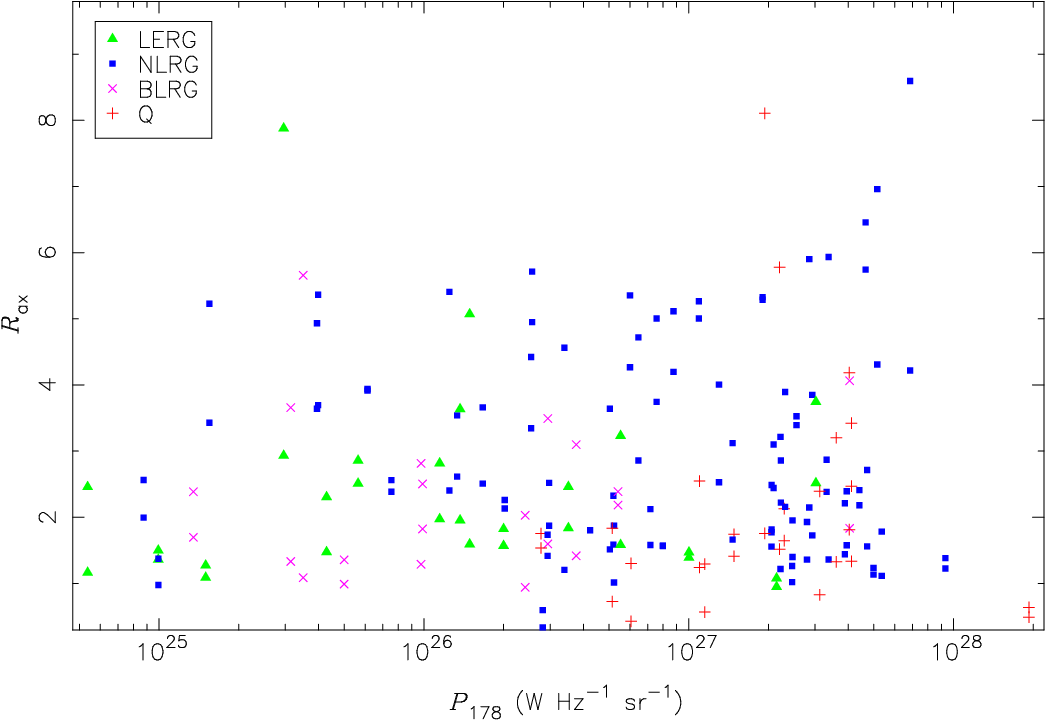,width=8cm}}
\caption{\label{fig:aspect_P}The lobe axial ratio, $R_{\rm ax}$, plotted against source
luminosity at 178 MHz, $P_{178}$, (two points per source).}
\centerline{\epsfig{file=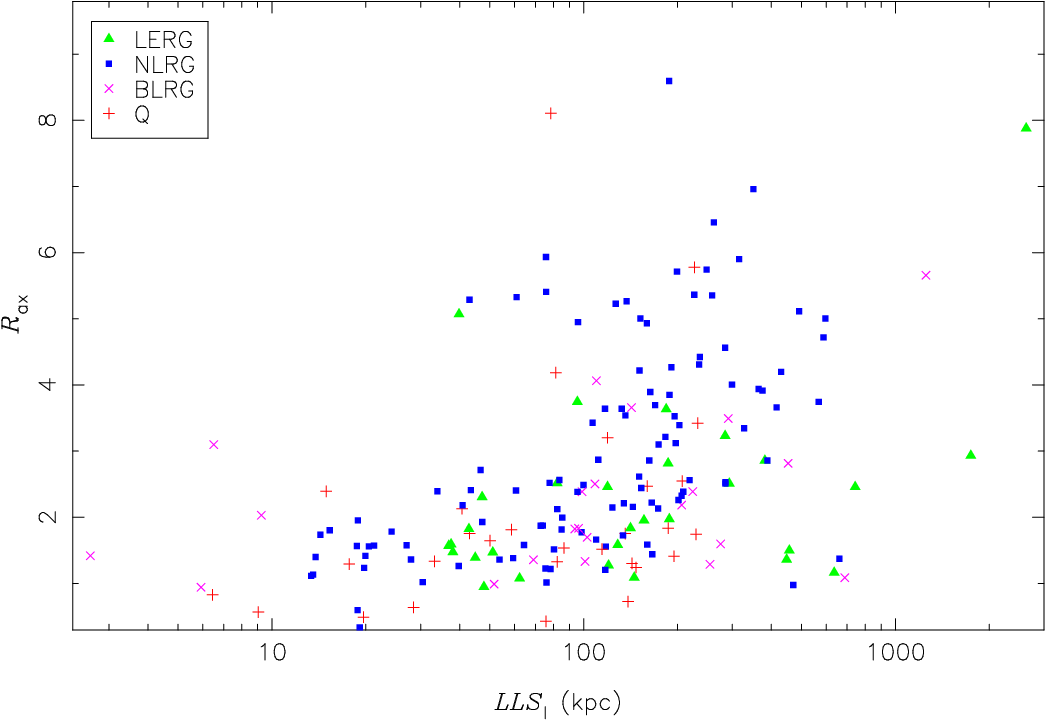,width=8cm}}
\caption{\label{fig:aspect_LLS}The lobe axial ratio, $R_{\rm ax}$, plotted against
the largest linear source size, $LLS_{\rm l}$, (two points per source).}
\end{figure}

%
%
A greater proportion of sources observed at low resolution are found
at high redshift, so that the high redshift sample may have a
systematic bias toward lower $R_{\rm ax}$ values. Another effect to
consider is the known correlation between spectral index and radio
power or redshift; the hotspots and lobes in higher-redshift
higher-power sources have steeper spectra, particularly at high
frequencies (e.g., Laing \& Peacock 1980; Blundell, Rawlings \&
Willott 1999). This correlation would mean that for high-redshift sources the low brightness lobe emission is harder to detect, resulting in smaller
$w$ values (and hence higher corresponding $R_{\rm ax}$) for
luminous sources. Dividing the $R_{\rm ax}$ distribution by $P_{178}$
(using $P_{\rm c}$) and $z$ (using a cutoff of 0.5), a
K-S test showed
no significant differences in the distribution of $R_{\rm ax}$ across
the power and redshift range. In Figs \ref{fig:aspect_z} and
\ref{fig:aspect_P} we plot $R_{\rm ax}$ (for each lobe) against
redshift and source luminosity respectively.

The plot of $R_{\rm ax}$ against $LLS_{\rm l}$ in Fig.
\ref{fig:aspect_LLS} shows that there is a trend with source size. The
distribution appears to change at $LLS_{\rm l}\sim100$ kpc, with a
much larger range in $R_{\rm ax}$ found above this size. A W-M-W test
shows that $R_{\rm ax}$ is significantly smaller in lobes with
$LLS_{\rm l}<100$ kpc at above the 99.9 per cent confidence level, for
sources of all spectral classes.

In a self-similar expansion model in which all sources in a sample are
subject to self-similar growth throughout their lifetime, $R_{\rm ax}$
should be independent of the lobe size. This is clearly not borne out
by the data. In fact it is possible that a source only grows
self-similarly in its early phases, on scale sizes of the order of
that of the associated galaxy or its hot-gas halo (e.g.\ Hardcastle \&
Worrall 2000); the data here lend support to this picture.
%
%
\subsubsection{Unification}\label{sec:Rax_uni}
%
%
The observed $R_{\rm ax}$ should be lower than the true physical value
due to the projection of the source length in the plane of the sky,
while the width should be less affected, notwithstanding any
orientation effects on lobe detectability as discussed in section
\ref{sec:obsaxialratio}. Assuming $w$ is unaffected by
orientation, the effect of projection on $R_{\rm ax}$ is the same form
as that for $LLS_{\rm s}$ (equation \ref{eq:LLS_theta}), that is
%
%
\begin{equation}
R_{\rm ax} = R^{'}_{\rm ax} \sin{\theta},\label{eq:R}
\end{equation}
%
%
where the prime indicates the true physical value of the parameters
and $\theta$ is the
angle subtended by the $LLS_{\rm s}$ axis with the observer's
line-of-sight.

The median $R_{\rm ax}$, $<R_{\rm ax}>$, is given for the different
spectral classes in Table \ref{tab:axialratio_median}. Unification
predicts the same ratio ($\sim0.6$) between the broad and narrow line
spectral classes as for $<LLS_{\rm s}>$ (given the model in equation (\ref{eq:R}) and ignoring scatter introduced by deviation of the lobes
from the common axis). For the sample data, $<R_{\rm ax_{Q}}>/<R_{\rm
ax_{N}}>=0.69$, while $<R_{\rm ax_{B}}>/<R_{\rm ax_{N}}>=0.75$.

The difference in $R_{\rm ax}$ between the Qs and high luminosity
NLRGs is statistically significant (at the 99.6 per cent confidence
level with a W-M-W test) whilst that between the BLRGs and low
luminosity NLRGs is not, though the difference between the low-power spectral classes is in the sense expected for unification. So are there intrinsic differences between
the Qs and high power NLRGs that have no correspondence in the low
luminosity sources? This may not be the case if some effect leads to
an observational bias that masks significant differences in $R_{\rm
ax}$ between the BLRGs and low-power NLRGs -- for
example, if lobes were more difficult to detect in these
low-power sources compared with high-power ones. However, despite the evidence that observing resolution does affect $R_{\rm ax}$ (section \ref{sec:obsaxialratio}), we concluded in section \ref{sec:trendaxialratio} that there was no evidence that this results in a systematic bias in lobe detectability across the power range.

While the tendency for Qs to have lower $R_{\rm ax}$ than high-power NLRGs is
consistent with projection effects, the orientation arguments predict
that this should be the case {\it as a result of} their lower values of $LLS_{\rm l}$. In fact we found no significant difference in lobe
length between the Qs and high luminosity NLRGs, which might suggest
that the Qs are associated with lower $R_{\rm ax}$ because they have
intrinsically broader lobes. A W-M-W test does not confirm that this is the case, however. 

\begin{figure}
\centerline{\epsfig{file=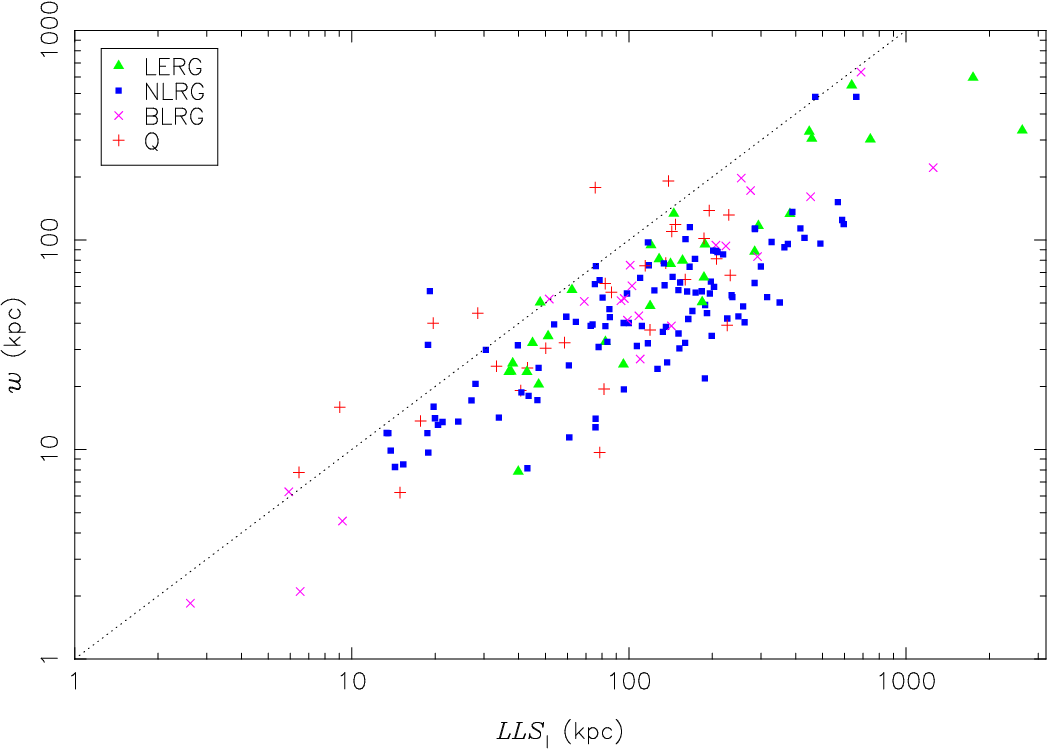,width=8cm}}
\caption{\label{fig:LLS_W_CLASS}The linear lobe width, $w$, plotted
  against the largest linear lobe size, $LLS_{\rm l}$. The dotted line is the line of $w =
  LLS_{\rm l}$.}
\centerline{\epsfig{file=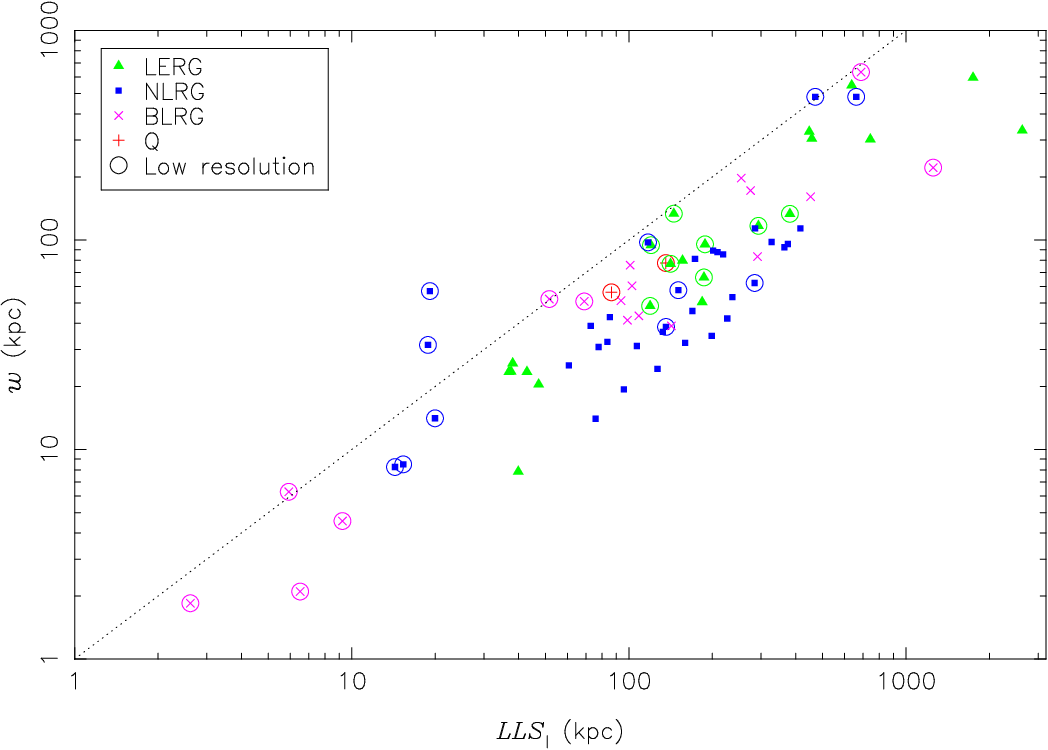,width=8cm}}
\caption{\label{fig:LLS_W_BN}The linear lobe width, $w$, plotted
  against the largest linear lobe size, $LLS_{\rm l}$, for the low
  power subsample. Circled points: low-resolution effective observing resolution $\le 40$. The dotted line is the line of $w =
  LLS_{\rm l}$.}
\end{figure}

In Fig. \ref{fig:LLS_W_CLASS} we
plot $w$ against $LLS_{\rm l}$ for the different spectral classes. From this plot it would appear that there is a tendency for the points corresponding to the broad line objects to lie to the left of those of the narrow line objects -- in other words, there is a tendency for Qs and BLRGs to have smaller $LLS_{\rm l}$ with respect to NLRGs of a similar $w$. This would suggest that differences are {\it consistent} with projection effects, although the $LLS_{\rm l}$ values in Qs and BLRGs are not significantly
lower than those in NLRGs. Our interpretation is therefore that the statistically lower $R_{\rm ax}$ found in Qs is consistent
with projection effects; the lack of a corresponding trend for BLRGs is not accounted for, though it is not strong evidence against unification in the lower-power subsample. Fig. \ref{fig:LLS_W_BN} shows that a number of the low-power sources with low effective observing resolution correspond to
either particularly small or particularly large sources and it is possible that the effects of observing resolution are more important for the BLRGs, though this is not clearly so.

As for the lobe size, we would predict no significant difference in
$R_{\rm ax}$ between the LERGs and low luminosity NLRGs and
BLRGs, if LERGs have the same intrinsic $R_{\rm ax}$ distribution. A
K-S test finds no significant difference between the $R_{\rm ax}$
distributions of the LERG and combined BLRG and low power NLRG
populations.
%
%
\begin{table}
\scriptsize
\centering
\begin{minipage}{8cm}
\caption{\label{tab:axialratio_median}The median lobe axial ratio, $<R_{\rm ax}>$, for each of the spectral class distributions.}
\begin{tabular}{lclcc}  \hline
Spectral class & $<R_{\rm ax}>$ &Spectral class & $<R_{\rm ax}>$ &$<R_{\rm ax}>$ ratio\\ \hline
Q          &  1.65  &  NLRG, high $P_{178 \rm MHz}$   & 2.38 &0.69\\
B          &  1.93  &  NLRG, low $P_{178 \rm MHz}$    & 2.56 &0.75\\
Q and BLRG &  1.75  &  NLRG                           & 2.49 &0.70\\
LERG       &  1.90  &  BLRG and low $P_{178 \rm MHz}$ & 2.40 &0.79\\\hline

\end{tabular}
\end{minipage}
\end{table}
%
%
%
\subsection{Lobe size asymmetry}
%
%
%
%
\subsubsection{Trends with $P_{178}$, $z$ and size}\label{sec:trendxlobe}
\begin{figure}
\centerline{\epsfig{file=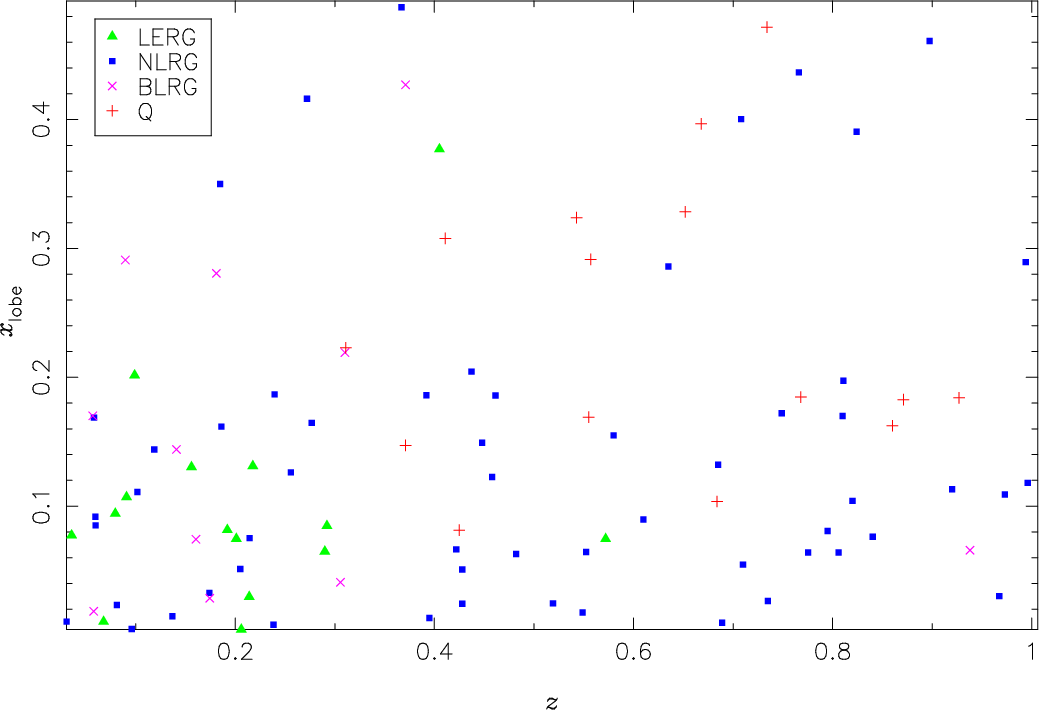,width=8cm}}
\caption{\label{fig:xlobe_z} The fractional separation difference as defined by the longer lobe side, $x_{\rm lobe}$, plotted against redshift, $z$.}
\centerline{\epsfig{file=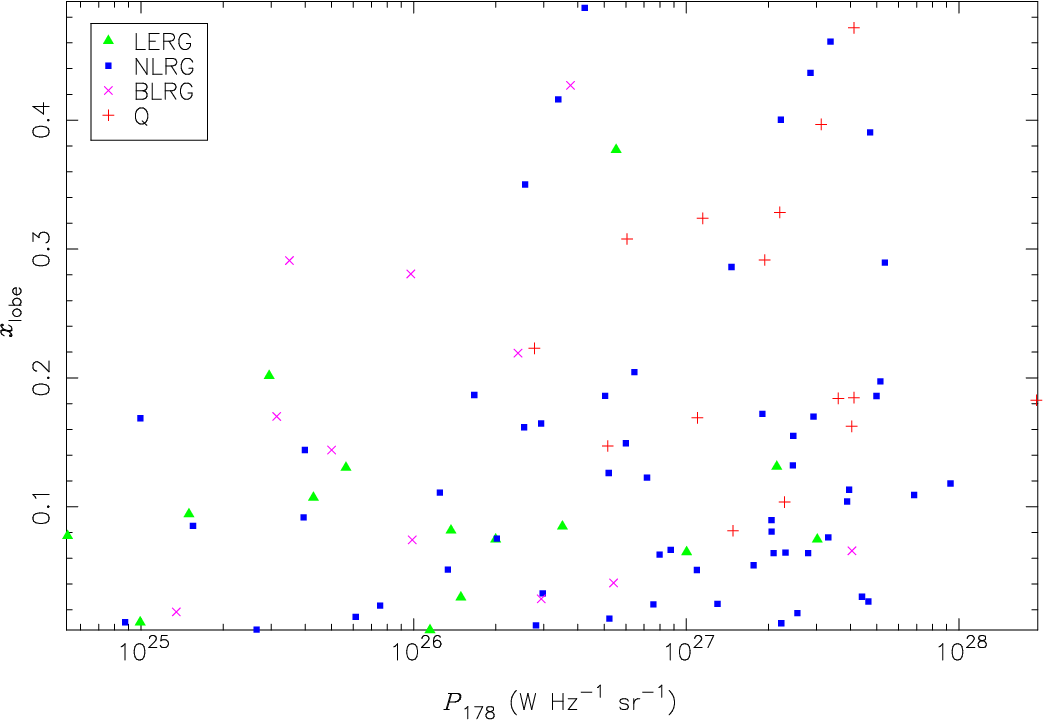,width=8cm}}
\caption{\label{fig:xlobe_P} The fractional separation difference as defined by the longer lobe side, $x_{\rm lobe}$, plotted against source luminosity at 178 MHz, $P_{178}$.}
\centerline{\epsfig{file=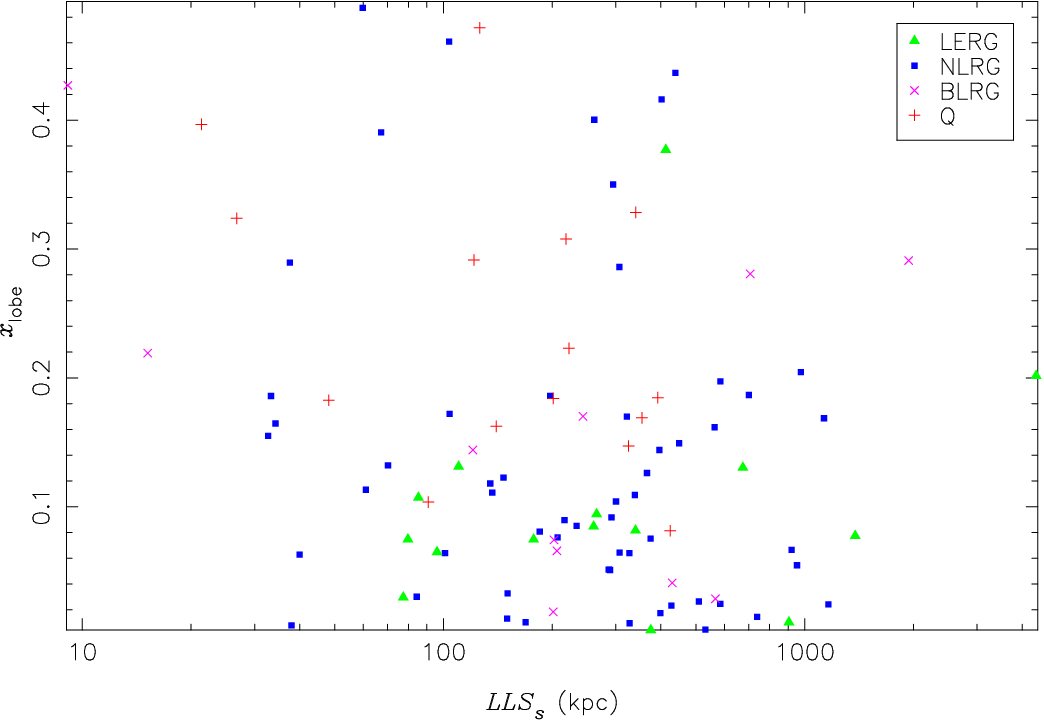,width=8cm}}
\caption{\label{fig:xlobe_LLS} The fractional separation difference as defined by the longer lobe side, $x_{\rm lobe}$, plotted against the largest linear source size, $LLS_{\rm s}$.}
\end{figure}

%
%
The resolution correction factor applied to the $LLS_{\rm l}$ data, as
discussed in section \ref{sec:lobesizobs}, should compensate for
systematic bias in lobe size asymmetry that might be introduced by observing
resolution. The fractional separation difference defined in terms of the
longer lobe, $x_{\rm lobe}$, is plotted as a function of redshift,
luminosity and source size in Figs \ref{fig:xlobe_z},
\ref{fig:xlobe_P} and \ref{fig:xlobe_LLS} respectively. There is no
trend in $x_{\rm lobe}$ with redshift but there is a tendency for the
high luminosity and smaller sources to have greater asymmetries.

A trend in $x_{\rm lobe}$ with redshift might have suggested
  environmental differences at different epochs; a tendency for
  greater asymmetries in higher power/smaller sources only could be
  consistent with asymmetries being imposed by environmental
  differences early in the source's development, if these sources are
  expected to be generally younger. That is, if relative environmental
  differences are not so great further out from the central engine the
  source's perceived asymmetry may be dominated by effects introduced
  while the source is still small; in this case the fractional asymmetry may decrease as the source expands, as any asymmetry represents a decreasing fraction of the source size. 

Broad-band studies of radio galaxies have demonstrated that in many
cases the the detected optical and/or infrared continuum emission from
the host galaxy is aligned with the radio axis (Chambers, Miley \& van
Breugel, 1987; McCarthy et al., 1987), a phenomenon known as the
`alignment effect'. This effect has been shown to be strong in sources
at redshifts $\gtrsim 0.6$, but less so for lower redshift samples. The
main processes implicated in the creation of the alignment effect are
photoionization from the central AGN and shock ionization from the
passage of the jet or lobes. Spectroscopic studies have suggested that
the photoionization mechanism dominates in more evolved, larger
sources but in smaller sources (especially those for which the radio
size of the sources is comparable to the emission line region) the
shock mechanism becomes important. This implies that source age is a
key factor when considering the extent to which the radio source will
affect its environment, with younger, less-evolved sources expanding
out through the host galaxy and gas environment and directly affecting
their kinematics (e.g., Inskip et al. 2002; Privon et al. 2008). But
the study of Inskip et al., which was made using multiple flux-limited
samples in order to break the redshift-luminosity degeneracy, has
suggested that the alignment depends on redshift as well as power,
implying that environmental differences at different epochs do
contribute to the overall picture.

The present sample cannot directly inform these latter results, as the redshift-luminosity degeneracy is not broken here. In fact, the tendency for asymmetry to be greater for higher power, smaller sources in this sample is not a result that contradicts or confirms any study of the alignment effect; binning the entire sample data by luminosity (using $P_{\rm
c}$) and source size (taking a cutoff of 200 kpc, corresponding to 100 kpc in lobe size, a somewhat arbitrary
choice based on the result of a trend in $R_{\rm ax}$ with lobe size
discussed in section \ref{sec:trendaxialratio}), K-S tests indicate
that the differences in the distributions of $x_{\rm lobe}$ across both
the luminosity and size ranges are not significant.
%
%
\subsubsection{Unification and beaming}\label{sec:x_lobe_beamin}
%
%
The median $x_{\rm lobe}$ for the different spectral classes, $<x_{\rm
lobe}>$, are given in Table \ref{tab:xlobe_median}. It can be seen that
the Qs are more asymmetric than the high luminosity NLRGs; a W-M-W
test shows that the difference is significant at the 99.7 per cent
confidence level. The $<x_{\rm lobe}>$ of the BLRGs and the low
luminosity NLRGs are not significantly different statistically and a K-S
test does not show any difference in the distribution of the
combined population of BLRGs and low-power NLRGs with respect to that of the LERGs. These findings
are generally consistent with those of H98 (with respect to the low
luminosity sources) and Best et al. (1995; with respect
to the high luminosity subsample), though both these studies defined $x_{\rm lobe}$
using $c-hs$.

In the case of the former result, Best et al. suggested that
relativistic effects might contribute to the greater asymmetry of Qs
relative to the corresponding NLRGs. If this were the case, we would
expect that the jet side would correlate with the longer lobe side.
The fractional separation difference defined by jet side, $x_{\rm
jet}$, uses the value of $LLS_{\rm l}$ on the straight jet side as $D_{1}$ and
$LLS_{\rm l}$ on the counterjet side as $D_{2}$. AL00 have studied
the observed distribution of $x_{\rm jet}$ for a sample of 3CR FRII
sources that includes the sources in this sample in addition to a
number of objects at $z>1$. They introduce an asymmetry parameter,
$\epsilon$, which is used to quantify the degree to which relativistic
effects contribute to the observed distribution of $x_{\rm jet}$ as
opposed to intrinsic and/or environmental effects. The asymmetry
parameter is defined as
%
%
\begin{equation}
\epsilon= 1-2 \frac{N(-FRII)}{N(+FRII)}
\end{equation}
%
%
where $N(-FRII)$ and $N(+FRII)$ are the numbers of sources with
positive and negative $x_{\rm jet}$ values. AL00 argued that an even
distribution of positive and negative $x_{\rm jet}$ about zero, giving
$\epsilon=-1$, implies that relativistic effects are not a significant
factor in the distribution. Where around 2/3 of the sample objects have
positive $x_{\rm jet}$ values, $\epsilon \sim 0$, which implies that
relativistic effects are as significant as intrinsic/environmental
ones. As relativistic effects become more important $\epsilon$ would
become increasingly positive. For their sample AL00 found an asymmetry
parameter of $-0.07 \pm 0.22$, for all the sources. For the radio
galaxies the result was $-0.3 \pm 0.32$ and for quasars, $0.33 \pm
0.36$. They concluded that the effects of relativistic motion on the
observed lobe size asymmetry distribution were not negligible and
that they were more important to the observed quasar asymmetries than
to the radio galaxies, consistent with unification models.

Here the sample is essentially the same as that of AL00 except for the
exclusion of those objects at $z>1$. The data have been reconsidered,
however, using only the jet-side information obtainable from the
sample maps. Where a definite or possible straight jet is detected, this is
taken as the jet side and no other information such as the
depolarization asymmetry associated with the source is used.
When determining $x_{\rm jet}$ in this way, it must be borne in mind that
the exclusion of those sources with no jet detections may bias the
data.

\begin{figure}
\centerline{\epsfig{file=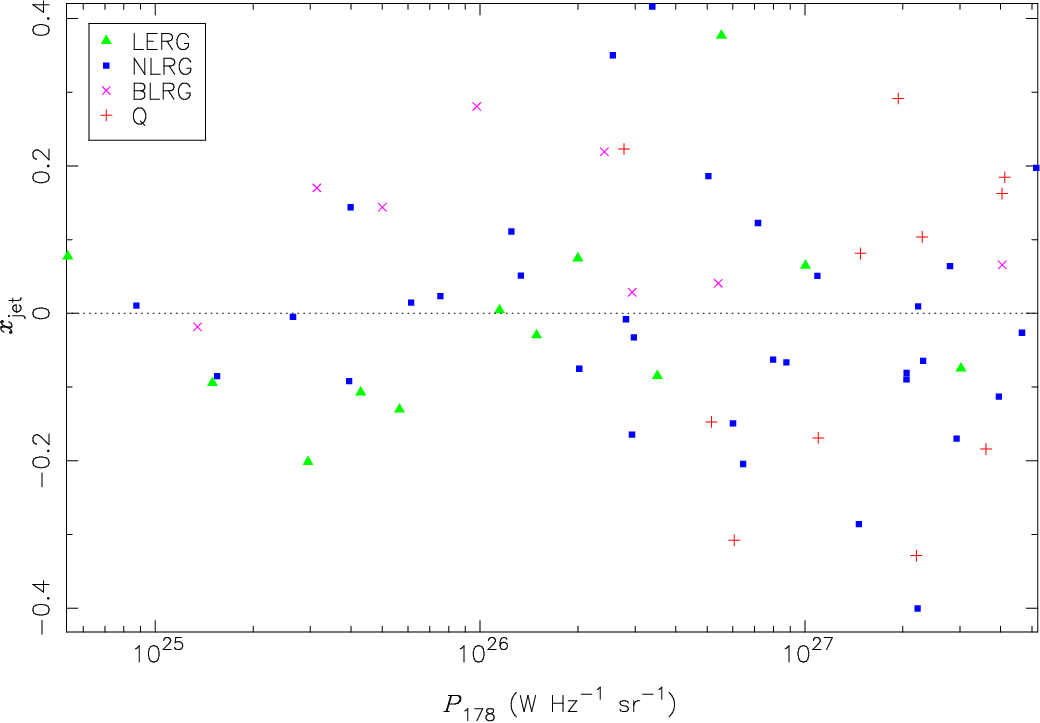,width=8cm}}
\caption{\label{fig:xjet_P} The fractional separation difference as
  defined by the straight jet side, $x_{\rm jet}$, plotted against source
  luminosity at 178 MHz, $P_{178}$. The dotted line
  shows a fractional separation difference of zero.}
\centerline{\epsfig{file=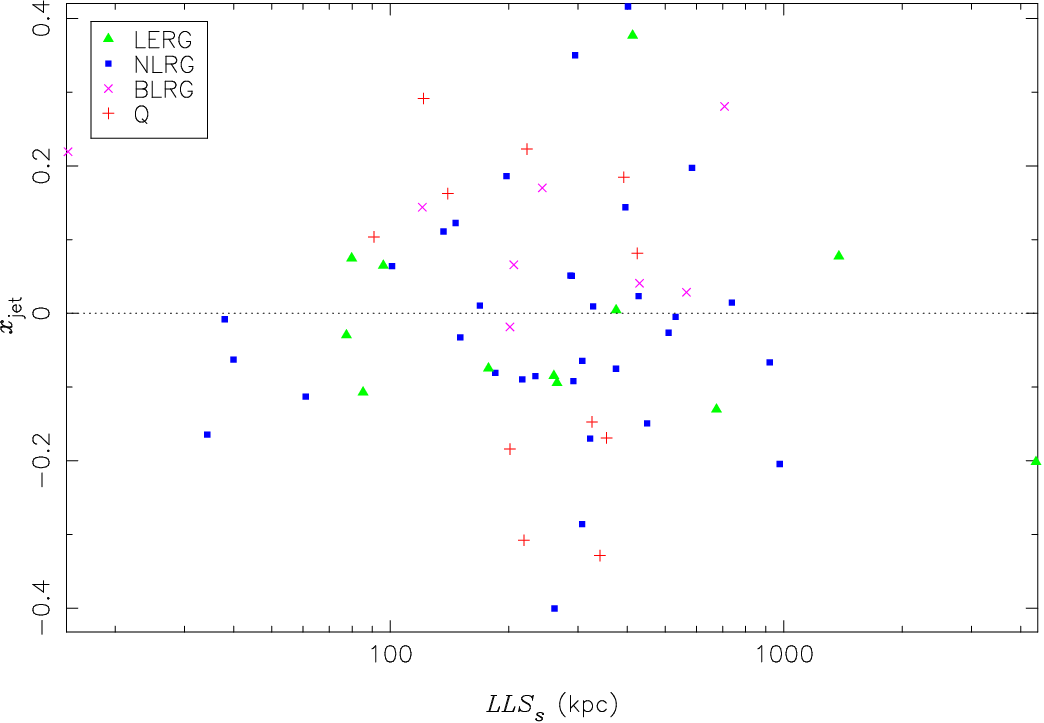,width=8cm}}
\caption{\label{fig:xjet_LLS}The fractional separation difference as
  defined by the straight jet side, $x_{\rm jet}$, plotted against the largest
  linear source size, $LLS_{\rm s}$. The dotted line
  shows a fractional separation difference of zero.}
\end{figure}

$x_{\rm jet}$ is plotted against $P_{178}$ in Fig. \ref{fig:xjet_P}.
Considering all sample sources with at least one definite or possible
straight jet, regardless of spectral class, the jet-side lobe is the
longer lobe in 49 per cent of the sources (using $LLS_{l}$). BLRGs
show the strongest apparent correlation of jet side and the longer
lobe with 7 out of 8 sources having positive $x_{\rm jet}$ values; a
marginally significant tendency at the 96.5 per cent confidence level.
Qs, NLRGs and LERGs do not have any significant tendency for positive
$x_{\rm jet}$ values. The distribution corresponds to an asymmetry
parameter of $\epsilon=-1.06$ for the combined Qs, BLRGs and NLRGs
sample, $\epsilon=0.08$ for the combined quasar and BLRG population
and $\epsilon=-1.86$ for the NLRGs, suggesting that relativistic
effects make a greater contribution to the observed asymmetry for Qs
and BLRGs than for NLRGs, which is in line with the prediction of
unification. This is consistent with the results of AL00, although the
evidence for relativistic effects making a significant contribution to
the $x_{\rm jet}$ distribution overall is weaker for this sample.

In Fig. \ref{fig:xjet_LLS}, $x_{\rm jet}$ is plotted against $LLS_{\rm
s}$. There is an apparent difference in the $x_{\rm jet}$ distribution
for smaller sources, ($LLS_{\rm s}\lesssim 200$ kpc), with fewer
negative values: $x_{\rm jet}$ is positive in 57 per cent of sources
with $LLS_{\rm s}\leq 200$ kpc; a
binomial test shows the tendency for $x_{\rm jet}$ to be positive in
the smaller sources is only weakly significant (although it can be noted that only one broad-line source has a negative value). A K-S test shows no significant difference between the large and small source $x_{\rm jet}$
distribution. 
%
%
\begin{table}
\scriptsize
\centering
\begin{minipage}{9cm}
\caption{\label{tab:xlobe_median}The median fractional separation difference as defined by lobe size, $<x_{\rm lobe}>$, for each of the spectral class distributions}
\begin{tabular}{lclc}  \hline
Spectral class & $<x_{\rm lobe}>$ & Spectral class & $<x_{\rm lobe}>$\\ \hline
Q          &  0.185  &  NLRG, high $P_{178 \rm MHz}$   & 0.111\\
B          &  0.144  &  NLRG, low $P_{178 \rm MHz}$    & 0.092\\
Q and BLRG &  0.184  &  NLRG                           & 0.109\\
LERG       &  0.082  &  BLRG and low $P_{178 \rm MHz}$ & 0.101\\

\hline
\end{tabular}
\end{minipage}
\end{table}
%
%
%
\begin{table*}
\caption{Summary of jet and hotspot detections for the sample \label{tab:summary}}

\begin{center}
\scriptsize
\begin{tabular}{llllllllllllllllll}  \hline
Source   &Class& \multicolumn{2}{c}{Straight jets} & \multicolumn{2}{c}{Hotspots} & Source   &Class& \multicolumn{2}{c}{Straight jets} & \multicolumn{2}{c}{Hotspots} & Source   &Class& \multicolumn{2}{c}{Straight jets} & \multicolumn{2}{c}{Hotspots} \\
         & & N lobe & S lobe & N lobe & S lobe &         & & N lobe & S lobe & N lobe & S lobe &         & & N lobe & S lobe & N lobe & S lobe \\\hline
4C12.03 & E & J & PCJ & 1 & 1 &3C196 & Q & none & none & 1 & 1 &3C289 & N & none & none & 0 & 1 \\
3C6.1 & N & none & none & 1 & 1 &3C200 & N & none & J & 1 & 1 &3C292 & N & none & none & 1 & 1 \\
3C16 & E & PJ & PCJ & 0 & 1 &4C14.27 & N & PJ & PCJ & 1 & 1 &3C295 & N & none & none & 1 & 1 \\
3C19 & N & none & PJ & 1 & 1 &3C207 & Q & none & J & 2 & 1 &3C299 & N & none & none & 3 & 1 \\
3C20 & N & J & none & 1 & 1 &3C215 & Q & none & PJ & 1 & 4 &3C300 & N & J & none & 1 & 2 \\
3C22 & B & J & none & 2 & 2 &3C217 & N & none & none & 2 & 1 &3C303 & B & J & none & 1 & 1 \\
3C33 & N & PCJ & PJ & 1 & 2 &3C216 & Q & none & none & 1 & 1 &3C319 & E & none & none & 1 & 0 \\
3C33.1 & B & none & J & 1 & 1 &3C219 & B & PCJ & J & 1 & 1 &3C321 & N & PJ & none & 1 & 1 \\
3C34 & N & PJ & PCJ & 1 & 1 &3C220.1 & N & J & none & 2 & 1 &3C325 & Q & PJ & none & 1 & 2 \\
3C35 & E & none & none & 1 & 0 &3C220.3 & N & none & none & 1 & 1 &3C326 & B & none & none & 0 & 1 \\
3C41 & N & none & J & 1 & 1 &3C223 & N & PJ & PCJ & 1 & 1 &3C330 & N & none & none & 1 & 3 \\
3C42 & N & none & none & 1 & 1 &3C225B & N & none & none & 1 & 1 &3C334 & Q & PCJ & J & 2 & 1 \\
3C46 & N & none & PJ & 2 & 2 &3C226 & N & none & none & 1 & 1 &3C336 & Q & none & J & 3 & 1 \\
3C47 & Q & none & J & 1 & 1 &4C73.08 & N & none & none & 1 & 1 &3C341 & N & none & J & 0 & 0 \\
3C55 & N & none & PJ & 1 & 2 &3C228 & N & none & J & 1 & 2 &3C337 & N & none & PJ & 1 & 1 \\
3C61.1 & N & none & none & 1 & 0 &3C234 & N & J & none & 1 & 1 &3C340 & N & none & none & 1 & 1 \\
3C67 & B & PJ & none & 2 & 1 &3C236 & E & PJ & none & 0 & 1 &3C349 & N & none & PJ & 1 & 1 \\
3C79 & N & none & none & 3 & 3 &4C74.16 & N & none & J & 1 & 1 &3C351 & Q & PJ & none & 1 & 1 \\
3C98 & N & J & none & 1 & 1 &3C244.1 & N & PJ & none & 2 & 1 &3C352 & N & J & none & 2 & 0 \\
3C109 & B & none & PJ & 1 & 1 &3C247 & N & none & none & 1 & 1 &3C381 & B & none & none & 1 & 1 \\
4C14.11 & E & PJ & none & 2 & 1 &3C249.1 & Q & J & none & 4 & 1 &3C382 & B & J & none & 1 & 1 \\
3C123 & E & none & none & 1 & 1 &3C254 & Q & none & none & 1 & 1 &3C388 & E & none & J & 1 & 1 \\
3C132 & E & none & PJ & 1 & 1 &3C263 & Q & none & J & 1 & 1 &3C390.3 & B & J & none & 1 & 1 \\
3C153 & N & PCJ & PJ & 1 & 3 &3C263.1 & N & none & none & 1 & 1 &3C401 & E & none & J & 1 & 1 \\
3C171 & N & CJ & J & 1 & 1 &3C265 & N & PJ & none & 1 & 1 &3C427.1 & E & PJ & none & 2 & 1 \\
3C172 & N & none & none & 1 & 2 &3C268.1 & N & none & none & 2 & 1 &3C433 & N & J & none & 0 & 4 \\
3C173.1 & E & J & none & 1 & 1 &3C268.3 & B & none & none & 1 & 1 &3C436 & N & none & J & 3 & 1 \\
3C175 & Q & none & J & 1 & 2 &3C274.1 & N & none & PJ & 1 & 1 &3C438 & E & J & none & 1 & 1 \\
3C175.1 & N & none & PJ & 1 & 1 &3C275.1 & Q & J & none & 1 & 1 &3C441 & N & J & none & 1 & 1 \\
3C184 & N & none & none & 3 & 1 &3C277.2 & N & none & PJ & 3 & 1 &3C452 & N & CJ & J & 1 & 1 \\
3C184.1 & N & PJ & none & 1 & 1 &3C280 & N & none & none & 1 & 1 &3C455 & Q & none & none & 1 & 1 \\
DA240 & E & none & PJ & 1 & 2 &3C284 & N & none & none & 1 & 1 &3C457 & N & none & none & 2 & 1 \\
3C192 & N & none & PJ & 1 & 2 &3C285 & E & J & none & 1 & 1 &\\\hline

\end{tabular}\\
\end{center}

Column [1]:3CR catalogue source name. Column [2]: Spectral class. L: low excitation galaxies, Q: quasars, B and N: broad and narrow line radio galaxies respectively. Column [3] \& [4]: Jet detections for north and south lobes respectively. J: definite jet, PJ: possible jet, CJ: counterjet, PCJ: possible counterjet. Column [5] \& [6]: Number of hotspots in the north and south lobe respectively. Columns [7] to [12] and columns [13] to [18] as for columns [1] to [6].\\                             
\end{table*}
%
%
%
\section{Cores and Jets}\label{sec:cores_jets}
%
%
%
\subsection{Observing effects}\label{sec:obscoresjets}
%
%
The effects of varying observing resolution should introduce little
bias into the core measurements as they are bright features that are
typically unresolved. For the jets, however, observing resolution will
have an effect on detectability, which we now consider for the case of the total jet features.

30 per cent of sources in the sample have a definite jet, and a further 34
per cent have a possible jet; a summary of straight jet detections for the
entire sample is given in Table \ref{tab:summary}, with the detection
rate broken down by spectral class in Table
\ref{tab:jet_detection_class}. The total jet classifications are the same as for the straight jets but for the following three exceptions: the definite total jet is in the northern lobe in source 3C171 with a definite total counterjet detected in the southern lobe, a possible total counterjet is detected in the southern lobe of 3C20 and a definite total counterjet is detected in the southern lobe of 3C438.

The appearance of the jet features
varies from source to source, and many of the detected jets do not
cover the entire length from the core to the hotspot feature. 
Observing resolution and sensitivity should be an
important factor in jet detectability; however, the nature of any
dependence on observational constraints is difficult to evaluate. The
variation of observing resolution across the sample is potentially a
source of observational bias, more so as this is accompanied by a
variation in observing sensitivity.

\begin{figure}
\centerline{\epsfig{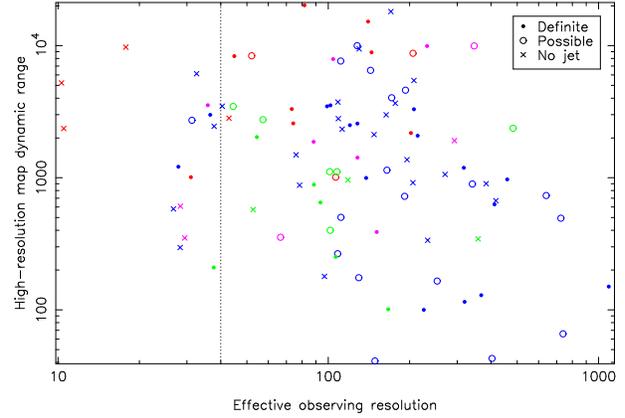}}
\caption{\label{fig:dyn_res} The dynamic range plotted against the
  high-resolution effective observing resolution, binned by total jet status. Filled circles: at least one
  definite jet detected, open circles: no definite jet detected but at
  least one possible jet, diagonal cross: no jet feature detected. Green: low excitation radio galaxies, blue: narrow line radio galaxies, magenta: broad line radio galaxies and red: quasars (on-line colour version). The
  dotted line shows a high-resolution effective observing resolution of 40.}
\end{figure}

To investigate the effect on total jet detectability, in Fig. \ref{fig:dyn_res} we plot the dynamic range (defined as the ratio
of the maximum intensity to the off-source root mean square noise)
against the effective observing resolution corresponding to the
highest-resolution map for all the sample sources, binning by jet status. Sources lacking total jet features entirely
are observed across the resolution and sensitivity range; thus there is
no simple trend for those sources observed with relatively high
resolution and high dynamic range to be associated with jet features.

\begin{figure}
\centerline{\epsfig{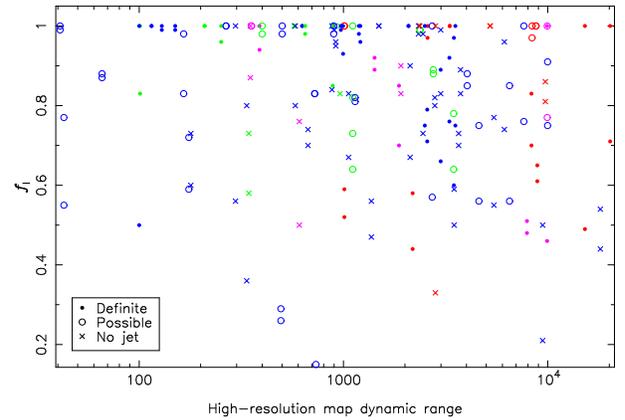}}
\caption{\label{fig:flob_dyn_jet} The fractional observed lobe length, $f_{\rm l}$, plotted against the dynamic range, binned by total jet status. Filled circles: at least one definite jet detected, open circles: no definite jet detected but at least one possible jet, diagonal cross: no jet feature detected. Green: low excitation radio galaxies, blue: narrow line radio galaxies, magenta: broad line radio galaxies and red: quasars (on-line colour version).}
\end{figure}

\begin{figure}
\centerline{\epsfig{file=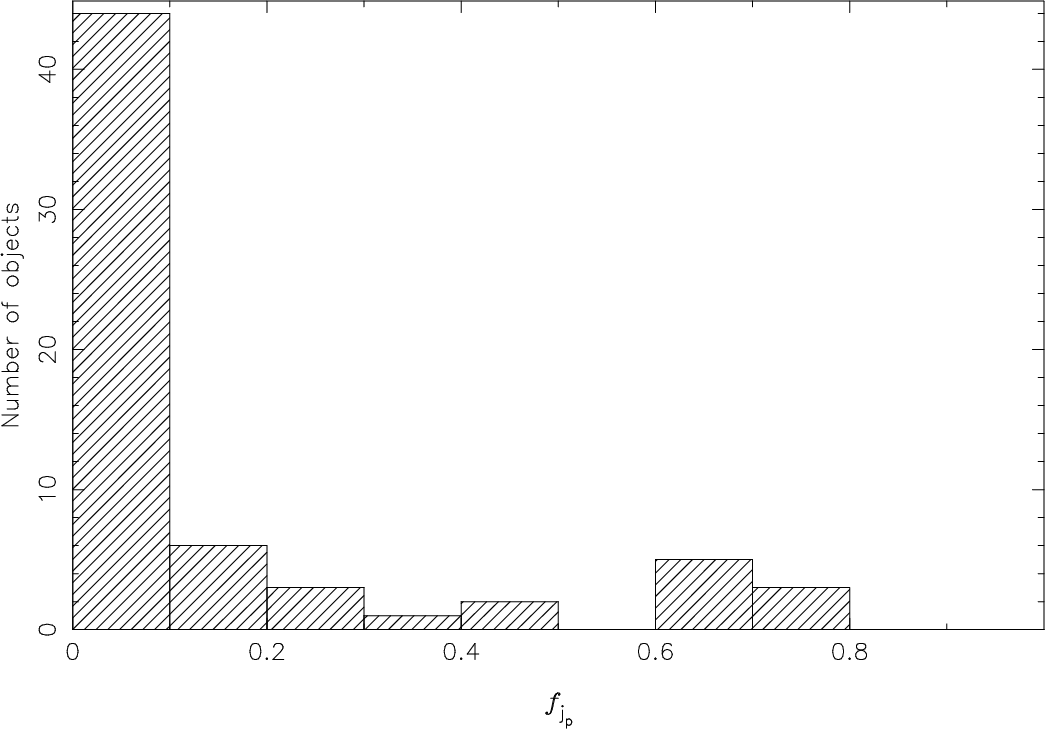,width=8cm}}
\caption{\label{fig:histjetpos} Histogram of the fractional jet position, $f_{\rm j_{p}}$, for the sample.}
\centerline{\epsfig{file=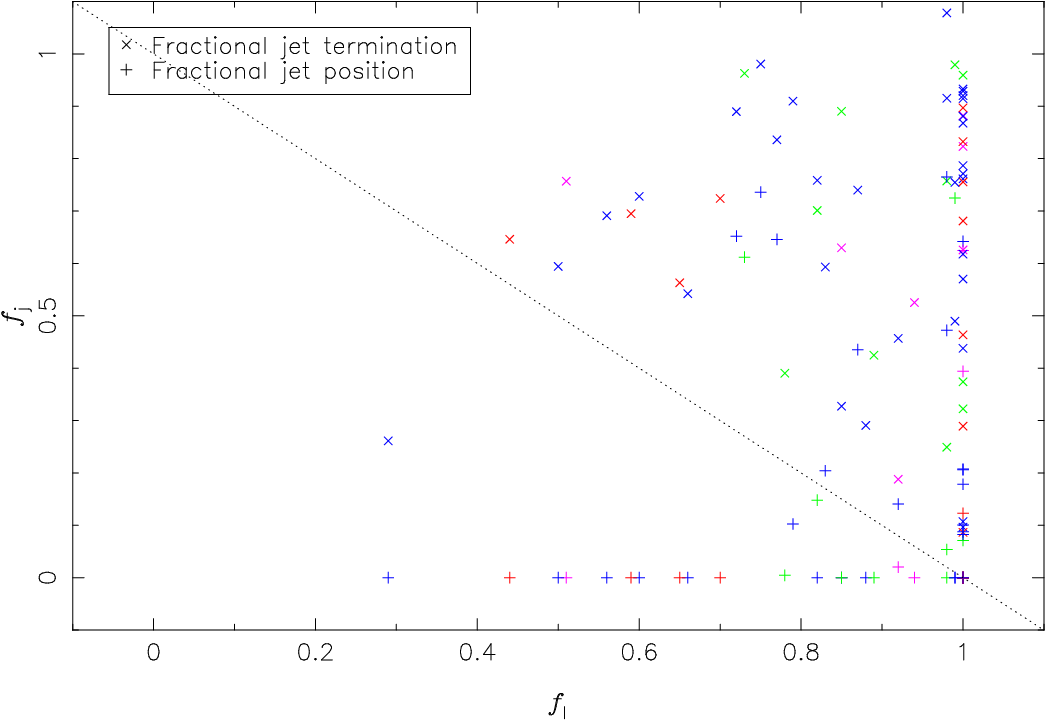,width=8cm}}
\caption{\label{fig:jetpos_flobs} The fractional jet position, $f_{\rm j_{p}}$, and fractional jet termination, $f_{\rm j_{t}}$, plotted against the fractional observed lobe length, $f_{\rm l}$, of the corresponding lobe. Green: low excitation radio galaxies, blue: narrow line radio galaxies, magenta: broad line radio galaxies and red: quasars (on-line colour version). The dotted line is the line of $f_{\rm j_{t}}=1-f_{\rm l}$.}
\centerline{\epsfig{file=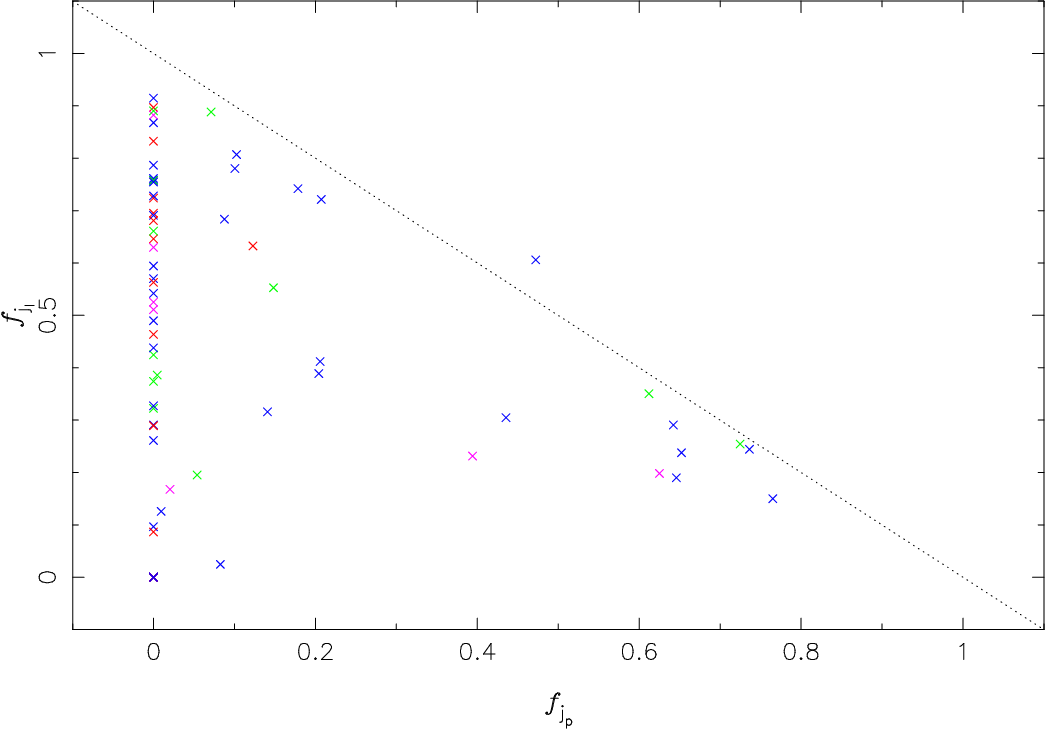,width=8cm}}
\caption{\label{fig:length_position} The fractional jet length, $f_{\rm j_{l}}$, plotted against the fractional jet position, $f_{\rm j_{p}}$, of the corresponding lobe. Green: low excitation radio galaxies, blue: narrow line radio galaxies, magenta: broad line radio galaxies and red: quasars (on-line colour version). The dotted line is the line of $f_{\rm j_{l}}=1-f_{\rm j_{p}}$.}
\end{figure}

Fig. \ref{fig:flob_dyn_jet} plots the fractional observed lobe length,
$f_{\rm l}$, against dynamic range, with data binned by total jet detection
status. It might be expected that if the detected lobe emission in a source is more extensive (with $f_{\rm l}$ values closer to 1), it would be more difficult to detect a jet feature, but this does not seem to be the case. Definite and possible jet features are detected across the range in $f_{\rm l}$. An additional aspect of jet detectability that can be considered is
the jet location within the lobe. As mentioned previously, while in
some sources a bright jet is observed to extend from the core to the
lobe extremity, in many sources the jet is detected along a fraction
of this length only. In Fig. \ref{fig:histjetpos}, a histogram shows
the distribution of the fractional jet position, $f_{\rm j_{p}}$, for the definite and possible
jets; 69 per cent of these of objects have a jet that is traced from
near the core, having $f_{\rm j_{p}} < 0.1$. In order to examine the
possibility that the jets are systematically becoming obscured as they
progress through the lobe, the jet termination, $f_{\rm j_{t}}$, is
considered alongside $f_{\rm j_{p}}$. $f_{\rm j_{p}}$ and $f_{\rm
j_{t}}$ are plotted against the fractional lobe length, $f_{l}$, in
Fig. \ref{fig:jetpos_flobs}. The dashed line represents $f_{\rm
j_{t}}$, corresponding to a given $f_{l}$, that would be obtained if
the observed total jet terminated on reaching the inner edge of the
lobe. As there is no tendency for the $f_{\rm j_{t}}$ data to crowd
toward this line, jets are generally observed to extend into the lobe.

If the emitting material in the beam decelerated as it progresses from
the core, and if the jet were detected more easily as this happens,
then it might be expected that once the jet becomes detectable it could
be traced to the hotspot region or the lobe extremity. This would be
consistent with an anticorrelation between jet length and jet position. In
Fig. \ref{fig:length_position} $f_{\rm j_{l}}$ is plotted against
$f_{\rm j_{p}}$ and the dashed line shows the jet length corresponding
to a given jet position that would indicate that the jet is observed
continously from its base to the lobe extremity. As there is no
crowding toward this line, there is no strong tendency for this to be
the case.

We conclude that, although the observing resolution should affect jet
detectability, there is no obvious systematic bias in the sample that
can be simply compensated for. The jet is not obviously less easily
detected in sources with extensive lobes and the lack of jet detection
does not appear to be a result of high lobe detectability.
%
%
\begin{table*}
\centering

\caption{\label{tab:jet_detection_class}The jet detection data for the sample based on spectral class}
\begin{tabular}{llclclclc}  \hline
Spectral class &\multicolumn{2}{c}{jet features}&\multicolumn{2}{c}{definite jets}&\multicolumn{2}{c}{possible jets}&\multicolumn{2}{c}{null detection}\\ \hline
Q        & 73.3\%&(11/15)& 53.3\%&( 8/15) & 20.0\%&( 3/15)  & 26.7\%&( 4/15)\\
B        & 72.7\%&( 8/11)& 54.5\%&( 6/11) & 18.2\%&( 2/11)  & 27.3\%&( 3/11)\\
N        & 59.6\%&(34/57)& 28.1\%&(16/57) & 31.6\%&(18/57)  & 40.4\%&(23/57)\\
E        & 80.0\%&(12/15)& 40.0\%&( 6/15) & 40.0\%&( 6/15)  & 20.0\%&( 3/15)\\

\hline
\end{tabular}
\end{table*}
%
%
%
\subsection{Trends with $P_{178}$, $z$ and size}
%
%
%
\subsubsection{Cores}\label{sec:cores}
\begin{figure}
\centerline{\epsfig{file=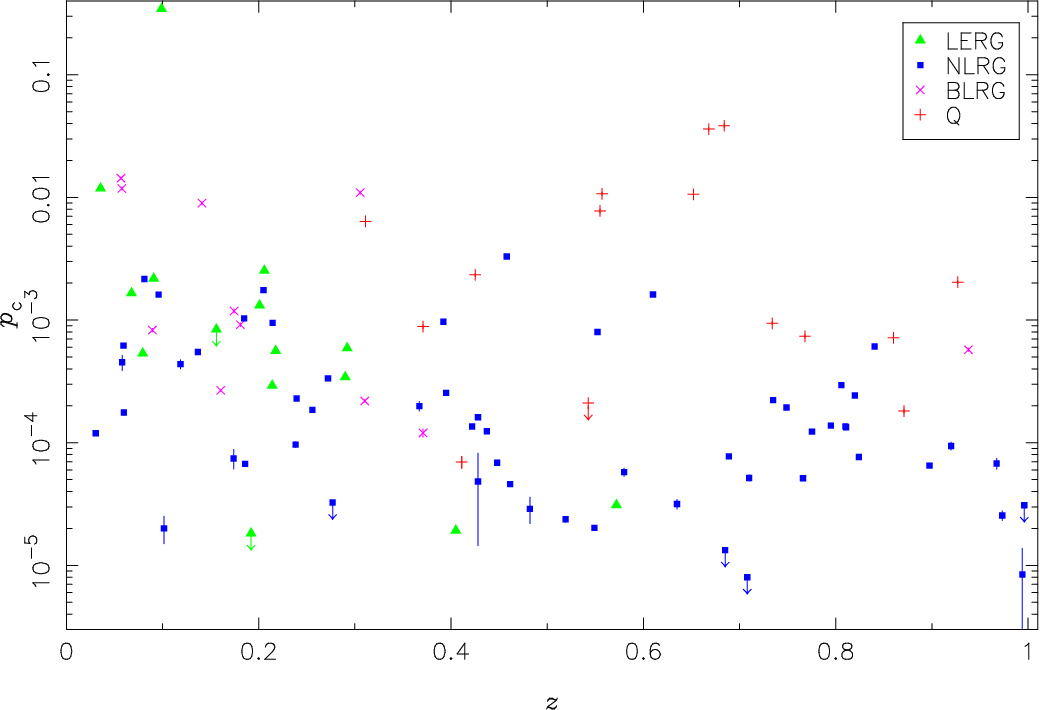,width=8cm}}
\caption{\label{fig:coreprom_z}The core prominence, $p_{\rm c}$, plotted against redshift, $z$. Vertical bars indicate errors, arrows indicate upper limits.}
\centerline{\epsfig{file=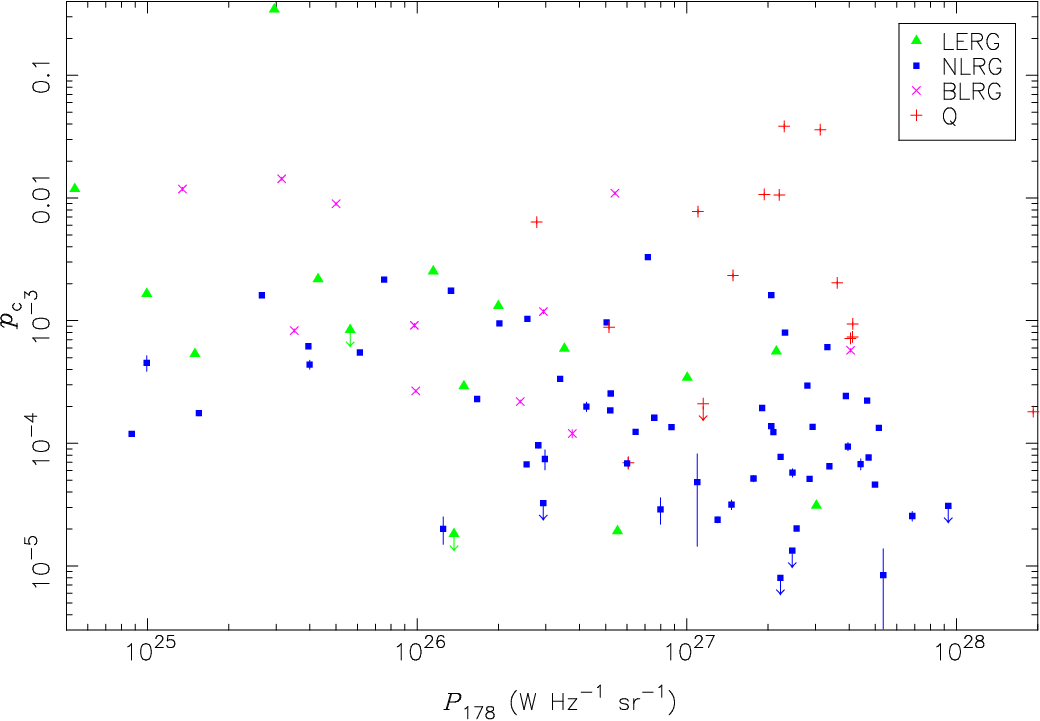,width=8cm}}
\caption{\label{fig:coreprom_P}The core prominence, $p_{\rm c}$, plotted against the source luminosity at 178~MHz, $P_{178}$. Vertical bars indicate errors, arrows indicate upper limits.}
\centerline{\epsfig{file=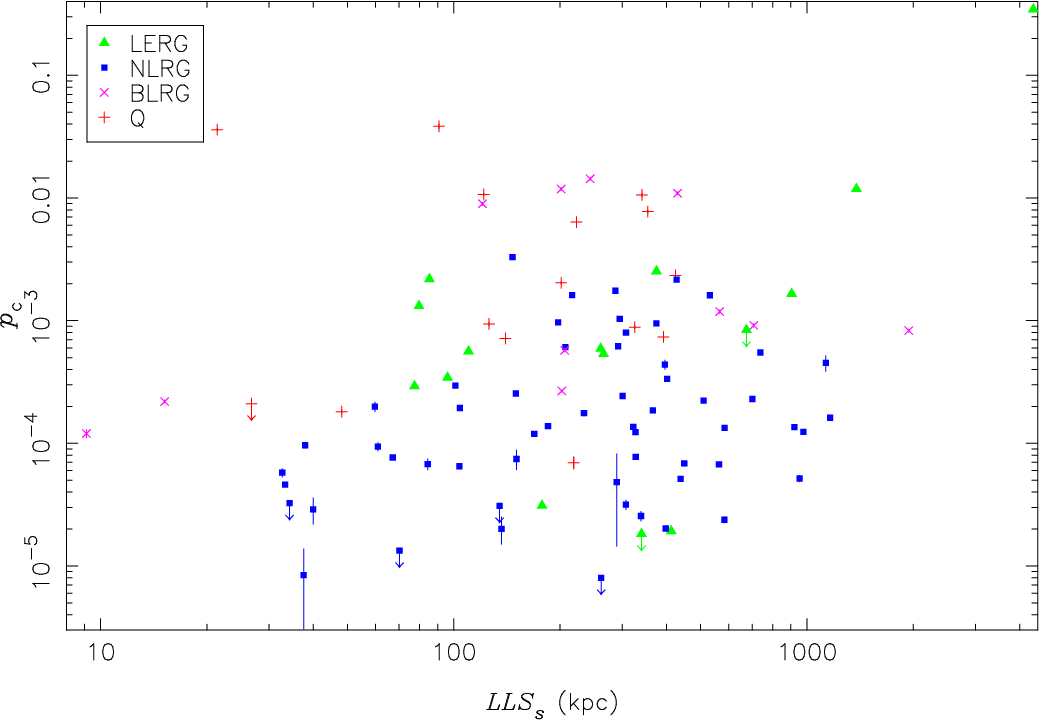,width=8cm}}
\caption{\label{fig:coreprom_LLS}The core prominence, $p_{\rm c}$, plotted against the largest linear source size, $LLS_{\rm s}$. Vertical bars indicate errors, arrows indicate upper limits.}
\end{figure}

%
%
The core prominence distribution is plotted as a function of redshift
in Fig. \ref{fig:coreprom_z} and as a function of luminosity in Fig.
\ref{fig:coreprom_P}. Any trend with redshift is weak but there is a tendency for the core prominence to decrease with
increasing source luminosity and there is a lack of low-luminosity
sources with faint cores. A Peto-Prentice test in which the sample objects
(inclusive of all spectral classes) are divided by source luminosity
at $P_{\rm c}$ shows that the trend for lower core prominence in higher-power sources is
significant at the 99.5 per cent confidence level. 

From Fig. \ref{fig:coreprom_LLS} it can be seen that there is a
tendency for smaller sources in general to be associated with lower
core prominence. Binning the sample data by size, including all
classes, a Peto-Prentice test between sources above and below
$LLS_{\rm s}=200$ kpc shows no significant difference. (The 200-kpc
size criterion was chosen as evidence was found for morphological
differences in lobe sizes above and below $\sim 100$ kpc; see Section
\ref{sec:trendaxialratio}.) However, if the sources are divided by
spectral class, there is a significant tendency (at the 96.5 per cent
confidence level) on a Peto-Prentice test for the smaller NLRGs to have
lower core prominences: this was also noted by H98. It is not clear
whether this is simply a result of the core-prominence/luminosity
inverse correlation noted above in combination with the known luminosity/size
inverse correlation, or whether (as suggested by H98) it is a genuine
physical effect that is masked in other spectral types by beaming effects.
%
%
\subsubsection{Jets}\label{sec:jets}
\begin{figure}
\centerline{\epsfig{file=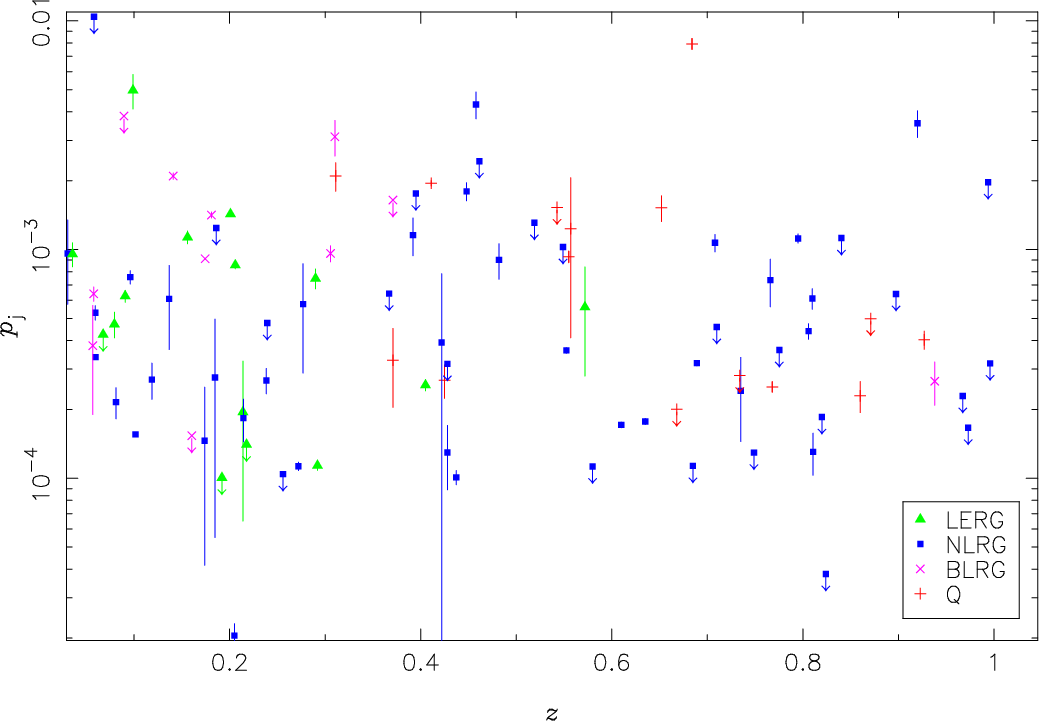,width=8cm}}
\caption{\label{fig:jetprom_z}The straight jet prominence, $p_{\rm j}$, plotted against $z$. Vertical bars indicate errors, arrows indicate upper limits.}
\centerline{\epsfig{file=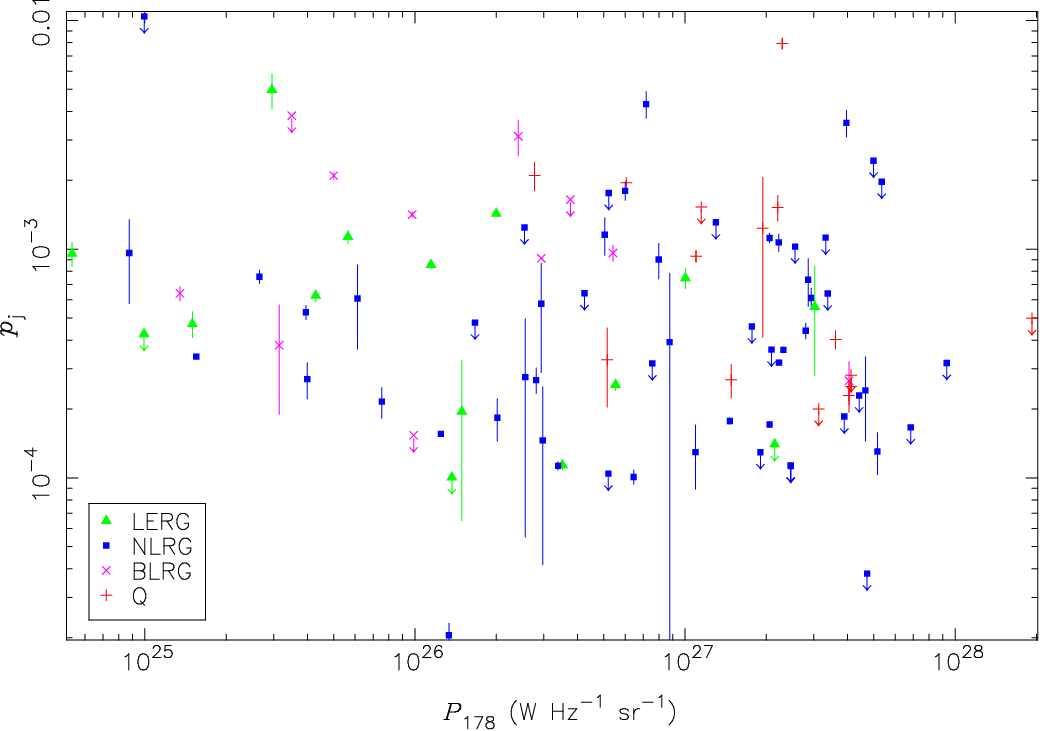,width=8cm}}
\caption{\label{fig:jetprom_P} The straight jet prominence, $p_{\rm j}$, plotted against the source luminosity at 178~MHz, $P_{178}$. Vertical bars indicate errors, arrows indicate upper limits.}
\end{figure}

%
%
In Figs \ref{fig:jetprom_z} and \ref{fig:jetprom_P} the straight jet
prominence is plotted against redshift and luminosity; there is no
trend in the distribution with either parameter. A slightly broader
range is found at $z\lesssim 0.3$; there is a marginally significant
(93.7 per cent confidence level) difference between these low redshift
sources and those with $z>0.3$ on a K-S test. However, taking into
account the upper limits in the data, a Peto-Prentice test does not
suggest any significant difference in the distribution of $p_{\rm j}$
between the high and low luminosity sources, nor is any trend found
with respect to redshift or source size. The many limits in the
$p_{\rm j}$ data (around one third of the sample sources) may mask any
trend.
%
%
\subsection{Unification and beaming}\label{sec:coresjetsuni}
%
%
%
\subsubsection{Cores}\label{sec:cores2}
%
%
The median core prominences for the different spectral classes are
given in Table \ref{tab:coreprom_median}. The core prominence
distribution of the Qs and BLRGs was compared with that of the NLRGs
using a Peto-Prentice test. The difference in the distributions of Qs and high-luminosity NLRGs is significant above the 99.9 per cent
confidence level, whilst that between the
BLRGs and low power NLRGs is significant at the 97.4 per cent level. We find no significant difference between the $p_{\rm c}$
distributions of the LERGs and the BLRGs and low power NLRGs on a
Peto-Prentice test.

In section \ref{sec:cores} it was shown that there is evidence that the higher-power sources are associated with lower core prominence. Considering the spectral classes separately, Peto-Prentice tests show that the high-power BLRGs and NLRGs are significantly lower than the corresponding low-power BLRGs and NLRGs (note that there are only 2 sources in the high-power BLRG subsample), but there is no significant trend in the Qs or LERGs. The core prominence data are consistent with the idea that higher
luminosity sources have higher Lorentz factors. The
beaming factor is $\propto \gamma(1-\beta\cos\theta)^{-2}$ (Scheuer \&
Readhead 1979, assuming the spectral index for the core features to be
0), and the range in this factor increases with $\gamma$. For a given
$\gamma$, above a certain threshold angle of orientation with respect to the
observer's line-of-sight, $\theta_{\rm t}$, the emission will be Doppler suppressed and the observed core
prominence will be lower than the intrinsic value. As $\gamma$
increases, $\theta_{\rm t}$ decreases and the suppression of parsec
scale jet emission at large $\theta$ becomes strong. In Qs, VLBI
observations have reported $\gamma\sim 5$--$10$ for some sources (e.g., Zensus 1997; Hough et al. 2002). This would result in Doppler-boosted cores in broad-line objects and
Doppler-suppressed cores for the equivalent NLRGs, assuming
$\theta_{\rm c}\approx40-50^{\circ}$. If sources of lower luminosity
were associated with lower $\gamma$, the core prominence of BLRG
sources would not be as strongly boosted as the Qs (though there is no
significant difference between the core prominence of Qs and BLRGs) and also the Doppler
suppression of the NLRG cores would be less strong. This latter point could lead to generally lower core prominence being found in {\it higher}-luminosity NLRGs.
%
%
\begin{table}
\scriptsize
\centering
\begin{minipage}{8cm}
\caption{\label{tab:coreprom_median}The median core prominence, $<p_{\rm c}>$, for the different spectral classes.}
\begin{tabular}{lclc}  \hline
Spectral class & $<p_{\rm c}> / 10^{-3}$ &Spectral class & $<p_{\rm c}> / 10^{-3}$ \\ \hline
Q          &  2.030  &  NLRG, high $P_{178 \rm MHz}$   & 0.086\\
B          &  0.915  &  NLRG, low $P_{178 \rm MHz}$    & 0.335\\
Q and BLRG &  1.061  &  NLRG                           & 0.134\\
LERG       &  0.592  &  BLRG and low $P_{178 \rm MHz}$ & 0.524\\\hline

\end{tabular}
\end{minipage}
\end{table}
%
%
%
\subsubsection{Jets}\label{sec:jets-asurv}
%
%
The one-sidedness of FRII jets is difficult
 to account for without beaming. There are very few counterjets
 detected in the sample but the general symmetry of the
 extended structure requires there to be bi-polar beams emanating from
 the central engine. The fact that so few counterjets are detected at
 all suggests that kiloparsec scale jet emission is beamed.

The straight jet detection statistics indicate a difference with spectral class
that is consistent with beaming models (see Table
\ref{tab:jet_detection_class}, previously discussed in section
\ref{sec:obscoresjets}). The Qs and BLRG sources have a similarly high
jet-feature detection rate ($\sim73$ per cent), with definite jets
detected in $\sim53$ per cent of the sources. (The Q sources with no
jet detected are observed at relatively low resolution, $\lesssim 40$
beams across the source in the high-resolution map; this is not true for the BLRGs). For the NLRG
class, jet features are found in 60 per cent of sources, with only 28
per cent of NLRGs having a definite jet detected. The upper limits of
the NLRG and LERG sources are distributed across the range in
observing resolution.

The median straight jet prominence {\it for detected jets only} for
each of the spectral classes is shown in Table
\ref{tab:jetprom_median}. When we compare the
straight jet prominences of the broad-line objects (Q and BLRG) to the
NLRG with a Peto-Prentice test, taking upper limits into account, we find a difference that is
significant at the 99.6 per cent confidence limit in the sense that the
median prominence of the broad-line objects is significantly higher.
This confirms the earlier result of H98. The difference in jet prominence is still
significant if the sample is divided into low-luminosity and
high-luminosity sub-samples, at the 98.7 per cent and 94.4 per cent
confidence levels respectively -- the marginal significance in the
high-luminosity bin presumably arises from the large fraction of upper
limits in the NLRG in this sample. There is no significant difference
between the LERG and other (NLRG/BLRG/Q) straight jet prominence
distributions. These differences in prominence are consistent with the
expectations from unified models and beaming.

In addition, the Laing-Garrington effect (as discussed
in section \ref{sec:intro}), in which the jet occurs in the
lobe that shows less depolarization, can be considered. Depolarization data were available in the literature (Table \ref{tab:depol}) for 60 of the sample sources, 41 of which have detected jets (possible
and definite). Of these 41 sources, 30 have the jet on the less
depolarized side (73 per cent). On a binomial test this is a
significant trend at the 99.8 per cent confidence limit. The effect
would be expected to be stronger for broad line sources and there were
depolarization data available for 16 of the 26 Qs and BLRGs: 75 per
cent of these showed correlation between the jet-side lobe and the
less depolarized lobe (marginally significant at the 96 per cent level
on a binomial test). Given that other properties of the sources and
their environments are known to affect source depolarization, and that
our data are necessarily heterogeneous, these results also seem to be
in good agreement with the expectation from beaming models.
\begin{table*}
\caption{Depolarisation data taken from the literature for our sample\label{tab:depol}}
\scriptsize
\begin{tabular}{lllllllllllllllllllll}  \hline
Source  & Jet side & \multicolumn{2}{c}{Depolarization} & \multicolumn{2}{c}{$\lambda$} & Reference&Source  & Jet side & \multicolumn{2}{c}{Depolarization} & \multicolumn{2}{c}{$\lambda$} & Reference\\
        &          & N lobe        & S lobe             & high [GHz] & low [GHz]        & &        &          & N lobe        & S lobe             & high [GHz] & low [GHz]        &  \\\hline
4C12.03 & N & -    & -    & -   & -   & -  &3C236   & N & -    & -    & -   & -   & -  \\
3C6.1   & - & 0.50 & 0.27 & 8.1 & 2.7 & 1  &4C74.16 & S & 0.23 &-0.01 & 5.0 & 1.5 & 6  \\
3C16    & N &-0.38 &-0.28 & 4.8 & 1.4 & 2  &3C244.1 & N & 0.27 & 0.30 & 8.1 & 1.4 & 1  \\
3C19    & S & -    & -    & -   & -   & -  &3C247   & - & 0.84 & 0.00 & 5.0 & 1.5 & 12 \\
3C20    & N & 0.50 & 0.67 & 8.1 & 1.4 & 1  &3C249.1 & N & 0.27 & 0.37& 5.0 & 1.5 & 10 \\
3C22    & N &-0.05 & 0.54 & 5.0 & 1.5 & 3  &3C254   & - & 0.27 & 0.68 & 5.0 & 1.5 & 12 \\
3C33    & S & -    & -    & -   & -   & -  &3C263   & S & -    & -    & -   & -   & -  \\
3C33.1  & S & 0.05 & 0.04 & 4.8 & 1.5 & 4  &3C263.1 & - & 0.91 & 0.56& 5.0 & 1.5 & 12 \\
3C34    & N & 0.21 & 0.38 & 4.8 & 1.5 & 5  &3C265   & N &-0.26 &-0.19 & 4.8 & 1.4 & 2  \\
3C35    & - & -    & -    & -   & -   & -  &3C268.1 & - &-0.27 & 0.19 & 4.8 & 1.4 & 2 \\
3C41    & S & 0.06 & 0.02 & 5.0 & 1.5 & 6  &3C268.3 & - & -    & -    & -   & -   & -  \\
3C42    & - &-0.01 & 0.00 & 4.8 & 1.4 & 2  &3C274.1 & S & 0.11 & 0.08 & 2.4 & 1.0 & 9 \\
3C46    & S &-0.12 &-0.07 & 4.8 & 1.4 & 2  &3C275.1 & N & 0.05 & 0.42 & 5.0 & 1.5 & 6  \\
3C47    & S & 0.82 & 0.05 & 4.9 & 1.5 & 7  &3C277.2 & S &-0.09 & 0.54 & 5.0 & 1.5 & 11\\
3C55    & S & 0.11 & 0.00 & 5.0 & 1.5 & 3  &3C280   & - &-0.21 &-0.29 & 4.8 & 1.4 & 2  \\
3C61.1  & - & 0.58 & 0.54 & 4.8 & 1.5 & 4  &3C284   & - & 0.04 & 0.31 & 2.4 & 1.0 & 9 \\
3C67    & N & 0.35 &-0.65 & 4.9 & 1.6 & 8  &3C285   & N & -    & -    & -   & -   & -  \\
3C79    & - & 0.44 & 0.05 & 2.4 & 1.0 & 9  &3C289   & - & 0.87 & 0.66 & 5.0 & 1.5 & 12\\
3C98    & N & -    & -    & -   & -   & -  &3C292   & - & -    & -    & -   & -   & -\\
3C109   & S & 0.53 & 0.52 & 4.8 & 1.5 & 4  &3C295   & - & -    & -    & -   & -   & -\\
4C14.11 & N & -    & -    & -   & -   & -  &3C299   & - &-0.30 &-0.09 & 4.8 & 1.4 & 2\\
3C123   & - & -    & -    & -   & -   & -  &3C300   & N & 0.11 & 0.77 & 2.4 & 1.0 & 9\\
3C132   & S & 0.31 & 0.14 & 2.4 & 1.5 & 9  &3C303   & N & -    & -    & -   & -   & -\\
3C153   & S & -    & -    & -   & -   & -  &3C319   & - & -    & -    & -   & -   & -\\
3C171   & S & 0.14 & 0.40 & 8.1 & 2.7 & 1  &3C321   & N & -    & -    & -   & -   & -\\
3C172   & - & 0.26 & 0.27 & 2.4 & 1.0 & 9  &3C325   & N & -    & -    & -   & -   & -\\
3C173.1 & N & -    & -    & -   & -   & -  &3C326   & - & -    & -    & -   & -   & -\\
3C175   & S & 1.00 & 1.00 & 5.0 & 1.5 & 10 &3C330   & - & 0.09 & 0.29 & 5.0 & 1.5 & 13\\
3C175.1 & S & 0.11 & 0.54 & 5.0 & 1.5 & 11 &3C334   & S & 0.23 & 0.09 & 5.0 & 1.5 & 6\\
3C184   & - & -    & -    & -   & -   & -  &3C336   & S & 1.00 & 0.74 & 2.4 & 1.0 & 10\\
3C184.1 & N & -    & -    & -   & -   & -  &3C341   & S &-0.18 &-0.21 & 4.8 & 1.4 & 2\\
DA240   & S & -    & -    & -   & -   & -  &3C337   & S & 0.16 & 0.47 & 5.0 & 1.5 & 11\\
3C192   & S & 0.58 & 0.82 & 2.4 & 1.0 & 9  &3C340   & - & 0.21 & 0.03 & 4.8 & 1.5 & 5\\
3C196   & - & -    & -    & -   & -   & -  &3C349   & - & -    & -    & -   & -   & -\\
3C200   & S & 0.39 & 0.08 & 5.0 & 1.5 & 6  &3C351   & N & 0.46 &-0.44 & 4.8 & 1.4 & 2\\
4C14.27 & N &-0.17 &-0.01 & 4.8 & 1.4 & 2  &3C352   & N & 0.28 & 0.74 & 5.0 & 1.5 & 6\\
3C207   & S & 0.60 & 0.16 & 5.0 & 1.5 & 6  &3C381   & - & 0.06 & 0.04 & 4.8 & 1.5 & 4\\
3C215   & S & 0.45 & 0.21 & 5.0 & 1.5 & 6  &3C382   & N & 0.09 & 0.16 & 2.5 & 1.5 & 10\\
3C217   & - & 0.20 & 0.67 & 5.0 & 1.5 & 11 &3C388   & S & -    & -    & -   & -   & -\\
3C216   & - & -    & -    & -   & -   & -  &3C390.3 & N & 0.01 & 0.03 & 1.5 & 0.3 & 10\\
3C219   & S & 0.53 & 0.55 & 4.8 & 1.5 & 4  &3C401   & S & -    & -    & -   & -   & -\\
3C220.1 & N & -    & -    & -   & -   & -  &3C427.1 & N & -    & -    & -   & -   & -\\
3C220.3 & - & -    & -    & -   & -   & -  &3C433   & N & -    & -    & -   & -   & -\\
3C223   & N & 0.07 & 0.11 & 2.4 & 1.0 & 9  &3C436   & S & -    & -    & -   & -   & -\\
3C225B  & - & -    & -    & -   & -   & -  &3C438   & N & -    & -    & -   & -   & -\\
3C226   & - & -    & -    & -   & -   & -  &3C441   & N & 0.28 & 0.06 & 5.0 & 1.5 & 6\\
4C73.08 & - & -    & -    & -   & -   & -  &3C452   & S & -    & -    & -   & -   & -\\
3C228   & S & 0.30 & 0.05 & 4.8 & 1.5 & 5  &3C455   & - &-0.45 & 0.03 & 4.9 & 1.6 & 8\\
3C234   & N & 0.55 & 0.56 & 4.8 & 1.5 & 4  &3C457   & - & 0.00 &-0.09 & 4.8 & 1.4 & 2\\\hline
\end{tabular}\\
\begin{minipage}{16cm}
Column [1]: 3CR catalogue source name. Column [2]: jet side. Column
[3] \& [4]: depolarization measure, $DPM$, for north and south lobe
respectively, where $DPM=(m_{\rm h} - m_{\rm l})/(m_{\rm h} + m_{\rm
  l})$, $m_{\rm h}$ and $m_{\rm l}$ being the fractional polarization
measured at the higher and lower frequency respectively. Column [5] \&
[6]: Frequency of high and low frequency maps respectively. Column
[7]: References for data. (1): Wright (1979), (2): Goodlet et al.
(2004), (3): Fernini et al. (1993), (4): Dennett-Thorpe, Barthel \&
van Bemmel (2000), (5): Johnson, Leahy \& Garrington (1995), (6):
Garrington, Conway \& Leahy (1991), (7): Fernini et al. (1991), (8):
Akujor \& Garrington (1995), (9): Conway et al. (1983), (10):
Garrington \& Conway (1991), (11): Pedelty et al. (1989), (12): Liu \&
Pooley (1991), (13): Fernini (2001). Columns [8] to [14] and [15] to
[21] as for [1] to [7].
\end{minipage}
\end{table*}

%
%
\subsubsection{Correlation between core and straight jet prominence}\label{sec:core_jet}
%
%
Considering all the data, including upper limits, the
correlation between jet prominence and core prominence is significant, at the 99.9 per cent confidence level, on a modified
Kendall's $\tau$ test as implemented in {\sc asurv}. This result is consistent with the
correlation found by H98 in a sample with substantially fewer upper limits.

\begin{figure}
\centerline{\epsfig{file=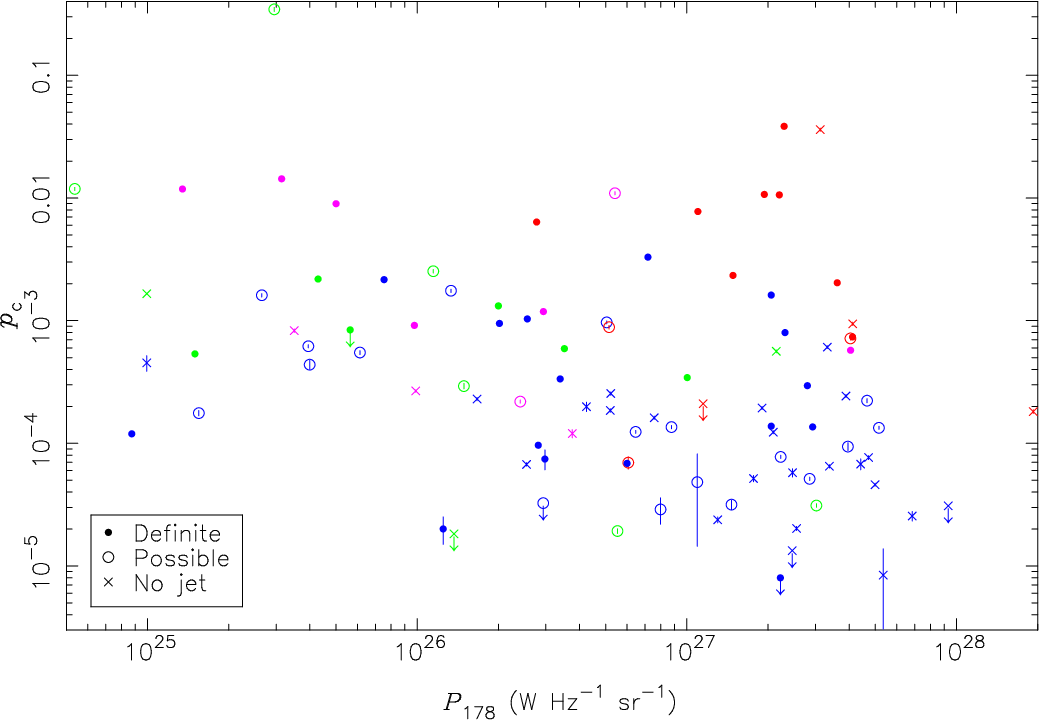,width=8cm}}
\caption{\label{fig:coreprom_P_jetstat}The core prominence, $p_{\rm c}$, for the entire
sample plotted against the source luminosity at 178~MHz, $P_{178}$, binned by straight jet detection status. Filled circles: at least one definite jet detected, open circles: no definite jet detected but at least one possible jet, diagonal cross: no jet feature detected. Vertical bars indicate errors, arrows indicate upper limits. Green: low excitation radio galaxies, blue: narrow line radio galaxies, magenta: broad line radio galaxies and red: quasars (on-line colour version).}
\centerline{\epsfig{file=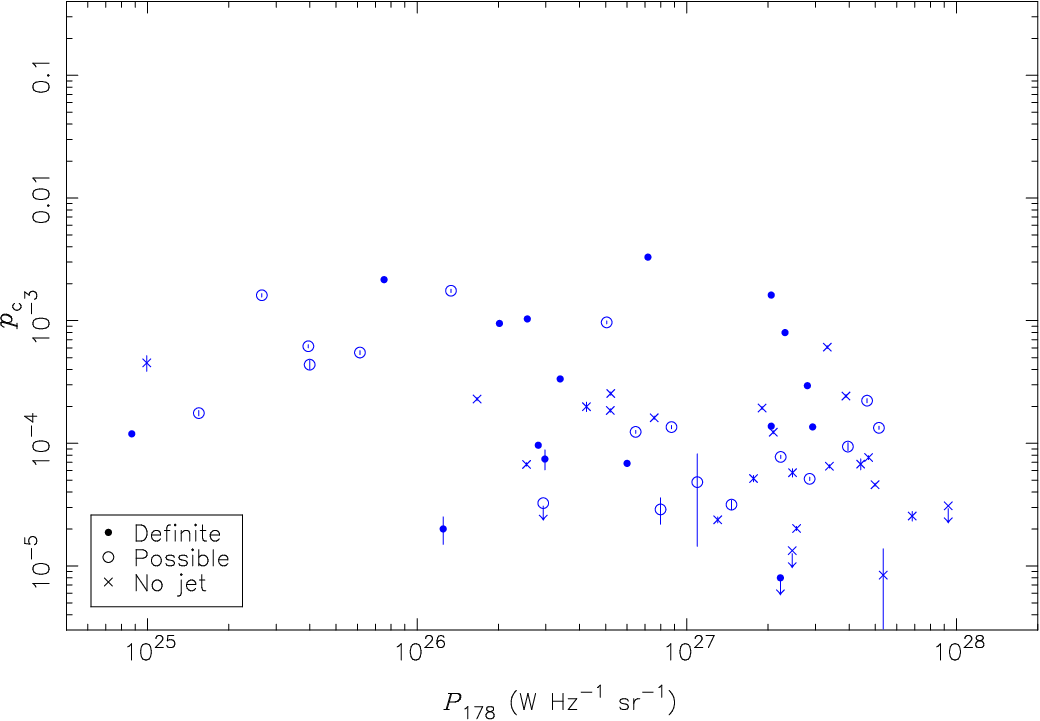,width=8cm}}
\caption{\label{fig:coreprom_P_Njetstat}The core prominence, $p_{\rm c}$, for the NLRGs
plotted against the source luminosity at 178~MHz, $P_{178}$, binned by straight jet detection status. Filled circles: at least one definite jet detected, open circles: no definite jet detected but at least one possible jet, diagonal cross: no jet feature detected. Vertical bars indicate errors, arrows indicate upper limits. Green: low excitation radio galaxies, blue: narrow line radio galaxies, magenta: broad line radio galaxies and red: quasars (on-line colour version).}
\centerline{\epsfig{file=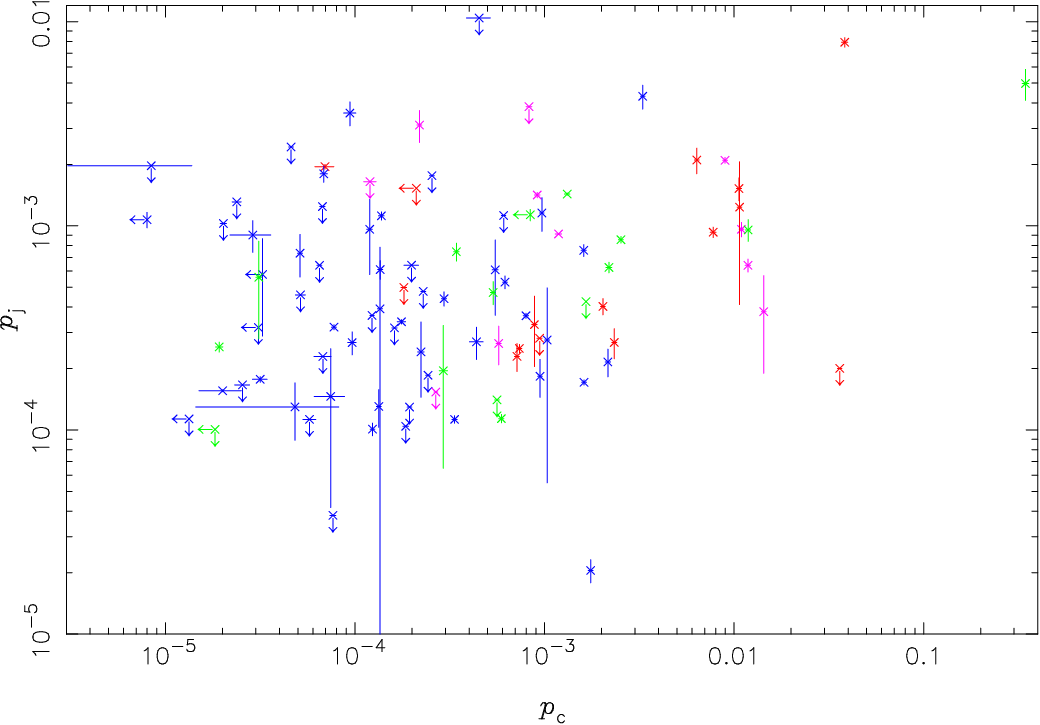,width=8cm}}
\caption{\label{fig:core_jet_prom}Jet prominence of definite and
possible straight jets, $p_{\rm j}$, plotted against core prominence, $p_{\rm c}$, for all spectral classes. Vertical bars indicate errors, arrows indicate upper limits. Green: low excitation radio galaxies, blue: narrow line radio galaxies, magenta: broad line radio galaxies and red: quasars (on-line colour version).}
\end{figure}

In Figs \ref{fig:coreprom_P_jetstat} and
\ref{fig:coreprom_P_Njetstat} we plot the core prominence data against source
luminosity for both the entire sample and the NLRG class respectively,
binning the data according to the straight jet detection status indicated as
before. For the sample as a whole (Fig. \ref{fig:coreprom_P_jetstat})
those sources with at least one jet detection have higher core
prominence. As Qs and BLRGs are generally observed to have
brighter jets this is to be expected and is further illustrated in Fig.
\ref{fig:core_jet_prom}. However, even when considering the NLRG population
separately (Fig. \ref{fig:coreprom_P_Njetstat}) it is clear that the NLRGs with definite jet detections are also associated with relatively
higher core prominence. A Peto-Prentice test applied to the core prominence
data between all sources with at least one definite or possible straight jet
and those with no detection indicates the jetted sources have higher core prominence, significant at the 99.9 per cent confidence level.
%
%
\begin{table}
\scriptsize
\centering
\begin{minipage}{8cm}
\caption{\label{tab:jetprom_median}The median straight jet prominence for detected jets only, $<p_{\rm j}>$, for each of the spectral class distributions.}
\begin{tabular}{lclc}  \hline
Spectral class & $<p_{\rm j}> / 10^{-3}$ &Spectral class & $<p_{\rm j}> / 10^{-3}$ \\ \hline
Q          &  0.930  &  NLRG, high $P_{178 \rm MHz}$   & 0.134\\
B          &  0.936  &  NLRG, low $P_{178 \rm MHz}$    & 0.440\\
Q and BLRG &  0.930  &  NLRG                           & 0.351\\
LERG       &  0.690  &  BLRG and low $P_{178 \rm MHz}$ & 0.619\\\hline

\end{tabular}
\end{minipage}
\end{table}
%
%

%
%
\section{Hotspots}\label{sec:hotspots}
%
%
\subsection{Hotspot prominence and size}
\subsubsection{Observing effects}
A summary of hotspot detections in the sample (as defined by the
hotspot criteria in section \ref{sec:def_hotspots}) is given in Table
\ref{tab:summary}. From this table it can be seen that 58 per cent of sources have one hotspot per lobe and 34 per cent have
at least one lobe with more than one hotspot feature. Only 9 sources have one lobe that lacks a hotspot, and only 1 source
of the 98 lacks hotspots entirely. Observations of sources with a single
bright, compact hotspot had led to the suggestion that hotspots
corresponded to enhanced emission associated with the beam termination
shock, but as multiple features are often detected at high resolution,
this interpretation is too simplistic. However, it is still thought
that the hotspots correspond to shocks at or near the beam
termination, although the exact relation between the observed emission and the
physical structure is not understood.

\begin{figure}
\centerline{\epsfig{file=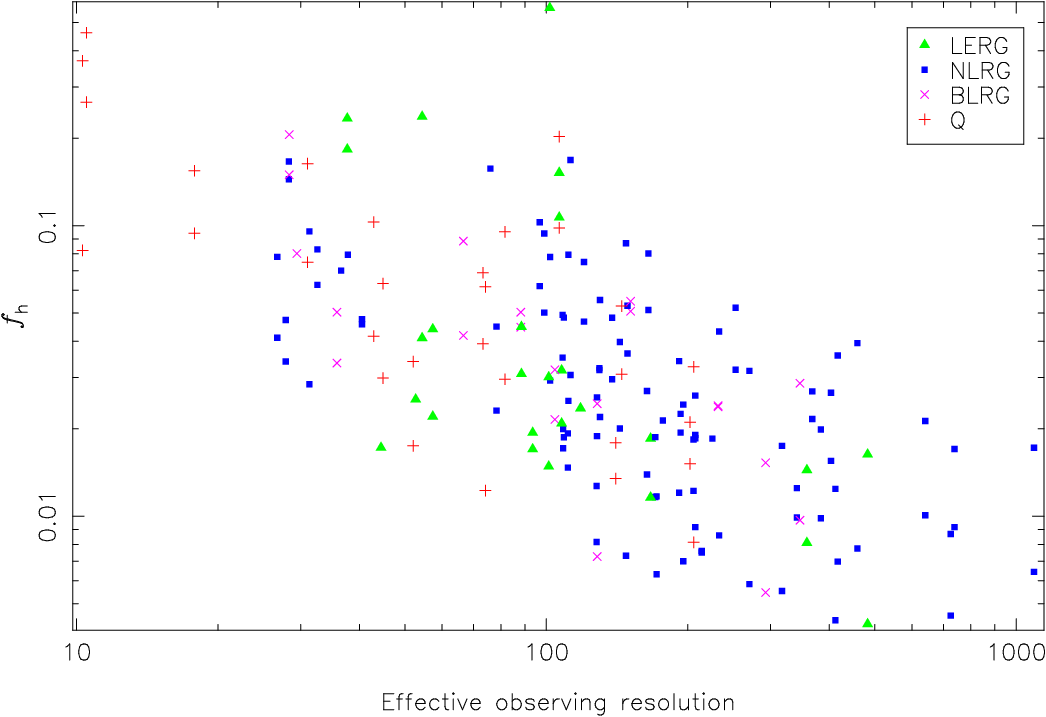,width=8cm}}
\caption{\label{fig:fhtsz_beams}The fractional hotspot size, $f_{\rm h}$, plotted against the corresponding high-resolution effective observing resolution.}
\centerline{\epsfig{file=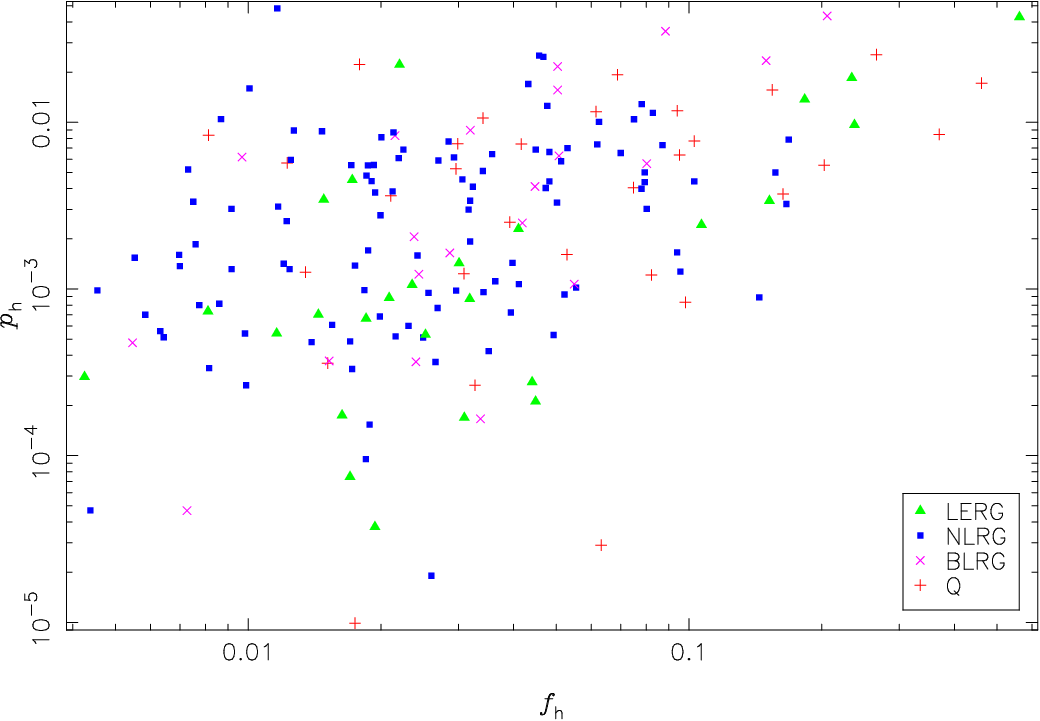,width=8cm}}
\caption{\label{fig:htprom_fhtsz} The hotspot prominence, $p_{\rm h}$ plotted against the fractional hotspot size, $f_{\rm h}$.} 
\end{figure}

From the sample
sources mapped at more than one resolution, it can be seen that
hotspot features generally appear more diffuse at the lower
resolution, with a larger size fitted by JMFIT. This effect of
observing resolution on apparent hotspot size can be quantitatively
considered by making use of the fractional hotspot size, $f_{\rm h}$,
defined in section \ref{sec:def_hotspots}. In Fig.
\ref{fig:fhtsz_beams} we plot $f_{\rm h}$ against the effective
observing resolution for all sources in the sample; from this figure
it can be seen that relatively smaller hotspots are indeed associated
with sources observed at relatively higher resolution. 

The larger the region that is identified as the hotspot (that is, the higher $f_{\rm h}$ is) the greater the hotspot
prominence may potentially be, as more flux is included in the hotspot
flux measurement. If low observing resolution results in larger hotspots, this may cause a bias for more prominent hotspots, if more lobe emission is included in the hotspot measurement. Fig.
\ref{fig:htprom_fhtsz} plots hotspot prominence against $f_{\rm
h}$, and it can be seen that sources with larger $f_{\rm h}$ correspond with higher prominence; very approximately $p_{\rm h} \propto f_{\rm h}^{2}$, though for
$f_{\rm h} < 0.1$ there is little correlation between the two
quantities.

We conclude that hotspot properties are strongly affected by observing
resolution and that this is difficult to compensate for. This should be
borne in mind when considering the following results.
\subsubsection{Trends with $P_{178}$, $z$ and size}\label{sec:trendhotspot}
\begin{figure}
\centerline{\epsfig{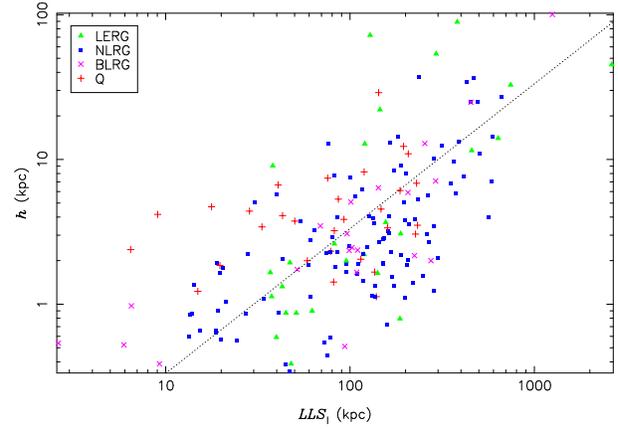}}
\caption{\label{fig:hth_LLS}The linear hotspot size, $h$, plotted against
  the largest linear size of the lobe, $LLS_{\rm l}$. The dotted line
  has slope unity.}
\end{figure}

No apparent trends in the sample can be seen in the plots of hotspot prominence, binned by high-resolution effective observing resolution,
against $z$, $P_{178}$ and $LLS_{\rm s}$ in Figs \ref{fig:htprom_z} to
\ref{fig:htprom_LLS} respectively. Various authors, including H98 and Kharb et al. (2008), have found a significant correlation between lobe linear size
and hotspot size, which is also apparent in our data (Fig.
\ref{fig:hth_LLS}). The correlation seen here as determined by Kendall's $\tau$ coefficient is significant above the 99.9 per cent confidence limit. 

\begin{figure}
\centerline{\epsfig{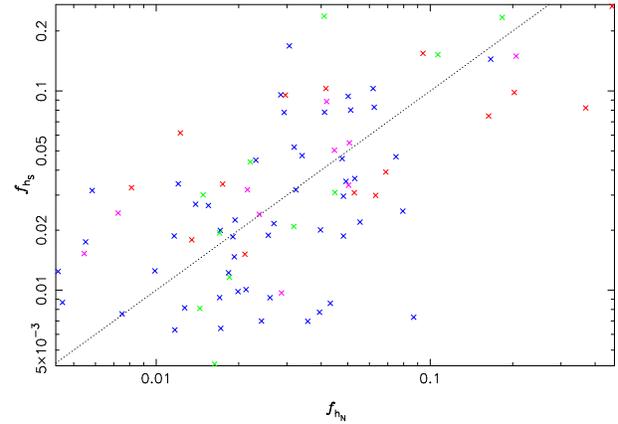}}
\caption{\label{fig:fh_NS}The S lobe fractional hotspot size, $f_{\rm
    h_{S}}$, plotted against the N lobe fractional hotspot size,
  $f_{\rm h_{N}}$. Green: low excitation radio galaxies, blue: narrow line radio galaxies, magenta: broad line radio galaxies and red: quasars (on-line colour version). The dotted line indicates $f_{\rm
    h_{S}}=f_{\rm h_{N}}$.}
\end{figure}

However, it can be noted that there are serious
potential biases in the hotspot-size lobe-size correlation result, given that the observing resolution
used is also strongly correlated with source angular size. One way of
circumventing this is to compare hotspot data from a single map. In
Fig. \ref{fig:fh_NS} we plot the {\it fractional} sizes ($f_{\rm h}$)
of the primary hotspot in each lobe against each other. If the two
fractional hotspot sizes were uncorrelated, this would suggest that
there is no tendency for the hotspots to `know about' the linear size
of the source, while a strong correlation would be consistent with the
notion that the hotspot size is proportional to lobe size and support
the hypothesis of self-similarity in the lobe. We find that the
Kendall's $\tau$ test shows a correlation significant above
the 99.9 per cent confidence level. Thus there is some support in the
data for a real physical correlation between hotspot and lobe size. We
also note that the hotspot size-linear size correlation is still
highly significant if we consider only the subsample of objects
(approximately half of the total) that are observed with more than 100
restoring beams across the source. A correlation between hotspot and lobe size supports models of self-similarity in which the beam's working surface maintains pressure balance as it extends (Carvalho \& O'Dea, 2002). 

In section \ref{sec:trendaxialratio} we
suggested that there was some evidence that the $R_{\rm ax}$
distribution is consistent with self-similar source expansion on
smaller scales. Considering the sample by binning with respect to
size, using a cutoff of $LLS_{\rm s}=200$ kpc as before, there is
a difference in the $f_{\rm h}$ distributions of the small and large sources significant above the 99.9 per cent confidence level on a K-S test. Binning
using luminosity (with cutoff $P_{\rm c}$) gives evidence of a difference in
the high and low power sources also as a K-S test suggests that they are
different at the 93 per cent confidence level. The difference appears
to be in the sense that the more powerful/smaller sources have a broader
distribution of hotspot fractional size. This may well simply be an
observational effect: many of the sources with the largest fractional
hotspot size are a) small, b) powerful and c) observed at low
effective resolution. We cannot draw strong conclusions about
self-similarity from these results.

\subsubsection{Beaming and unification}
If beaming affects the observed hotspot prominence then the brighter
hotspot might be expected to be correlated with the straight jet side. The most
compact feature may also show such a correlation, if the approaching
and receding hotspot emission corresponds to a different physical
region within the flow. Laing (1989) suggested such a model, whereby
the approaching hotspot emission originated in a region of higher
$\beta$ flow, closer to the core of the beam. The model predicted that the
most compact hotspot would be correlated with the jet side.

\begin{figure}
\centerline{\epsfig{file=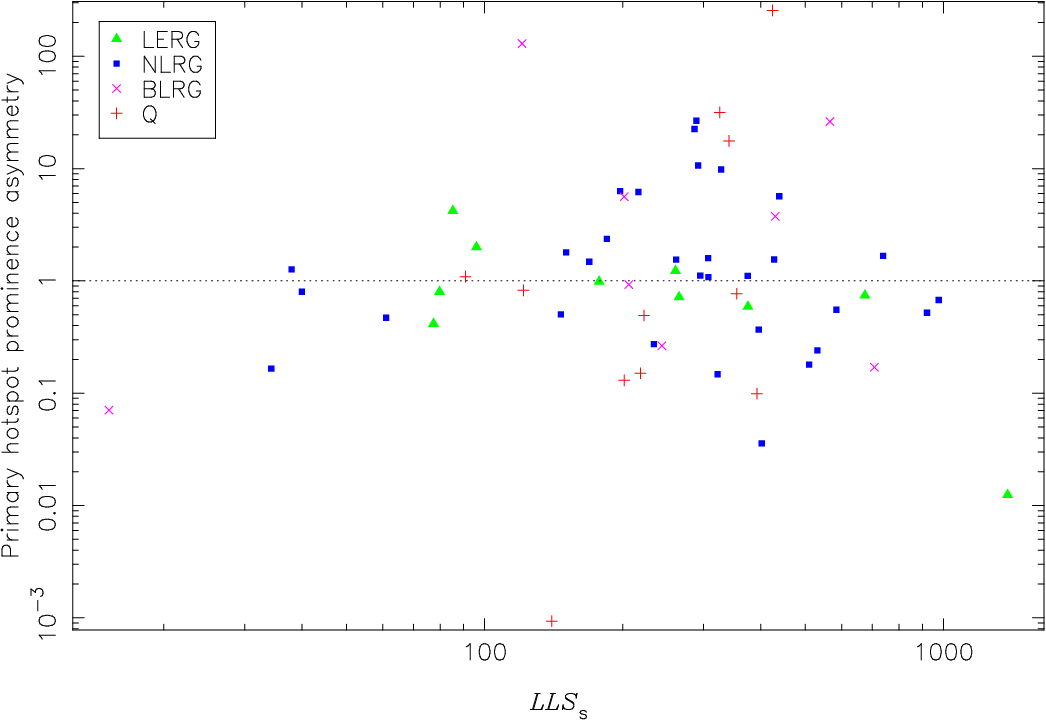,width=8cm}}
\caption{\label{fig:htfl_LLS}The hotspot flux ratio as defined by
  jet side plotted against the largest linear lobe size, $LLS_{\rm
    s}$. The dotted line shows a ratio of unity.}
\centerline{\epsfig{file=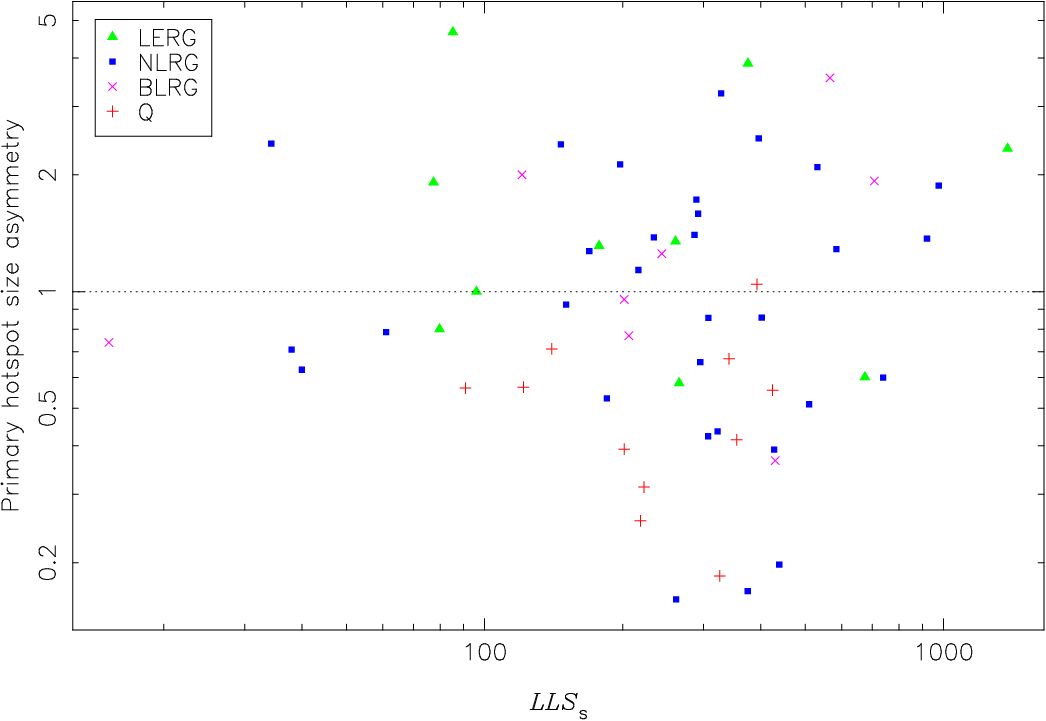,width=8cm}}
\caption{\label{fig:htsz_LLS}The ratio of the fractional hotspot size,
  $f_{\rm h}$, for the primary hotspot in each lobe as defined by jet side, plotted against the largest linear lobe size, $LLS_{\rm s}$. The dotted line shows a ratio of unity.}
\end{figure}

In our analysis we consider the ratio of $f_{\rm h}$ in each lobe rather than the
hotspot size alone and both this ratio and the corresponding flux
ratio are defined by taking the ratio of the straight jet-side measurement to
that of the counterjet. The hotspot size ratio will therefore be less
than one if the most relatively compact hotspot is on the jet side and
the hotspot flux ratio will be more than one if the more prominent
hotspot is on the jet side. The hotspot flux and $f_{\rm h}$ ratios as
defined by jet side are plotted against source size in Fig. \ref{fig:htfl_LLS} and
\ref{fig:htsz_LLS}. As not all sources have a straight jet detection this introduces a bias when we consider the data quantitatively and this should be borne in mind.

We find that there are no significant tendencies for brighter hotspots
to be found on the jet side, for any spectral class. However, the
$f_{\rm h}$ data suggest that the more relatively compact hotspot is
correlated with the jet side in quasars: 10/11 jetted Qs have the more
compact hotspot on the jet side, a result significant at the 99.4 per
cent level on a binomial test.
\subsection{Hotspot recession}
\subsubsection{Observing resolution}
The effect of observing resolution on the measured hotspot recession
has been addressed by the application of a resolution-correction
factor (G99) as discussed in section \ref{sec:def_lobe_size}. This correction
factor should ensure that the hotspot recession parameter $\eta$ will
be 1 for hotspots located at the lobe extremity.
\subsubsection{Trends with $P_{178}$, $z$ and size}
$\zeta$, the source hotspot recession parameter (defined in section
\ref{sec:def_hotspotrec}), is plotted against redshift, source
luminosity and size in Figs \ref{fig:zeta_z} to \ref{fig:zeta_LLS}
respectively. There is no significant tendency for higher-luminosity sources to be more recessed, but a K-S test binning the sample using a divide of 200 kpc does indicate that smaller sources are more recessed with significance at the 99.7 per cent confidence level.
\subsubsection{Unification}\label{sec:zeta}
For sources not lying close to the plane of the sky geometric effects may cause a hotspot that is intrinsically positioned
near the lobe edge to appear set back in the lobe. Thus orientation
effects may contribute to the observed range in $\zeta$. Gilbert (2001) considered hotspot recession for the sub-sample of sources with $z \le
0.5$. Sources were modelled as an expanding ellipse with a Gaussian emission
density, which was rotated with respect to the observer's
line-of-sight. Taking different expansion speeds (into the
relativistic regime) and different orientations the model was compared
to the data by predicting number counts for recessed sources. Gilbert concluded that effects other than simple geometric effects contributed
to the observed recession distribution. Around 15 per cent of sources
were expected to have $\zeta < 0.9$ from the model when in fact 26 per
cent of his sample were observed to have this degree of recession.
Furthermore, only around 3 per cent of sources were predicted to show
strong recession, with $\zeta < 0.8$, whereas close to 13 per cent does. When we consider our
current sample, 32 per cent of sources (including all spectral
classes) have $\zeta < 0.9$ while 10 per cent have $\zeta < 0.8$. The
conclusion that effects other than
simple geometric effects from source orientation contribute to the observed hotspot recession is thus valid
for our sample too.

\begin{figure}
\centerline{\epsfig{file=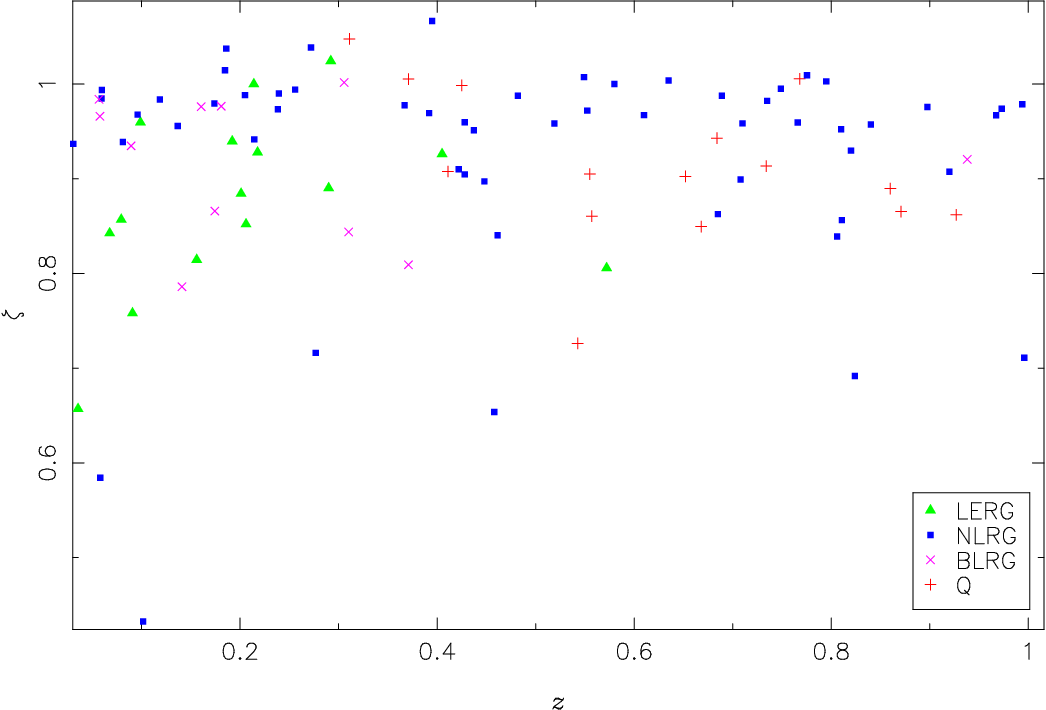,width=8cm}}
\caption{\label{fig:zeta_z}The source hotspot recession coefficient, $\zeta$, plotted against $z$.}
\centerline{\epsfig{file=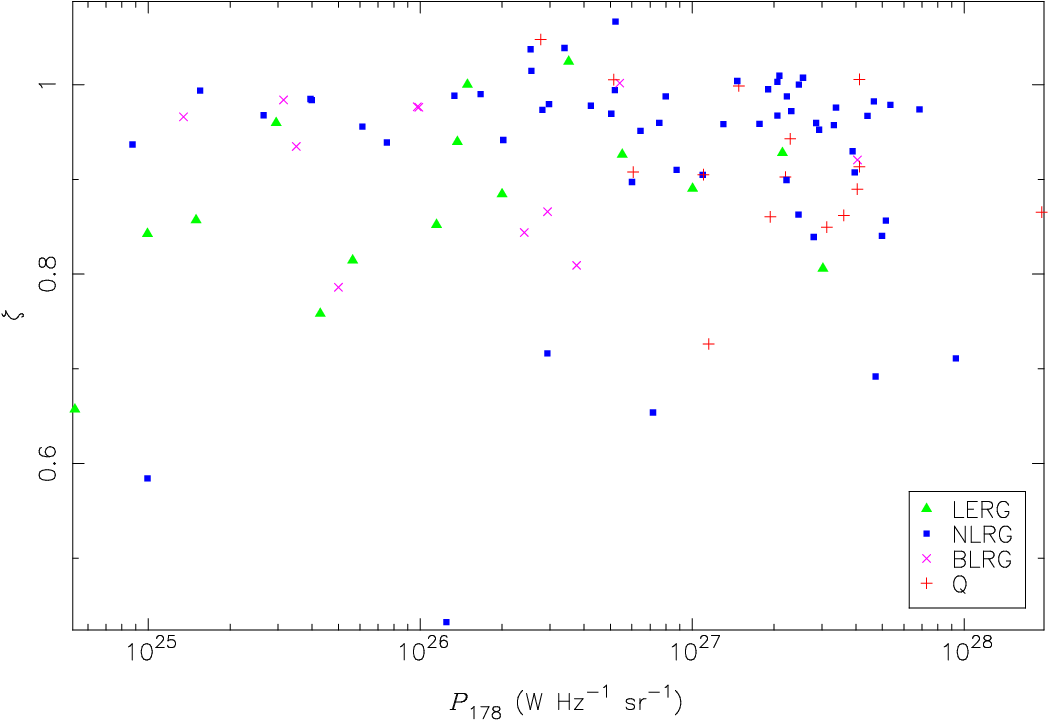,width=8cm}}
\caption{\label{fig:zeta_P}The source hotspot recession coefficient, $\zeta$, plotted against $P_{178}$.}
\centerline{\epsfig{file=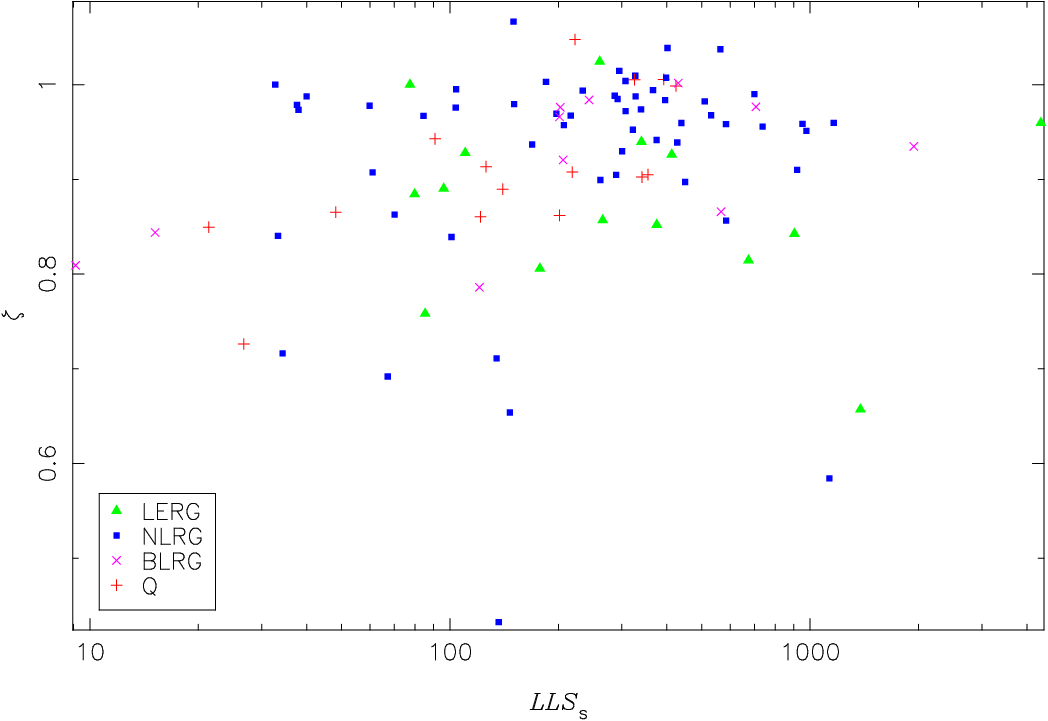,width=8cm}}
\caption{\label{fig:zeta_LLS}The source hotspot recession coefficient, $\zeta$, plotted against $LLS_{\rm s}$.}
\end{figure}

A W-M-W test comparing the Qs and high-luminosity NLRGs shows that the
difference in median $\zeta$ between the spectral classes is not
significant, although a K-S test shows a difference in the
distributions for the samples at the 94.5 per cent level, consistent
with the observed broader spread of the Qs (Figs \ref{fig:zeta_z} to
\ref{fig:zeta_LLS}) but formally not significant. There are no
significant differences between the low luminosity spectral classes
and there is no significant tendency for the straight jet side to be
any more or less recessed than that of the counterjet.

%
%
%
\begin{figure}
\centerline{\epsfig{file=figures/htprom_z.ps,width=8cm}}
\caption{\label{fig:htprom_z}The hotspot prominence, $p_{\rm h}$, of the primary hotspot plotted against $z$, binned by high-resolution effective observing resolution. Diagonal cross: effective observing resolution $>100$, vertical cross: effective observing resolution $\le 100$. Green: low excitation radio galaxies, blue: narrow line radio galaxies, magenta: broad line radio galaxies and red: quasars (on-line colour version).}
\centerline{\epsfig{file=figures/htprom_P.ps,width=8cm}}
\caption{\label{fig:htprom_P}The hotspot prominence, $p_{\rm h}$, of the primary hotspot plotted against the source luminosity, binned by high-resolution effective observing resolution. Diagonal cross: effective observing resolution $>100$, vertical cross: effective observing resolution $\le 100$. Green: low excitation radio galaxies, blue: narrow line radio galaxies, magenta: broad line radio galaxies and red: quasars (on-line colour version).}
\centerline{\epsfig{file=figures/htprom_LLS.ps,width=8cm}}
\caption{\label{fig:htprom_LLS}The hotspot prominence, $p_{\rm h}$, of the primary hotspot plotted against the lobe size, binned by high-resolution effective observing resolution. Diagonal cross: effective observing resolution $>100$, vertical cross: effective observing resolution $\le 100$. Green: low excitation radio galaxies, blue: narrow line radio galaxies, magenta: broad line radio galaxies and red: quasars (on-line colour version).} 
\end{figure}

%
%
\section{Discussion}\label{sec:discussion}
%
%
\subsection{Summary of results}
Table \ref{tab:resultssum} contains a summary of all statistical
results from the preceding sections. The consequences of these for
physical models of FRII sources are discussed in the following sections. 
\begin{table*}
\centering
\caption{Summary of statistical tests.}\label{tab:resultssum}
\begin{tabular}{llll}  \hline
Proposition & Conclusion & Details & Section\\\hline
Are high power sources statistically smaller? &No  &W-M-W test compared high and low luminosity sources    & section \ref{sec:size_trends}\\
                                              &    &using a cutoff of $5\cdot~10^{26}\ \rm{W\ Hz^{-1}\ sr^{-1}}$, including& \\  
                                              &    &all sources. Null hypothesis only rejected at 86 per cent& \\
                                              &    & confidence level.\\                                
& & & \\
Are Qs and BLRGs statistically smaller than   &No  &W-M-W test compared Qs with high power NLRGs, and      & section \ref{sec:size_unif}\\
NLRGs of equivalent luminosity?               &    &BLRGs with low power NLRGs. Although medians of Qs      & \\
                                              &    &and BLRGs lower, difference not significant. Null & \\
                                              &    &hypothesis only rejected at 70 and 35 per cent confidence& \\
                                              &    &level.                                                   & \\
& & & \\
Are LERGs statistically smaller than BLRGs    &No  &W-M-W test compared LERGs with BLRGs and low power       & section \ref{sec:size_unif}\\
and low power NLRGs                           &    &NLRGs. Null hypothesis only rejected at 12 per cent     & \\
                                              &    &confidence level.                                       & \\\hline                              
Are lobes broader in sources observed at low  &Yes &W-M-W test compared sources observed with 40 or fewer  &section \ref{sec:obsaxialratio}\\
resolution?                                   &    &restoring beams across $LLS_{\rm s}$ with those observed& \\          
                                              &    &at higher resolution. Significant above 99.9 per cent     & \\
                                              &    &confidence level.                                       & \\
& & & \\
Is there a systematic tendency towards lower $R_{\rm ax}$&Yes &W-M-W test compared sources observed with 40 or fewer  & section \ref{sec:obsaxialratio}\\
in sources observed at relatively low resolution?&    &restoring beams across $LLS_{\rm s}$ with those observed& \\
                                              &    &at higher resolution. Significant above the 99.9 per cent  & \\
                                              &    &confidence level                                        & \\
& & & \\
Is there a significant difference in the      &No  &K-S test between sources binned by $P_{178}$ using a cutoff& section \ref{sec:trendaxialratio} \\
$R_{\rm ax}$ distribution across the          &    &of $5\cdot~10^{26}\ \rm{W\ Hz^{-1}\ sr^{-1}}$, and binned by& \\
power and redshift range?                     &    &$z$ using a cutoff of 0.5, including all sources. Null hypothesis only& \\
                                              &    &rejected at 4 and 22 per cent confidence levels respectively. & \\
& & & \\
Is there a significant tendency towards lower $R_{\rm ax}$ &Yes&W-M-W test compared small and large sources using a     & section \ref{sec:trendaxialratio}\\
in smaller sources?                           &   &cutoff of 100 kpc, including all sources. Significant    & \\
                                              &   &above 99.9 per cent confidence level.                    & \\
& & & \\
Is there a significant tendency towards lower $R_{\rm ax}$ &Yes&W-M-W test compared Qs with high power NLRGs.& section \ref{sec:Rax_uni}\\
in Qs than in NLRGs?                         &   &Significant at the 99.6 per cent confidence level.& \\  
& & & \\
Is there a significant tendency towards lower $R_{\rm ax}$ &No &W-M-W test compared BLRGs with low power NLRGs.      & section \ref{sec:Rax_uni}\\
in BLRGs than in NLRGs?         &   &Null hypothesis only rejected at 86 per cent confidence& \\
                                              &   &level.                                                 &  \\
& & & \\
Is there a significant tendency towards lower &No &W-M-W test compared Qs with high-power NLRGs.              & section \ref{sec:Rax_uni}\\
$w$ in Qs than in high-power NLRGs            &   &Null hypothesis only rejected at 48 per cent confidence level.& \\
& & & \\
Is there a significant difference in the      &No &K-S test compared LERG distribution with      & section \ref {sec:Rax_uni}\\
$R_{\rm ax}$ distribution of LERGs compared   &   &that of combined BLRGs and low power NLRGs. Null  & \\
to BLRGs and NLRGs?                           &   &hypothesis only rejected at 64 per cent confidence level. & \\\hline
Is there a significant difference in the $x_{\rm lobe}$&No  &K-S test between sources binned by $P_{178}$  & section \ref{sec:trendxlobe}\\
distribution across the power and size range?  &   &using a cutoff of $5\cdot~10^{26}\ \rm{W\ Hz^{-1}\ sr^{-1}}$, and binned by& \\
                                              &   &$LLS_{\rm s}$ using a cutoff of 200 kpc, including all sources. Null hypothesis& \\
                                              &   &only rejected at 41 and 73 per cent confidence level respectively. & \\ 
& & & \\
Is there a significant tendency towards higher $x_{\rm lobe}$&Yes&W-M-W test compared Qs with high power NLRGs.&section \ref{sec:x_lobe_beamin}\\
in Qs than in NLRGs?                           &   &Significant at the 99.7 per cent confidence level.        & \\
& & & \\
Is there a significant tendency for $x_{\rm lobe}$&No &W-M-W test compared BLRGs with low power NLRGs.         &section \ref{sec:x_lobe_beamin}\\
to be higher in BLRGs than in NLRGs?          &   &Null hypothesis only rejected at 55 per cent confidence & \\
                                             &   &level. & \\
& & & \\
Is there a significant difference in the $x_{\rm lobe}$&No &K-S test compared LERG distribution with & section \ref{sec:x_lobe_beamin}\\
distribution of LERGs compared to BLRGs& &that of combined BLRGs and low power NLRGs. Null  & \\
and NLRGs?                             & &hypothesis only rejected at 83 per cent confidence level. & \\\hline
\end{tabular}
\end{table*}
\begin{table*}
\centering
{\bf Table \ref{tab:resultssum}} continued.
\begin{tabular}{lcll}  \hline
Proposition & Conclusion & Details & Section\\\hline
Is there a significant tendency for $x_{\rm jet}$ to be&Yes  &Binomial test shows tendency for BLRGs to have positive& section \ref{sec:x_lobe_beamin}\\
positive in BLRGs?                               &    & $x_{\rm jet}$ values; significant at the 96.5 per cent confidence level.& \\
& & & \\
Is there a significant tendency for $x_{\rm jet}$ to be&No  &Binomial test shows no significant tendency for Qs,& section \ref{sec:x_lobe_beamin}\\
positive in Qs, NLRGs and LERGs?                 &    &NLRGs or LERGs to have positive $x_{\rm jet}$ values; null hypothesis & \\
                                                 &    &only rejected at the 50, 20 and 19 per cent confidence level respectively.& \\
& & & \\
Is there a significant tendency for $x_{\rm jet}$ to be&No &Binomial test suggests no significant tendency for small& section \ref{sec:x_lobe_beamin}\\
positive in smaller sources?                     &    &sources to have positive $x_{\rm jet}$ values; null hypothesis & \\
                                                 &    &only rejected at 81 per cent confidence level. & \\
& & & \\
Is there a significant difference in the $x_{\rm jet}$&No  &K-S test between sources binned by $LLS_{\rm s}$ using a cutoff of& section \ref{sec:x_lobe_beamin}\\
distribution across size range?                  &    &200 kpc, including all sources. Null hypothesis only rejected& \\
                                                 &    &at 77 per cent confidence level.& \\
\hline
Is there a significant tendency toward lower     &Yes &Peto-Prentice test compared high and low power sources& section \ref{sec:cores}\\
$p_{\rm c}$ in higher-power sources?             &    &using a cutoff of $5\cdot~10^{26}\ \rm{W\ Hz^{-1}\ sr^{-1}}$, including all& \\
                                                 &    &sources. Significant at 99.5 per cent  confidence level.& \\
& & & \\
Is there a significant tendency toward lower     &No &Peto-Prentice test compared large and small sources using a& section \ref{sec:cores} \\
$p_{\rm c}$ in smaller sources?                  &    &cutoff of $LLS_{\rm s}=200$ kpc, including all sources.   & \\
                                                 &    &Null hypothesis only rejected at 85 per cent confidence level.          & \\
& & & \\
Is there a significant tendency toward lower     &Yes &Peto-Prentice test compared large and small NLRG using a& section \ref{sec:cores} \\
$p_{\rm c}$ in smaller NLRGs?                  &    &cutoff of $LLS_{\rm s}=200$ kpc. Null hypothesis rejected at 96.5   & \\
                                                 &    &per cent confidence level.          & \\
& & & \\
Is there a significant tendency for the $p_{\rm j}$&No &Peto-Prentice test compared high and low-power sources dividing & section \ref{sec:jets} \\
distribution to vary across the power, redshift and &    &at $L = 5\times~10^{26}\ \rm{W\ Hz^{-1}\ sr^{-1}}$, at $z=0.5$ and at $LLS_{\rm s}=200$ kpc & \\
size range?                                      &    &including all sources. Null hypotheses only rejected at& \\
                                                 &    &77, 19 and 34 per cent confidence levels respectively.& \\
& & & \\
Is there a significant tendency toward higher    &Yes &Peto-Prentice test compared Qs and low power NLRGs. & section \ref{sec:cores2} \\
$p_{\rm c}$ in Qs than in NLRGs?                 &    &Significant above 99.9 per cent confidence level.              & \\
& & & \\
Is there a significant tendency toward higher    &Yes &Peto-Prentice test compared BLRGs and low power& section \ref{sec:cores2} \\
$p_{\rm c}$ in BLRGs than in NLRGs?              &    &NLRGs. Significant at 97.4 per cent confidence level.& \\
& & & \\
Is there a significant tendency for the LERGs    &No  &Peto-Prentice test compared LERGs and BLRGs and low & section \ref{sec:cores2}\\
$p_{\rm c}$ distribution to differ?              &    &power NLRGs. Null hypothesis only rejected at 34 per cent\\
                                                 &    & confidence level.& \\\hline                                            
\end{tabular}
\end{table*}
\begin{table*}
\centering
{\bf Table \ref{tab:resultssum}} continued.
\begin{tabular}{lcll}  \hline
Proposition & Conclusion & Details & Section\\\hline                                           
Is there a significant tendency toward higher   &Yes &Peto-Prentice test compared Qs and BLRGs with NLRGs.& section \ref{sec:jets-asurv}\\
$p_{\rm j}$ in Qs and BLRGs than in NLRGs?      &    &Significant at the 99.6 per cent confidence level.  & \\
& & & \\
Is there a tendency for $p_{\rm j}$ to be        &Yes &Linear regression gives Kendall's $\tau$ correlation coefficient& section \ref{sec:core_jet}\\
correlated with $p_{\rm c}$?                     &    &as 3.5, significant correlation above the 99.9 per cent confidence level.\\
& & & \\
Do jetted sources have significantly more        &Yes &Peto-Prentice test compared sources with definite or& section \ref{sec:core_jet}\\
prominent cores?                                 &    &possible jets to those with no jet detection. Significant& \\
                                                 &    &at 99.9 per cent confidence level.& \\\hline
& & & \\
Is there a tendency for the fractional hotspot   &Yes & Linear regression gives Kendall's $\tau$ correlation coefficient as & section \ref{sec:trendhotspot}\\
size to be similar for each lobe within a source?&    &5.5, significant above 99.9 per cent confidence level.& \\
& & & \\
Is there a significant difference in the $f_{\rm h}$&Yes &K-S test of $f_{\rm h}$ binned by $LLS_{\rm s}$ & section \ref{sec:trendhotspot}\\
distribution across the size range?                      &    &using cutoff of 200 kpc showed difference above    &\\
                                                    &    &99.9 per cent confidence level. & \\
& & & \\
Is there a significant difference in the $f_{\rm h}$&No &K-S test of $f_{\rm h}$ binned by $P_{178}$ using a &  section \ref{sec:trendhotspot}\\
distribution across the luminosity range?                  &    &cutoff of $5\cdot~10^{26}\ \rm{W\ Hz^{-1}\ sr^{-1}}$    &\\
                                                     &  & showed a difference significant at the 93.2 per cent  & \\
                                                     &  &confidence level                                      & \\
& & & \\
Is there a tendency for the more compact hotspot&Yes&Binomial test shows relatively more compact hotspot tends  &  section \ref{sec:trendhotspot} \\
in a source to be correlated with jet side in Qs?       & &to be on the jet-side at the 99.4 per cent confidence level.  & \\
& & & \\
Is there a tendency for smaller sources to have lower &Yes &K-S test  $\zeta$ significantly lower &  section \ref{sec:zeta}\\
source recession coefficients?                     &    &in small source 99.7 per cent confidence level & \\
& & & \\
Is there a tendency for Qs and BLRGs to have lower &No  &W-M-W test compared Qs and high-power NLRGs and BLRGs&  section \ref{sec:zeta}\\
source recession coefficients $\zeta$?             &    &and low-power NLRGs. Null hypothesis rejected at 74\\
                                                   &    &and 84 per cent confidence level repectively.& \\
& & & \\
Is there a significant difference in the distribution &No  &K-S test compared Qs and high-power NLRGs and BLRGs&  section \ref{sec:zeta}\\
of $\zeta$ between different spectral classes?        &    &and low-power NLRGs. Null hypothesis rejected at 94.5\\ 
                                                      &    & and 81 per cent confidence level.\\
\hline
\end{tabular}
\end{table*}
\subsection{Jets and evidence for beaming on kiloparsec scales}\label{sec:beaming}
On kiloparsec scales, jets have been detected in 30 per cent of
sources, with a further 34 per cent having a feature that is
classified as a possible jet. While observing resolution and
sensitivity are clearly factors in jet detectability, we found that
jets are more commonly detected in Qs and BLRGs, with more
definite jets associated with these classes than for the NLRGs of any
luminosity. This is consistent with the expectations from unification
and beaming models. Statistical tests taking into account the large
number of upper limits on jet prominence show that the broad-line and
narrow-line objects have jet prominences that differ in the sense
expected from beaming if the broad-line objects make smaller angles to
the line of sight.

Further evidence in support of beaming in the kpc-scale jets is the
correlation between core and jet prominence; we showed that there are
significantly more prominent cores in those sources with a detected
kpc-scale jet feature, while core prominence and jet prominence are
correlated in our data even in the presence of upper limits. Since we
know beaming is important in the cores from VLBI observations of
apparent superluminal motion (e.g., Zensus 1997; Hough et al.
2002), these results require beaming to be important in the
kiloparsec-scale jets as well. In addition, we found that the
Laing-Garrington effect, in which the less depolarized lobe is
correlated with the approaching (jet-side) lobe, is detected in the
sample for all jetted sources for which depolarization data were
available. This was the case for 41 sources, 30 of which had the jet
side corresponding to the less depolarized lobe, significant on a
binomial test at the 99.8 per cent confidence level.

Thus the data strongly support the idea that
the jets remain relativistic on kpc scales. We will explore the
implications of our measurements for bulk speeds in the jets and cores
in a future paper.

Some evidence was found for beaming effects in the hotspot data, in
that there was a tendency for the most compact hotspot to be on the
same side as the jet feature in quasars (parametrizing hotspot
compactness by the fractional hotspot size), consistent with the
results of Bridle et al.\ (1994). However, no corresponding correlation
between the jet side and the brighter hotspot was found in any
emission-line class, which is inconsistent with the results of Laing
(1989). The data thus provide only limited support to the idea that
relativistic beaming plays an important role in the appearance of
hotspots: most likely the varied appearance of these features is
dominated by the local conditions, with beaming playing a secondary role.
\subsection{Source morphology}
A significant trend in $R_{\rm ax}$ with source size is found across
the sample (Section \ref{sec:trendaxialratio}). The observed range in
$R_{\rm ax}$ is much greater for sources larger than $\approx 200$
kpc. While observational effects were found to be a source of bias in
the $R_{\rm ax}$ data, with lower $R_{\rm ax}$ values in sources
observed at lower resolution, we concluded that the trend with source
size does not result from such a bias but represents a real physical
trend. 

Early work on source expansion models assumed that the cocoon
  would remain overpressured as the source evolved, which would result
  in self-similar expansion (e.g., Begelman \& Cioffi, 1989),
  but subsequently it was demonstrated that this would only be the
  case for sources in an ambient medium with a decreasing density
  profile (e.g., Falle 1991; Kaiser \& Alexander 1997).
  However, numerical simulation has showed that the lateral expansion
  of the source will slow as the cocoon comes into pressure balance
  with the ambient medium and that this will occur in a typical source
  before it has grown to any considerable size (e.g., Carvalho
  \& O'Dea 2002), a result consistent with the known X-ray
  properties of the environments of FRIIs (e.g. Hardcastle \& Worrall
  2000). The $R_{\rm ax}$ data are consistent with the idea that
  radio sources in general go through an early self-similar expansion
  phase where $R_{\rm ax}$ is approximately constant on size scales of
  the order of the size of the host galaxy, after which lateral
  expansion slows and $R_{\rm ax}$ will increase as the source
  continues to expand linearly.

We also found that Qs are significantly more asymmetric than NLRGs.
The data suggest environmental factors are a predominant cause as
there was no strong evidence for a contribution from relativistic effects, in contrast to the findings of AL00. However, stronger support for relativistic contributions to the BLRG $x_{\rm jet}$ distribution despite the finding that the BLRGs are not significantly more asymmetric than the low-power NLRGs. The implications for unification are discussed below.
\subsection{Unification}\label{sec:diss_uni}
%
%
We begin by noting that the classification of the sources into the
broad and narrow line types is dependent on high-quality spectra. For
example, Laing et al. (1994) have shown that the classifications may
change significantly with improved observations. The classifications
that we use are the best possible with the available data, but
incorrect identification of some sources may introduce a bias that is
difficult to estimate.

The unification model for Qs, NLRGs and BLRGs makes a number of simple
predictions. We expect Qs and BLRGs to be seen at smaller angles to
the line of sight; this means that they should be more commonly
associated with brighter, one-sided jets and brighter cores and should
be statistically smaller, with lower $R_{\rm ax}$ as a consequence. There
is no expectation that lobe size asymmetries should be significantly
different from class to class, unless source expansion speeds are relativistic in which case the Qs and BLRGs would be expected to appear more asymmetrical.
%
%

As discussed in Section \ref{sec:beaming}, Qs and BLRGs do have higher
detection rates for kpc-scale jets, consistent with the expectation
from unification. On the other hand, there is no significant trend for either spectral
class to have quantitatively brighter jets than those in the NLRG, but
the effects of observing resolution and sensitivity are not negligible
and are difficult to account for.

The core prominence is found to be statistically higher in Qs than in
high power NLRGs (Section \ref{sec:cores2}) though the results for the
much smaller sample of BLRGs and low-power NLRGs were less clear-cut.
In fact, we found that the core prominence in NLRGs decreased with
increasing source luminosity, which explains the quantitative
difference between the high-luminosity spectral classes and the lack
of it between those at low luminosity. This trend in the NLRG core
prominence data is not obviously predicted from unification. However, we argued
in Section \ref{sec:cores2} that this is evidence that the
higher-luminosity sources may have higher nuclear bulk Lorentz factors
($\gamma$), leading to greater Doppler suppression of core emission:
if the parsec-scale bulk-flow speeds of the emitting material are
greater in the higher luminosity sources, the cores of a greater
proportion of the NLRGs could be Doppler suppressed, as the
angle to the observer's line-of-sight needed to detect
beamed emission would be smaller. This would represent a minor
modification to the standard unification picture.

Considering source morphology, we found the Qs to have significantly
lower $R_{\rm ax}$ values than the NLRGs (Section \ref{sec:Rax_uni}).
However, in unified models we expect this to be a result of
projection effects giving systematically lower source linear sizes. We
found no evidence that either Qs or BLRGs are significantly smaller
than the NLRGs. Although this could indicate that the statistically
lower $R_{\rm ax}$ values in Qs are a result of relatively broader
lobes (unexpected in the unified model), this does not appear to be the case and we concluded in section
\ref{sec:Rax_uni} that the difference in $R_{\rm ax}$ between the Qs
and NLRGs was not inconsistent with projection effects, with smaller
lobe sizes in Qs for a similar lobe width (Fig. \ref{fig:LLS_W_CLASS}).
The fact that we did not obtain a similar result for the
low-luminosity spectral classes might indicate a real difference between
the high and low luminosity classes, although there was no trend in
$R_{\rm ax}$ across the luminosity or redshift range and we found that the low power distribution is possibly more strongly affected by observational effects, with data from a few particularly small and large sources observed at low resolution (Fig. \ref{fig:LLS_W_BN}).

On the other hand, evidence for real differences in the Q environments
is provided by the distribution of the fractional separation
difference as defined by the longer lobe, $x_{\rm lobe}$ (Section \ref{sec:x_lobe_beamin}). These data
show that Qs are significantly more asymmetric than the high power
NLRGs. This is not expected from unification directly. It could be
consistent with the scheme if relativistic effects were contributing
to the observed lobe size asymmetry, which would require relativistic
source advance speeds; however, assuming that the kpc-scale jet
indicates the approaching lobe and re-defining the fractional
separation in terms of the jet-side lobe, $x_{\rm jet}$, no
significant differences were found between any of the spectral
classes, so that there is no evidence for the hypothesis that the greater asymmetry of the quasars is due to relativistic effects. This then suggests that the effect is environmental, though of course, as not all sources have jets, the $x_{\rm jet}$ data do not include all sample sources and so will be biased. 
%
%
\section{Summary and Conclusion}
A large complete sample of FRII type radio sources has been studied
with high sensitivity, high resolution observations, allowing standard
models of unification and relativistic beaming to be tested. The
sample consists of 98 sources from the 3CRR sample with $z<1$,
including 15 Qs, 11 BLRGs and 57 NLRGs, as well as 15 LERGs, and
covers a large range in source luminosity (from $5 \times 10^{24}$ to
$2 \times 10^{28}$ W Hz$^{-1}$ sr$^{-1}$ at 178 MHz). The high
quality of the maps has allowed a comprehensive search for trends and
correlations between source observables, with source sizes, axial
ratios, core, jet and hotspot properties measured from the same
observed data.

We have searched for differences in the distributions of the various
source observables with respect to the sample's range in power,
redshift and source size, and carried out tests of the predictions of
the standard model of unification and relativistic beaming. These
predictions are that Qs and BLRGs will be statistically smaller, with
higher jet detection rates and brighter jets and cores. In addition, there is some weaker evidence that hotspot properties, such as
compactness, may be correlated with the jet side, which implies
that there is continued relativistic flow in the hotspot regions.

Some evidence for differences in the sample as a function of
luminosity were found:
\begin{itemize}
\item core prominence was found to decrease with source luminosity. We proposed that a greater proportion of higher luminosity
sources have higher parsec-scale bulk flow speeds and experience
stronger Doppler suppression: this can be accommodated as a
modification to standard unified models.

\item Qs are found to be more asymmetric than the high power NLRGs and
  the evidence is that this is not due to relativistic effects; also
  no such difference is found between BLRGs and NLRGs. This is
  possible evidence that there is a systematic difference between the
  environments of Qs and NLRGs at high radio luminosities, which would
  not be consistent with simple
  unification models.
\end{itemize}
The principal conclusions {\it consistent} with the predictions of the standard model can be summarized as follows.
\begin{itemize}
\item evidence for beaming on kiloparsec scales was found across the
  sample; jet detection rates as a function of source class,
  correlation between core and jet prominence and detections of the
  Laing-Garrington effect were all consistent with relativistic speeds
  in the kpc-scale jets. 
\item cores were found to be statistically brighter in Qs and BLRGs than in the corresponding NLRGs, consistent with
  expectations. 
\item $R_{\rm ax}$ values were found to be lower in Qs (although not
  BLRGs) than those in the corresponding population of NLRG,
  consistent with the expected projection effects.
\end{itemize}
A further result somewhat independent from the expectations of the standard model was that
\begin{itemize}
\item there is evidence from the distribution of $R_{\rm ax}$ that source development has an initial phase where expansion is self-similar or close to being so, possibly on the scale of the host galaxy, before lateral expansion slows or ceases while the expansion along the source axis continues.
\end{itemize}
%
%

We will consider the implications of our measurements for quantitative
estimates of the relativistic bulk speeds in cores and jets in a
future paper (Mullin \& Hardcastle, in prep.).

\section{Acknowledgments}
LMM thanks PPARC for a research studentship that supported the early
parts of this work. MJH thanks the Royal Society for a research fellowship.
%
%

%
\onecolumn 
\begin{longtable}{lrrrrrrrrrrr}
\caption{Core and jet properties}\\
\hline
Source&\multicolumn{2}{c}{Core prominence}&Jet
side&\multicolumn{2}{c}{Jet prominence}&\multicolumn{2}{c}{C'jet
  prominence}&\multicolumn{3}{c}{Fractional jet}\\
name&\multicolumn{2}{c}{$(\times 1000)$}&&\multicolumn{2}{c}{$(\times
1000)$}&\multicolumn{2}{c}{$(\times 1000)$}&position&length&termination\\
&value&error&&value&error&value&error\\
\hline\endhead
4C12.03 & $<$9 & & N & 31 & 2 &42 & 4 &-- & 0.3223 & 0.3223 \\
3C6.1 & 7.46 & 0.2 &S & $<$18.6 & & $<$1.0 & & -- & -- & -- \\
3C16 & 0.23 & 0.010 &N & 3.6 & 0.2 &2.4 & 0.2 &0.0539 & 0.195 & 0.249 \\
3C19 & 0.33 & 0.08 &S & 17 & 3 &$<$23.0 & & 0.642 & 0.29 & 0.93 \\
3C20 & 3.3 & 0.6 &N & 7 & 5 &$<$9.0 & & 0.178 & 0.7421 & 0.921 \\
3C22 & 6.54 & 0.2 &N & 4.2 & 0.9 &$<$4.0 & & -- & 0.7568 & 0.7568 \\
3C33 & 36.2 & 1 &S & 42 & 3 &20 & 7 &-- & 0.2611 & 0.2611 \\
3C33.1 & 12.2 & 0.4 &S & 27.0 & 0.8 &$<$37.0 & & -- & 0.5252 & 0.5252 \\
3C34 & 1.04 & 0.03 &N & 7.3 & 0.2 &1.1 & 0.10 &0.472 & 0.606 & 1.08 \\
3C35 & 18.6 & 0.05 &N & $<$12 & & $<$15 & & -- & -- & -- \\
3C41 & 1.2 & 0.04 &S & 13 & 0.6 &$<$21.0 & & 0.207 & 0.721 & 0.929 \\
3C42 & 3.05 & 0.09 &S & $<$24.8 & & $<$17.8 & & -- & -- & -- \\
3C46 & 1.44 & 0.04 &S & 1.4 & 0.10 &$<$5.9 & & 0.20 & 0.39 & 0.59 \\
3C47 & 66.8 & 2 &S & 12 & 2 &$<$1.1 & & -- & 0.7238 & 0.7238 \\
3C55 & 5.33 & 0.2 &S & 10 & 4 &$<$36.0 & & -- & 0.291 & 0.291 \\
3C61.1 & 2.2 & 0.07 &N & $<$58.2 & & $<$26.9 & & -- & -- & -- \\
3C67 & 2.13 & 0.06 &N & 45 & 8 &$<$14.8 & & 0.39 & 0.23 & 0.63 \\
3C79 & 6.04 & 0.2 &N & $<$3.8 & & $<$1.3 & & -- & -- & -- \\
3C98 & 6.10 & 0.2 &N & 50 & 20 &$<$13.0 & & -- & 0.8676 & 0.8676 \\
3C109 & 247 & 7 &S & 25 & 2 &$<$8.8 & & 0.6250 & 0.198 & 0.823 \\
4C14.11 & 29.7 & 0.9 &N & 11 & 0.4 &$<$11.0 & & 0.7247 & 0.2543 & 0.9790 \\
3C123 & 109 & 3 &S & $<$30.0 & & $<$2.1 & & -- & -- & -- \\
3C132 & 4.1 & 0.2 &S & 3 & 2 &$<$13.0 & & 0.148 & 0.553 & 0.701 \\
3C153 & $<$0.5 & & S & 10 & 5 &8.0 & 0.2 &0.21 & 0.412 & 0.62 \\
3C171 & 2.0 & 0.10 &S & 6.3 & 0.8 &6.0 & 0.2 &-- & 0.762 & 0.762 \\
3C172 & 0.37 & 0.02 &S & $<$25.0 & & $<$3.6 & & -- & -- & -- \\
3C173.1 & 9.64 & 0.3 &N & 2.1 & 0.10 &$<$0.3 & & -- & 0.3739 & 0.3739 \\
3C175 & 14.0 & 0.4 &S & 6.3 & 0.3 &$<$10.6 & & -- & 0.6460 & 0.6460 \\
3C175.1 & 1.1 & 0.08 &S & 75 & 10 &$<$15.1 & & 0.652 & 0.24 & 0.89 \\
3C184 & 0.11 & 0.07 &N & $<$47.2 & & $<$37.4 & & -- & -- & -- \\
3C184.1 & 6.0 & 0.5 &N & 3.9 & 0.7 &$<$5.9 & & 0.4354 & 0.3048 & 0.7401 \\
DA240 & 273 & 8 &S & 83 & 10 &$<$23.0 & & -- & 0.4248 & 0.4248 \\
3C192 & 4.0 & 0.2 &S & 8.0 & 0.2 &$<$3.2 & & 0.765 & 0.150 & 0.915 \\
3C196 & 11.8 & 0.4 &S & $<$58.4 & & $<$46.3 & & -- & -- & -- \\
3C200 & 38.2 & 1 &S & 60 & 8 &$<$1.0 & & -- & 0.7279 & 0.7279 \\
4C14.27 & 11.4 & 0.3 &N & 16 & 3 &6.1 & 0.9 &-- & 0.4896 & 0.4896 \\
3C207 & 539 & 20 &S & 190 & 10 &$<$43.9 & & -- & 0.695 & 0.695 \\
3C215 & 0.88 & 0.10 &S & 38.4 & 1 &$<$2.8 & & -- & 0.8325 & 0.8325 \\
3C217 & 0.69 & 0.02 &N & $<$12.3 & & $<$2.1 & & -- & -- & -- \\
3C216 & 732 & 20 &S & $<$5.3 & & -- & & -- & -- & -- \\
3C219 & 51.6 & 2 &S & 56.5 & 0.3 &2.1 & 0.10 &0.0203 & 0.1677 & 0.188 \\
3C220.1 & 26.9 & 0.8 &N & 3.6 & 0.10 &$<$9.5 & & 0.103 & 0.8071 & 0.910 \\
3C220.3 & $<$0.2 & & N & $<$2.9 & & $<$1.5 & & -- & -- & -- \\
3C223 & 8.50 & 0.3 &N & 10 & 4 &6 & 9 &0.6459 & 0.1899 & 0.8358 \\
3C225B & 1.3 & 0.10 &N & $<$4.2 & & $<$2.9 & & -- & -- & -- \\
3C226 & 3.71 & 0.10 &S & $<$3.8 & & $<$3.1 & & -- & -- & -- \\
4C73.08 & 7 & 1 &S & $<$615.3 & & $<$87.2 & & -- & -- & -- \\
3C228 & 19.0 & 0.6 &S & 10.7 & 0.3 &$<$15.0 & & 0.141 & 0.316 & 0.457 \\
3C234 & 34.5 & 1 &N & 10 & 8 &$<$19.0 & & -- & 0.0965 & 0.0965 \\
3C236 & 5170 & $2 \times 10^{2}$ &N & 290 & 50 &$<$61.9 & & 0.00450 & 0.3860 & 0.390 \\
4C74.16 & 1.61 & 0.05 &S & 9.7 & 1 &$<$5.4 & & -- & 0.5702 & 0.5702 \\
3C244.1 & 1 & 0.7 &N & 3.2 & 1 &$<$13.4 & & 0.7361 & 0.244 & 0.980 \\
3C247 & 1.81 & 0.05 &N & $<$2.1 & & $<$1.6 & & -- & -- & -- \\
3C249.1 & 70.7 & 2 &N & 35 & 5 &$<$9.9 & & -- & 0.289 & 0.289 \\
3C254 & 20.0 & 0.6 &N & $<$10.3 & & $<$6.3 & & -- & -- & -- \\
3C263 & 161 & 5 &S & 39 & 5 &$<$1.7 & & -- & 0.8970 & 0.8970 \\
3C263.1 & 1.4 & 0.05 &N & $<$1.2 & & $<$0.5 & & -- & -- & -- \\
3C265 & 2.78 & 0.08 &N & 4.8 & 1 &$<$5.3 & & -- & 0.6910 & 0.6910 \\
3C268.1 & 0.45 & 0.04 &S & $<$4.1 & & $<$1.9 & & -- & -- & -- \\
3C268.3 & 1.2 & 0.09 &S & $<$25.0 & & $<$7.5 & & -- & -- & -- \\
3C274.1 & 2.33 & 0.07 &S & 8 & 8 &$<$20.9 & & -- & 0.7585 & 0.7585 \\
3C275.1 & 209 & 6 &N & 30 & 20 &$<$4.6 & & -- & 0.681 & 0.681 \\
3C277.2 & 0.68 & 0.02 &S & 17 & 4 &$<$2.2 & & -- & 0.327 & 0.327 \\
3C280 & $<$0.7 & & S & $<$13.8 & & $<$0.4 & & -- & -- & -- \\
3C284 & 2.79 & 0.08 &S & $<$6.6 & & $<$2.8 & & -- & -- & -- \\
3C285 & 6.49 & 0.2 &N & 7.8 & 1 &$<$19.9 & & 0.071 & 0.89 & 0.96 \\
3C289 & 0.78 & 0.08 &N & $<$4.8 & & 2 & 1 &-- & -- & -- \\
3C292 & 0.51 & 0.03 &S & $<$5.9 & & $<$3.9 & & -- & -- & -- \\
3C295 & 3.64 & 0.10 &N & $<$230.5 & & $<$118.9 & & -- & -- & -- \\
3C299 & 2.3 & 0.2 &N & $<$11.8 & & $<$21.2 & & -- & -- & -- \\
3C300 & 6.20 & 0.2 &N & 2.4 & 0.10 &$<$0.2 & & -- & 0.9148 & 0.9148 \\
3C303 & 106 & 3 &N & 63.0 & 2 &$<$13.0 & & -- & 0.5108 & 0.5108 \\
3C319 & $<$0.3 & & S & $<$1.8 & & $<$0.10 & & -- & -- & -- \\
3C321 & 23.1 & 0.7 &N & 15 & 1 &-- & & 0.00942 & 0.1258 & 0.135 \\
3C325 & 10.1 & 0.3 &N & 5.8 & 0.9 &$<$25.3 & & 0.123 & 0.633 & 0.755 \\
3C326 & 18.2 & 0.6 &S & $<$215.3 & & $<$25.8 & & -- & -- & -- \\
3C330 & 0.54 & 0.02 &S & $<$34.0 & & $<$21.5 & & -- & -- & -- \\
3C334 & 86.8 & 3 &S & 17 & 1 &0.2 & 0.2 &-- & 0.5630 & 0.5630 \\
3C336 & 21.3 & 0.6 &S & 7.7 & 0.7 &$<$14.9 & & -- & 0.464 & 0.464 \\
3C341 & 0.70 & 0.03 &S & 22 & 2 &$<$7.4 & & -- & 0.438 & 0.438 \\
3C337 & 0.34 & 0.03 &S & 3.2 & 0.10 &$<$4.4 & & 0.0876 & 0.6836 & 0.771 \\
3C340 & 1.16 & 0.03 &S & $<$6.0 & & $<$3.6 & & -- & -- & -- \\
3C349 & 24.2 & 0.7 &S & 0.31 & 0.04 &$<$2.1 & & 0.0826 & 0.0245 & 0.107 \\
3C351 & 12.1 & 0.4 &N & 5.3 & 2 &$<$0.9 & & -- & 0.0864 & 0.0864 \\
3C352 & 3.38 & 0.10 &N & 8.9 & 0.7 &$<$9.4 & & -- & 0.542 & 0.542 \\
3C381 & 4.70 & 0.10 &N & $<$2.9 & & $<$1.3 & & -- & -- & -- \\
3C382 & 251 & 8 &N & 14 & 1 &$<$1100.0 & & -- & 0.88 & 0.88 \\
3C388 & 57.1 & 2 &S & 17 & 1 &$<$5.3 & & -- & 0.661 & 0.661 \\
3C390.3 & 733 & 20 &N & 20 & 10 &$<$650.0 & & -- & 0.6298 & 0.6298 \\
3C401 & 28.5 & 0.9 &S & 33.8 & 0.4 &$<$5.7 & & -- & 0.8903 & 0.8903 \\
3C427.1 & 0.89 & 0.03 &N & 20 & 10 &$<$16.5 & & 0.612 & 0.351 & 0.963 \\
3C433 & 1.2 & 0.3 &N & 9.8 & 0.10 &$<$8.3 & & -- & 0.594 & 0.594 \\
3C436 & 17.9 & 0.5 &S & 3.8 & 0.8 &$<$0.3 & & 0.101 & 0.7804 & 0.881 \\
3C438 & 16.2 & 0.5 &N & 40 & 4 &$<$9.8 & & -- & 0.757 & 0.757 \\
3C441 & $<$0.10 & & N & 23 & 2 &$<$10.4 & & -- & 0.787 & 0.787 \\
3C452 & 126 & 4 &S & 13 & 2 &9.0 & 2 &-- & 0.7547 & 0.7547 \\
3C455 & $<$2.6 & & S & $<$30.8 & & $<$18.1 & & -- & -- & -- \\
3C457 & 2.32 & 0.07 &S & $<$5.4 & & $<$5.2 & & -- & -- & -- \\

\hline
\end{longtable}
\clearpage

\begin{longtable}{lrrrrrrrrrr}
\caption{Hotspot properties}\\
\hline
Source&\multicolumn{4}{c}{N hotspot properties}&\multicolumn{4}{c}{S
  hotspot properties}&\multicolumn{2}{c}{Recession properties}\\
&prominence&size&frac. size&$\eta$&prominence&size&frac. size&$\eta$&$\zeta$&$\delta$\\
&$(\times 1000)$&(arcsec)&&&$(\times 1000)$&(arcsec)\\
\hline\endhead
4C12.03 & 13.7 & 19.88 & 0.1831 & 0.9587 & 18.5 & 33.03 & 0.2340 & 0.6186 & 0.8145 & 0.6452 \\
3C6.1 & 5.53 & 0.25 & 0.017 & 0.9493 & 2.77 & 0.25 & 0.020 & 0.9665 & 0.9572 & 1.018 \\
3C16 & -- & -- & -- & 0.881 & 42.9 & 13.3 & 0.562 & 1.027 & 0.926 & 1.17 \\
3C19 & 0.5 & 0.17 & 0.049 & 1.12 & 0.42 & 0.1 & 0.04 & 0.838 & 0.988 & 0.746 \\
3C20 & 3.3 & 0.18 & 0.0075 & 0.9670 & 1.9 & 0.20 & 0.0076 & 0.9926 & 0.9794 & 1.026 \\
3C22 & 8.31 & 0.30 & 0.022 & 0.9762 & 8.97 & 0.39 & 0.032 & 0.8576 & 0.9206 & 0.8785 \\
3C33 & 0.979 & 0.63 & 0.0045 & 0.9865 & 10.4 & 1.0 & 0.0087 & 0.9829 & 0.9849 & 0.9964 \\
3C33.1 & 6.29 & 4.23 & 0.0507 & 0.978 & 1.07 & 8.16 & 0.0549 & 0.9758 & 0.977 & 1.00 \\
3C34 & 5.01 & 1.85 & 0.0794 & 0.966 & 0.511 & 0.57 & 0.025 & 1.01 & 0.988 & 1.05 \\
3C35 & 0.64 & -- & -- & 0.9355 & 0.53 & 8.90 & 0.0252 & 0.7511 & 0.8426 & 1.245 \\
3C41 & 10.4 & 1.0 & 0.075 & 0.9701 & 24.7 & 0.53 & 0.047 & 1.040 & 1.003 & 1.072 \\
3C42 & 4.54 & 0.42 & 0.031 & 1.040 & 7.87 & 2.4 & 0.17 & 1.092 & 1.066 & 0.9528 \\
3C46 & 1.42 & 1.2 & 0.012 & 0.92 & 0.958 & 2.34 & 0.034 & 0.99 & 0.95 & 1.1 \\
3C47 & 0.029 & 2.21 & 0.0632 & 1.034 & 7.44 & 1.2 & 0.030 & 0.9815 & 0.9986 & 1.054 \\
3C55 & 3.12 & 0.42 & 0.012 & 0.9903 & 0.559 & 0.21 & 0.0063 & 0.9734 & 0.9822 & 0.9830 \\
3C61.1 & 5.00 & 11.9 & 0.157 & 1.11 & -- & -- & -- & 0.9882 & 1.04 & 1.12 \\
3C67 & 2.5 & 0.08 & 0.04 & 0.882 & 35.2 & 0.12 & 0.088 & 0.785 & 0.844 & 0.890 \\
3C79 & 0.683 & 0.80 & 0.020 & 1.013 & 0.541 & 0.51 & 0.0098 & 0.9792 & 0.9941 & 1.035 \\
3C98 & 0.77 & 3.75 & 0.0269 & 0.9558 & 0.52 & 2.95 & 0.0216 & 0.9173 & 0.9368 & 0.9597 \\
3C109 & 1.65 & 1.31 & 0.0287 & 1.016 & 6.18 & 0.48 & 0.0097 & 0.9887 & 1.002 & 1.027 \\
4C14.11 & 0.175 & 0.91 & 0.016 & 1.035 & 0.297 & 0.24 & 0.0043 & 0.6680 & 0.8522 & 0.6455 \\
3C123 & 0.70 & 0.26 & 0.014 & 0.913 & 0.734 & 0.1 & 0.008 & 0.941 & 0.928 & 0.970 \\
3C132 & 3.44 & 0.17 & 0.015 & 1.002 & 1.4 & 0.33 & 0.030 & 0.9981 & 1.000 & 0.9964 \\
3C153 & 7.68 & 0.1 & 0.03 & 0.51 & 1.3 & 0.33 & 0.096 & 1.01 & 0.72 & 2.0 \\
3C171 & 5.1 & 0.2 & 0.03 & 0.996 & 4.04 & 0.24 & 0.047 & 0.950 & 0.973 & 0.954 \\
3C172 & 6.44 & 1.64 & 0.0357 & 1.01 & 1.60 & 0.34 & 0.0070 & 0.911 & 0.958 & 1.11 \\
3C173.1 & 0.666 & 0.51 & 0.019 & 1.072 & 0.54 & 0.38 & 0.012 & 0.9886 & 1.024 & 1.084 \\
3C175 & 3.62 & 0.46 & 0.021 & 1.003 & 0.358 & 0.47 & 0.015 & 1.007 & 1.006 & 0.9960 \\
3C175.1 & 4.10 & 0.14 & 0.032 & 0.982 & 1.93 & 0.11 & 0.032 & 0.814 & 0.907 & 0.830 \\
3C184 & 0.601 & 0.07 & 0.02 & 0.974 & 6.86 & 0.08 & 0.04 & 0.988 & 0.979 & 1.01 \\
3C184.1 & 0.48 & 1.80 & 0.0170 & 0.9814 & 1.3 & 0.72 & 0.0092 & 0.987 & 0.984 & 1.01 \\
DA240 & 22.2 & 19.81 & 0.02203 & 0.7572 & 0.28 & 46.25 & 0.04405 & 0.571 & 0.657 & 1.33 \\
3C192 & 3.4 & 3.50 & 0.0319 & 0.9840 & 0.927 & 4.83 & 0.0522 & 1.01 & 0.994 & 1.02 \\
3C196 & 11.7 & 0.24 & 0.094 & 0.902 & 15.7 & 0.57 & 0.15 & 0.840 & 0.865 & 1.07 \\
3C200 & 3.30 & 0.55 & 0.050 & 0.55 & 1.66 & 1.33 & 0.0940 & 0.7375 & 0.65 & 0.74 \\
4C14.27 & 7.01 & 1.2 & 0.053 & 0.9479 & 1.11 & 0.55 & 0.036 & 1.001 & 0.9694 & 1.056 \\
3C207 & 3.71 & 0.94 & 0.16 & 0.925 & 4.05 & 0.53 & 0.075 & 0.958 & 0.943 & 0.965 \\
3C215 & 5.53 & 5.30 & 0.203 & 0.8588 & 0.832 & 1.36 & 0.0982 & 0.9978 & 0.9076 & 1.162 \\
3C217 & 9.28 & -- & -- & 1.00 & 4.37 & 0.29 & 0.079 & 0.911 & 0.976 & 0.911 \\
3C216 & 8.46 & 0.34 & 0.37 & 0.94 & 1.21 & 0.17 & 0.082 & 0.78 & 0.85 & 1.2 \\
3C219 & 0.047 & 0.68 & 0.0073 & 0.8405 & 1.23 & 2.40 & 0.0244 & 0.8870 & 0.8660 & 0.9476 \\
3C220.1 & 0.947 & 0.38 & 0.026 & 1.009 & 0.153 & 0.33 & 0.019 & 0.9323 & 0.9672 & 1.082 \\
3C220.3 & 3.23 & 0.71 & 0.17 & 0.800 & 0.890 & 0.81 & 0.14 & 0.911 & 0.863 & 0.878 \\
3C223 & 0.61 & 2.40 & 0.0155 & 0.9236 & 0.4 & 4.00 & 0.0266 & 0.9883 & 0.9558 & 1.070 \\
3C225B & 7.37 & 0.13 & 0.062 & 0.930 & 4.42 & 0.29 & 0.10 & 1.05 & 1.00 & 0.886 \\
3C226 & 0.479 & 0.31 & 0.014 & 1.017 & 5.89 & 0.48 & 0.027 & 0.8560 & 0.9297 & 1.189 \\
4C73.08 & 1.07 & 24.17 & 0.04114 & 0.3239 & 12.9 & 32.66 & 0.07814 & 0.9323 & 0.5843 & 2.879 \\
3C228 & 4.44 & 0.48 & 0.019 & 0.9214 & 4.79 & 0.42 & 0.019 & 1.033 & 0.9719 & 1.121 \\
3C234 & 1.54 & 0.35 & 0.0055 & 1.007 & 1.4 & 0.54 & 0.017 & 1.031 & 1.015 & 1.024 \\
3C236 & -- & -- & -- & 0.9141 & 4.51 & 24.77 & 0.01722 & 0.9966 & 0.9597 & 0.9172 \\
4C74.16 & 6.64 & 1.2 & 0.048 & 0.9948 & 0.979 & 0.53 & 0.030 & 0.8880 & 0.9524 & 0.8927 \\
3C244.1 & 8.93 & 0.34 & 0.013 & 1.009 & 0.335 & 0.20 & 0.0081 & 0.7867 & 0.9046 & 0.7798 \\
3C247 & 12.5 & 0.28 & 0.048 & 1.00 & 25.1 & 0.38 & 0.046 & 0.989 & 0.995 & 1.01 \\
3C249.1 & 5.70 & 0.36 & 0.012 & 1.74 & 11.6 & 1.17 & 0.0615 & 0.7924 & 1.05 & 2.19 \\
3C254 & 7.41 & 0.53 & 0.042 & 0.9199 & 7.72 & 0.47 & 0.10 & 0.877 & 0.913 & 0.954 \\
3C263 & 1.26 & 0.44 & 0.013 & 0.8526 & 22.3 & 0.30 & 0.018 & 1.001 & 0.9025 & 1.174 \\
3C263.1 & 7.30 & 0.23 & 0.087 & 0.937 & 5.20 & 0.04 & 0.007 & 0.584 & 0.692 & 1.60 \\
3C265 & 3.80 & 0.90 & 0.019 & 0.7714 & 6.87 & 0.70 & 0.023 & 0.9823 & 0.8564 & 1.273 \\
3C268.1 & 48.2 & 0.28 & 0.012 & 0.9945 & 1.71 & 0.35 & 0.019 & 0.9484 & 0.9739 & 0.9536 \\
3C268.3 & 43.5 & 0.1 & 0.2 & 0.67 & 23.4 & 0.19 & 0.15 & 0.869 & 0.81 & 0.77 \\
3C274.1 & 5.83 & 4.54 & 0.0512 & 0.8866 & 3.03 & 6.21 & 0.0802 & 0.9375 & 0.9101 & 1.057 \\
3C275.1 & 5.26 & 0.36 & 0.030 & 0.804 & 6.37 & 0.64 & 0.095 & 0.960 & 0.861 & 1.19 \\
3C277.2 & 1.43 & 1.7 & 0.040 & 0.9439 & 8.13 & 0.33 & 0.020 & 0.9976 & 0.9595 & 1.057 \\
3C280 & 0.700 & 0.06 & 0.006 & 0.594 & 2.99 & 0.24 & 0.032 & 0.849 & 0.711 & 1.43 \\
3C284 & 0.983 & 2.02 & 0.0184 & 0.9756 & 2.55 & 0.92 & 0.012 & 1.011 & 0.9900 & 1.036 \\
3C285 & 2.43 & 8.55 & 0.11 & 0.79 & 3.38 & 14.74 & 0.152 & 0.91 & 0.86 & 0.86 \\
3C289 & -- & -- & -- & 0.963 & 8.72 & 0.1 & 0.02 & 0.971 & 0.967 & 1.01 \\
3C292 & 1.02 & 3.48 & 0.0554 & 0.9195 & 6.10 & 1.54 & 0.0220 & 0.9921 & 0.9586 & 0.9268 \\
3C295 & 10.1 & 0.15 & 0.062 & 0.841 & 11.4 & 0.28 & 0.083 & 0.840 & 0.840 & 1.00 \\
3C299 & 17.0 & 0.1 & 0.04 & 0.980 & 0.82 & 0.07 & 0.009 & 0.977 & 0.978 & 1.00 \\
3C300 & 0.05 & 0.30 & 0.0044 & 1.027 & 1.3 & 0.35 & 0.012 & 1.068 & 1.039 & 1.039 \\
3C303 & 21.6 & 1.40 & 0.0504 & 0.648 & 0.17 & 0.70 & 0.034 & 1.00 & 0.786 & 1.54 \\
3C319 & 1.1 & 1.1 & 0.024 & 1.006 & -- & -- & -- & 0.8832 & 0.9397 & 1.139 \\
3C321 & 3.84 & 3.16 & 0.0213 & 1.000 & 16.0 & 1.51 & 0.0101 & 0.9349 & 0.9677 & 0.9349 \\
3C325 & 0.0099 & 0.18 & 0.017 & 0.894 & 10.6 & 0.26 & 0.034 & 0.883 & 0.890 & 0.988 \\
3C326 & -- & -- & -- & 0.9827 & 5.64 & 60.16 & 0.08020 & 0.9094 & 0.9348 & 1.081 \\
3C330 & 0.019 & 0.79 & 0.026 & 0.9817 & 3.03 & 0.29 & 0.0092 & 1.032 & 1.007 & 0.9510 \\
3C334 & 1.61 & 1.7 & 0.053 & 0.8915 & 1.24 & 0.70 & 0.031 & 0.9256 & 0.9049 & 1.038 \\
3C336 & 19.3 & 1.0 & 0.069 & 1.005 & 2.51 & 0.41 & 0.039 & 0.654 & 0.862 & 0.650 \\
3C341 & -- & -- & -- & 0.867 & -- & -- & -- & 0.938 & 0.897 & 1.08 \\
3C337 & 5.54 & 0.56 & 0.019 & 1.042 & 8.84 & 0.23 & 0.015 & 0.9342 & 1.004 & 0.8968 \\
3C340 & 4.42 & 1.1 & 0.048 & 0.9865 & 5.51 & 0.39 & 0.019 & 1.035 & 1.009 & 1.049 \\
3C349 & 0.26 & 0.40 & 0.0099 & 1.006 & 5.9 & 0.56 & 0.012 & 0.9721 & 0.9884 & 1.035 \\
3C351 & 8.36 & 0.22 & 0.0081 & 0.9937 & 0.264 & 1.2 & 0.033 & 1.013 & 1.005 & 0.9807 \\
3C352 & 6.54 & 0.50 & 0.070 & 0.699 & -- & -- & -- & 0.989 & 0.839 & 1.41 \\
3C381 & 0.48 & 0.18 & 0.0055 & 0.9911 & 0.37 & 0.60 & 0.015 & 0.9631 & 0.9761 & 1.029 \\
3C382 & 2.1 & 2.10 & 0.024 & 0.97 & 0.37 & 2.20 & 0.024 & 0.96 & 0.97 & 0.98 \\
3C388 & 2.29 & 1.15 & 0.0410 & 0.719 & 9.66 & 5.34 & 0.237 & 0.800 & 0.758 & 1.11 \\
3C390.3 & 4.12 & 5.78 & 0.0447 & 0.9846 & 15.6 & 4.61 & 0.0504 & 0.983 & 0.984 & 0.998 \\
3C401 & 0.2 & 0.50 & 0.045 & 0.825 & 0.2 & 0.40 & 0.031 & 0.9351 & 0.884 & 0.882 \\
3C427.1 & 0.873 & 0.40 & 0.032 & 0.930 & 0.885 & 0.30 & 0.021 & 0.697 & 0.806 & 1.33 \\
3C433 & -- & -- & -- & 0.793 & 0.10 & 0.60 & 0.018 & 0.17 & 0.43 & 4.8 \\
3C436 & 0.72 & 2.28 & 0.0394 & 0.9803 & 0.800 & 0.39 & 0.0077 & 0.8948 & 0.9416 & 0.9128 \\
3C438 & 0.07 & 0.20 & 0.017 & 0.9454 & 0.04 & 0.20 & 0.019 & 0.830 & 0.890 & 0.878 \\
3C441 & 6.17 & 0.32 & 0.029 & 0.715 & 4.00 & 1.99 & 0.0780 & 0.973 & 0.899 & 0.735 \\
3C452 & 0.3 & 2.35 & 0.0172 & 0.9969 & 0.51 & 0.92 & 0.0064 & 0.8857 & 0.9389 & 1.125 \\
3C455 & 17.2 & 0.66 & 0.46 & 0.63 & 25.5 & 0.74 & 0.27 & 0.773 & 0.73 & 0.82 \\
3C457 & 1.59 & 2.58 & 0.0242 & 0.9604 & 1.37 & 0.71 & 0.0070 & 0.9587 & 0.9596 & 0.9982 \\

\hline
\end{longtable}

\clearpage

\begin{longtable}{lrrrrrrrrrrr}
\caption{Lobe properties}\\
\hline
Source&\multicolumn{5}{c}{North lobe}&\multicolumn{5}{c}{South
  lobe}&$x_{\rm lobe}$\\
&$LLS_{\rm l}$&$l$&RC&$f_{\rm l}$&$R_{\rm ax}$&$LLS_{\rm
  l}$&$l$&RC&$f_{\rm l}$&$R_{\rm ax}$\\
&(kpc)&(kpc)&(kpc)&&&(kpc)&(kpc)&(kpc)\\
\hline\endhead
4C12.03 & 293.3 & 293.3 & 16.5 & 1.00 & 2.511 & 381.2 & 215.7 & 8.26 & 1.00 & 2.856 & 0.1304\\
3C6.1 & 111.4 & 111.4 & 9.08 & 0.89 & 2.87 & 95.62 & 95.62 & 8.78 & 0.83 & 2.38 & 0.07630\\
3C16 & 284 & 284 & 13.5 & 0.98 & 3.23 & 128.6 & 128.6 & 20.3 & 1.00 & 1.58 & 0.377\\
3C19 & 21.2 & 20.5 & 2.8 & 1.00 & 1.57 & 18.7 & 18.4 & 3.6 & 1.00 & 1.57 & 0.0629\\
3C20 & 72.83 & 71.50 & 5.02 & 1.00 & 1.871 & 77.76 & 67.63 & 4.87 & 1.00 & 2.520 & 0.03275\\
3C22 & 109.9 & 109.1 & 6.5 & 0.51 & 4.06 & 96.30 & 96.30 & 7.3 & 0.48 & 1.83 & 0.06575\\
3C33 & 159.4 & 159.4 & 1.64 & 0.26 & 4.93 & 132.6 & 132.6 & 1.64 & 0.29 & 3.640 & 0.09179\\
3C33.1 & 254 & 254 & 12.7 & 1.00 & 1.29 & 452.8 & 436.8 & 9.54 & 0.94 & 2.81 & 0.281\\
3C34 & 165 & 165 & 8.5 & 0.98 & 2.22 & 162 & 151 & 8.88 & 1.00 & 2.9 & 0.00953\\
3C35 & 447.7 & 447.7 & 60.81 & 1.00 & 1.360 & 457.0 & 454.1 & 48.49 & 1.00 & 1.498 & 0.01032\\
3C41 & 99.87 & 95.15 & 7.87 & 0.75 & 2.49 & 84.96 & 84.96 & 8.69 & 1.00 & 1.81 & 0.08066\\
3C42 & 74.01 & 74.01 & 20.8 & 1.00 & 1.87 & 75.98 & 75.98 & 20.8 & 0.98 & 1.013 & 0.01315\\
3C46 & 588 & 588 & 12.3 & 0.83 & 4.7 & 390 & 390 & 19.4 & 0.83 & 2.9 & 0.20\\
3C47 & 194.8 & 105.7 & 6.80 & 0.83 & 1.410 & 229.4 & 219.4 & 12.0 & 0.70 & 1.741 & 0.08143\\
3C55 & 261.5 & 261.5 & 10.9 & 0.85 & 6.46 & 248.1 & 238.3 & 4.4 & 0.88 & 5.74 & 0.02644\\
3C61.1 & 236 & 236 & 13.1 & 1.00 & 4.42 & 327.3 & 327.3 & 13.9 & 1.00 & 3.35 & 0.162\\
3C67 & 9.25 & 9.25 & 0.3 & 1.00 & 2.03 & 5.93 & 5.93 & 0.4 & 1.00 & 0.942 & 0.219\\
3C79 & 160.0 & 160.0 & 11.3 & 1.00 & 1.585 & 206.2 & 206.2 & 11.0 & 0.99 & 2.324 & 0.1262\\
3C98 & 85.38 & 85.38 & 1.37 & 1.00 & 1.992 & 83.62 & 83.62 & 0.43 & 0.99 & 2.561 & 0.01040\\
3C109 & 206.2 & 206.2 & 16.0 & 0.77 & 2.183 & 223.8 & 223.8 & 18.3 & 1.00 & 2.388 & 0.04081\\
4C14.11 & 188.2 & 188.2 & 12.4 & 0.99 & 1.972 & 186.6 & 186.6 & 10.1 & 0.99 & 2.817 & 0.004324\\
3C123 & 62.30 & 28.7 & 3.81 & 0.58 & 1.077 & 47.85 & 34.43 & 4.62 & 0.73 & 0.9483 & 0.1312\\
3C132 & 39.85 & 39.85 & 2.5 & 1.00 & 5.07 & 37.56 & 37.56 & 2.4 & 0.82 & 1.59 & 0.02965\\
3C153 & 20 & 20 & 1.1 & 0.57 & 1.4 & 14.3 & 14.3 & 1.6 & 1.00 & 1.73 & 0.16\\
3C171 & 18.8 & 18.6 & 2.9 & 1.00 & 0.596 & 19.1 & 18.3 & 2.8 & 0.96 & 0.335 & 0.00795\\
3C172 & 285 & 285 & 23.3 & 0.70 & 2.53 & 299 & 299 & 14.6 & 0.74 & 4.00 & 0.0246\\
3C173.1 & 119.3 & 106.4 & 9.05 & 1.00 & 2.460 & 141.3 & 141.3 & 9.79 & 0.83 & 1.836 & 0.08472\\
3C175 & 159.9 & 159.9 & 9.77 & 0.58 & 2.47 & 232.3 & 214.4 & 10.8 & 0.44 & 3.42 & 0.1846\\
3C175.1 & 33.9 & 33.9 & 4.3 & 0.59 & 2.39 & 27.0 & 27.0 & 4.3 & 0.72 & 1.58 & 0.113\\
3C184 & 24.2 & 24.2 & 4.6 & 1.00 & 1.78 & 13.4 & 13.4 & 4.6 & 0.84 & 1.11 & 0.289\\
3C184.1 & 226.4 & 226.4 & 5.01 & 0.87 & 5.36 & 169 & 169 & 5.01 & 0.88 & 3.69 & 0.144\\
DA240 & 636.5 & 636.5 & 17.02 & 0.88 & 1.163 & 743.4 & 743.4 & 8.50 & 0.89 & 2.46 & 0.07744\\
3C192 & 126.8 & 126.8 & 3.17 & 0.83 & 5.23 & 107 & 107 & 2.3 & 0.98 & 3.43 & 0.0851\\
3C196 & 19.7 & 19.7 & 4.2 & 0.81 & 0.490 & 28.5 & 28.5 & 3.9 & 0.86 & 0.636 & 0.183\\
3C200 & 64.34 & 63.64 & 1.3 & 0.97 & 1.6 & 82.31 & 80.22 & 2.2 & 0.60 & 2.12 & 0.1226\\
4C14.27 & 117.1 & 117.1 & 6.74 & 0.99 & 3.64 & 80.35 & 80.35 & 6.47 & 1.00 & 1.514 & 0.1860\\
3C207 & 40.7 & 40.6 & 3.3 & 0.52 & 2.13 & 50.1 & 47.3 & 3.0 & 0.59 & 1.65 & 0.104\\
3C215 & 142.9 & 139.6 & 4.3 & 1.00 & 1.301 & 75.65 & 75.65 & 6.28 & 1.00 & 0.4254 & 0.3078\\
3C217 & 75.75 & 75.05 & 4.2 & 0.73 & 5.93 & 27.9 & 27.9 & 3.9 & 0.98 & 1.36 & 0.461\\
3C216 & 6.45 & 6.24 & 1.7 & 1.00 & 0.829 & 14.9 & 8.20 & 1.2 & 1.00 & 2.4 & 0.397\\
3C219 & 275.0 & 240.0 & 2.1 & 0.89 & 1.597 & 291.1 & 291.1 & 2.1 & 0.92 & 3.491 & 0.02848\\
3C220.1 & 98.49 & 98.49 & 10.1 & 0.79 & 1.77 & 117.9 & 117.4 & 8.89 & 0.71 & 1.556 & 0.08966\\
3C220.3 & 30.5 & 30.5 & 2.6 & 1.00 & 1.02 & 39.7 & 39.7 & 2.8 & 0.56 & 1.26 & 0.132\\
3C223 & 374.6 & 368.1 & 4.79 & 0.77 & 3.915 & 363.8 & 363.8 & 5.66 & 0.55 & 3.938 & 0.01461\\
3C225B & 13.8 & 13.2 & 4.1 & 0.73 & 1.40 & 18.9 & 18.6 & 3.8 & 0.60 & 1.95 & 0.155\\
3C226 & 165.9 & 113.3 & 8.55 & 0.83 & 1.440 & 134.6 & 134.6 & 8.55 & 0.99 & 2.21 & 0.1041\\
4C73.08 & 661.1 & 516.2 & 23.86 & 0.80 & 1.371 & 470.3 & 386.1 & 44.64 & 1.00 & 0.9753 & 0.1687\\
3C228 & 163.6 & 163.6 & 11.9 & 0.76 & 3.89 & 143.8 & 136.4 & 11.9 & 0.92 & 2.156 & 0.06438\\
3C234 & 199.2 & 199.2 & 10.1 & 1.00 & 5.711 & 95.87 & 95.87 & 10.7 & 0.98 & 4.95 & 0.3501\\
3C236 & 1746 & 1746 & 87.66 & 0.78 & 2.932 & 2628 & 2151 & 111.4 & 0.64 & 7.880 & 0.2016\\
4C74.16 & 188.4 & 188.4 & 11.3 & 0.93 & 3.85 & 133.7 & 123.9 & 10.6 & 1.00 & 1.726 & 0.1699\\
3C244.1 & 152.0 & 152.0 & 7.16 & 0.75 & 5.00 & 137.3 & 134.3 & 6.10 & 0.91 & 5.27 & 0.05086\\
3C247 & 43.1 & 43.1 & 4.2 & 0.59 & 5.29 & 60.9 & 60.9 & 4.3 & 0.50 & 5.33 & 0.172\\
3C249.1 & 136.0 & 32.1 & 7.58 & 1.00 & 1.752 & 86.43 & 86.43 & 4.89 & 0.97 & 1.535 & 0.2229\\
3C254 & 92.66 & 92.66 & 3.0 & 0.33 & -- & 33.3 & 16.6 & 3.1 & 1.00 & 1.33 & 0.472\\
3C263 & 226.5 & 226.5 & 8.81 & 0.49 & 5.78 & 114.5 & 114.5 & 12.2 & 1.00 & 1.518 & 0.3284\\
3C263.1 & 20.5 & 20.5 & 3.8 & 0.67 & 1.56 & 46.7 & 46.7 & 2.8 & 0.90 & 2.71 & 0.391\\
3C265 & 350.0 & 347.7 & 8.45 & 0.56 & 6.96 & 234.6 & 234.6 & 12.6 & 0.75 & 4.31 & 0.1974\\
3C268.1 & 188.0 & 188.0 & 12.9 & 0.54 & 8.59 & 151.0 & 151.0 & 10.1 & 0.44 & 4.22 & 0.1091\\
3C268.3 & 2.6 & 2.6 & 0.4 & 0.76 & 1.4 & 6.51 & 6.26 & 0.4 & 0.50 & 3.1 & 0.43\\
3C274.1 & 491.0 & 491.0 & 1.9 & 0.81 & 5.113 & 429.8 & 423.4 & 5.55 & 0.82 & 4.196 & 0.06652\\
3C275.1 & 78.46 & 75.42 & 1.4 & 1.00 & 8.11 & 43.0 & 43.0 & 2.1 & 0.71 & 1.76 & 0.292\\
3C277.2 & 315.2 & 300.3 & 7.84 & 0.56 & 5.90 & 123.6 & 123.6 & 12.0 & 0.85 & 2.15 & 0.4367\\
3C280 & 75.37 & 70.4 & 4.6 & 0.83 & 1.22 & 59.4 & 59.4 & 3.4 & 0.67 & 1.38 & 0.118\\
3C284 & 415.7 & 415.7 & 22.4 & 0.96 & 3.659 & 284.9 & 284.9 & 23.4 & 0.95 & 2.508 & 0.1867\\
3C285 & 120 & 110 & 7.72 & 1.00 & 1.3 & 145 & 145 & 8.25 & 0.96 & 1.1 & 0.094\\
3C289 & 43.5 & 43.5 & 4.2 & 0.73 & 2.41 & 40.9 & 40.9 & 4.6 & 0.70 & 2.18 & 0.0301\\
3C292 & 450.5 & 432.3 & 5.7 & 0.50 & -- & 502.5 & 502.5 & 9.49 & 0.21 & -- & 0.05459\\
3C295 & 13.6 & 13.6 & 1.2 & 0.96 & 1.13 & 19.7 & 19.7 & 1.2 & 0.74 & 1.23 & 0.186\\
3C299 & 15.3 & 14.9 & 3.1 & 0.80 & 1.80 & 44.5 & 44.5 & 2.8 & 0.36 & -- & 0.487\\
3C300 & 284.5 & 284.5 & 19.5 & 1.00 & 4.562 & 117.3 & 109.5 & 26.9 & 1.00 & 1.203 & 0.4162\\
3C303 & 69.01 & 64.8 & -- & -- & 1.36 & 51.63 & 42.0 & -- & -- & 0.9881 & 0.1440\\
3C319 & 156.2 & 156.2 & 8.95 & 1.00 & 1.955 & 183.9 & 183.9 & 2.5 & 0.83 & 3.636 & 0.08156\\
3C321 & 264.3 & 264.3 & -- & -- & -- & 266.8 & 259.5 & -- & 0.15 & -- & 0.004689\\
3C325 & 81.40 & 81.40 & 4.4 & 1.00 & 4.19 & 58.6 & 58.6 & 4.5 & 0.97 & 1.81 & 0.162\\
3C326 & 688.2 & 665.8 & 72.84 & 0.87 & 1.087 & 1253 & 1253 & 76.42 & 1.00 & 5.657 & 0.2910\\
3C330 & 195.8 & 195.8 & 26.3 & 0.55 & 3.52 & 202.7 & 202.7 & 23.6 & 0.77 & 3.39 & 0.01752\\
3C334 & 207.1 & 197.7 & 9.08 & 0.61 & 2.546 & 147.2 & 128.2 & 7.41 & 0.65 & 1.241 & 0.1690\\
3C336 & 119.2 & 119.2 & 13.0 & 1.00 & 3.20 & 82.17 & 82.17 & 8.80 & 1.00 & 1.33 & 0.1841\\
3C341 & 258 & 258 & 10.6 & 1.00 & 5.4 & 191 & 191 & 12.2 & 1.00 & 4.3 & 0.149\\
3C337 & 197.5 & 197.5 & 10.8 & 0.76 & 3.12 & 109.7 & 107.3 & 11.4 & 1.00 & 1.66 & 0.2860\\
3C340 & 174.0 & 170.7 & 10.8 & 0.82 & 3.10 & 153.1 & 153.1 & 12.0 & 0.80 & 2.44 & 0.06403\\
3C349 & 136.1 & 136.1 & 12.9 & 0.98 & 3.539 & 150.8 & 150.8 & 15.0 & 1.00 & 2.615 & 0.05125\\
3C351 & 138.8 & 129.7 & 24.1 & 1.00 & 0.7257 & 186.7 & 186.7 & 17.0 & 1.00 & 1.833 & 0.1471\\
3C352 & 53.7 & 49.8 & 4.1 & 0.66 & 1.36 & 47.3 & 46.5 & 4.2 & 0.89 & 1.93 & 0.0641\\
3C381 & 93.65 & 93.65 & 7.08 & 0.90 & 1.824 & 108.7 & 108.7 & 5.86 & 0.83 & 2.502 & 0.07434\\
3C382 & 99 & 99 & 0.78 & 1.00 & 2.4 & 100 & 96 & 0.43 & 0.46 & 1.7 & 0.018\\
3C388 & 47.2 & 37.4 & -- & -- & 2.31 & 38.1 & 34.7 & -- & -- & 1.47 & 0.107\\
3C390.3 & 142.4 & 142.4 & 3.65 & 0.85 & 3.658 & 101 & 101 & 4.14 & 0.70 & 1.33 & 0.170\\
3C401 & 36.90 & 36.27 & 1.7 & 1.00 & 1.57 & 42.86 & 42.86 & 2.0 & 0.85 & 1.82 & 0.07475\\
3C427.1 & 82.2 & 82.2 & 4.1 & 0.73 & 2.5 & 95.5 & 93.8 & 2.9 & 0.64 & 3.7 & 0.0747\\
3C433 & 75.8 & 44.8 & 0.47 & 0.5 & 5.4 & 60.7 & 60.7 & 0.47 & 1.0 & 2.40 & 0.111\\
3C436 & 201.4 & 201.4 & 9.58 & 0.99 & 2.262 & 173.2 & 166.6 & 9.58 & 1.00 & 2.131 & 0.07523\\
3C438 & 51.12 & 48.64 & 7.31 & 0.98 & 1.47 & 44.90 & 44.90 & 7.31 & 1.00 & 1.39 & 0.06479\\
3C441 & 78.39 & 72.72 & 6.5 & 1.00 & 1.22 & 183 & 183 & 7.90 & 0.75 & 3.22 & 0.400\\
3C452 & 208.8 & 199.4 & 4.88 & 1.00 & 2.385 & 218.8 & 217.5 & 3.58 & 0.99 & 2.559 & 0.02324\\
3C455 & 9.04 & 9.04 & 3.7 & 1.00 & 0.568 & 17.7 & 17.7 & 3.6 & 1.00 & 1.29 & 0.324\\
3C457 & 596.0 & 558.1 & 5.3 & 0.56 & 5.004 & 567.8 & 536.5 & 4.9 & 0.47 & 3.745 & 0.02428\\

\hline
\end{longtable}

\label{lastpage}

\end{document}